\documentclass[nofootinbib]{revtex4-2}
\usepackage[utf8]{inputenc}
\usepackage[english]{babel}

\usepackage[hidelinks]{hyperref}

\usepackage{comment}

\usepackage{graphicx}
\usepackage{subcaption}
\usepackage{parskip}

\usepackage{float}
\usepackage{booktabs}
\usepackage{makecell}
\usepackage{multirow}

\usepackage[figurename=Fig.]{caption}
\usepackage[tablename=Tab.]{caption}
\usepackage[font=small]{caption}

\usepackage{amsmath}
\usepackage{amssymb}
\usepackage{amsfonts}
\usepackage{mathtools}

\usepackage{upgreek}
\usepackage{xcolor}
\usepackage[normalem]{ulem}
\usepackage{mathrsfs}

\newcommand{\bh}[1]{{\textcolor{blue}{{#1}}}}

\begin{document}

\title{Photon rings and shadows of Kerr black holes immersed in a swirling universe}

\date{\today}

\author{Rog\'erio Capobianco}
 \email{rogerio.capobianco@gmail.com}
\affiliation{Instituto de F\'isica de S\~ao Carlos, Universidade de S\~ao Paulo, S\~ao Carlos, S\~ao Paulo 13560-970, Brazil $\&$\\
Institut f\"ur Physik, Carl-von-Ossietzky Universit\"at Oldenburg, 26111 Oldenburg, Germany}

\author{Betti Hartmann}
 \email{b.hartmann@ucl.ac.uk}
\affiliation{Department of Mathematics, University College London, Gower Street, London, WC1E 6BT, UK}

\author{Jutta Kunz}
 \email{jutta.kunz@uni-oldenburg.de}
\affiliation{Institut f\"ur Physik, Carl-von-Ossietzky Universit\"at Oldenburg, 26111 Oldenburg, Germany}

\author{Jo\~ao Novo}
 \email{j.novo@ua.pt}
\affiliation{Departamento de Matem\'atica da Universidade de Aveiro and Center for Research and Development in Mathematics and Applications (CIDMA),
Campus de Santiago, 3810-193 Aveiro, Portugal}

\author{Nikhita Vas}
 \email{nikhita.vas.18@ucl.ac.uk}
\affiliation{Department of Mathematics, University College London, Gower Street, London, WC1E 6BT, UK}

\begin{abstract}
   We discuss photon rings around as well as shadows of Kerr black holes immersed in a swirling spacetime (KBHSU). We find that the spin-spin interaction between the angular momentum of the black hole and the swirling of the background leads to new interesting effects as it breaks the symmetry between the upper and lower hemispheres. We find that a pair of light rings exists for all values of the parameter space. Using a topological argument, we prove that there should be, indeed, two light rings and that, additionally, these light rings are unstable.  In comparison to the Schwarzschild black hole immersed in a swirling universe, the light rings typically all possess different radii. Interestingly, as the value of the swirling parameter is increased at fixed angular momentum of the black hole the two disconnected patches of the ergoregions eventually merge. The light ring at this merger possesses no angular velocity (as measured by an observer at infinity) and is called a \textit{light point}. To our knowledge, this is the first time the existence of such a light point in a black hole space-time is reported. Finally, we also present the shadows of KBHSU for various parameter values and observe that, due to the presence of the swirling background, the shadows are twisted. 
\end{abstract}

\maketitle

\section{Introduction}

Obtaining exact solutions to the Einstein equation is no easy task.
Such solutions can only be found once symmetries are imposed. 
Exact solutions of the Einstein equation are of particular importance, since these allow us to study the effects of gravity using analytical techniques. Rotation is a characteristic observed in all celestial objects in our universe, and the Kerr spacetime is an exact solution representing a stationary rotating black hole (BH). 
Other exact, rotating solutions are, for instance,
the G\"odel Universe \cite{Godel:1949ga}, where homogeneous dust is rotating around every point, or the Taub-NUT solution \cite{Taub:1950ez,Newman:1963yy}, where the NUT charge gives a rotational sense to the spacetime. 

Recently, a new exact, rotating solution has been constructed using the Ernst formalism \cite{Astorino:2022aam}. 
This solution describes a rotating Kerr black hole in a swirling universe \cite{Harrison:1968wue, Gibbons:2013yq}.
The swirling universe, i.e. the solution without the black hole, is a vacuum solution equipped with an odd-$\mathbb{Z}_2$ symmetry: 
the upper ($\theta\in [0:\pi/2[$ ) and lower hemispheres ($\theta\in \
]\pi/2:\pi]$) of the spacetime rotate in opposite directions. 
The swirling universe can, in fact, be obtained by using the Lie point symmetry through the application of the Ehlers transformation. Note that the geodesic motion can be decoupled and completely integrated in this spacetime \cite{Capobianco:2023kse}. 
An interesting question is whether this solution is of 
potential observational relevance, and which new features appear that allow distinction from the ``pure" Kerr spacetime.
An analysis of accretion discs in this spacetime has already been performed, revealing their bowl-like structure \cite{Gjorgjieski:2024cko}. 
Recently, the interest in generating techniques has grown again, and different new solutions including the swirling background have been constructed. In particular, external electromagnetic fields were considered in a composition of Harrison and Ehlers' transformation \cite{DiPinto:2024axv, Barrientos:2024pkt}, as well as charged black holes, i.e. Kerr-Newman black holes immersed in the swirling background \cite{Astorino:2022prj,DiPinto:2024axv,DiPinto:2025yaa}. Furthermore, new solutions can be obtained by exploring different symmetries. Recently, the \textit{inversion transformation} has been used to produce the Kerr-Levi-Civita spacetime \cite{Barrientos:2025rjn}.

In this work we aim to contribute to the exploration of the phenomenological imprints of black holes immersed in the swirling universe.
A main objective of this paper is to study the existence and location of circular photon orbits, also known as \textit{light rings} (LRs), in the spacetime of a Kerr black hole immersed in a swirling universe (KBHSU). 
The motion of light-like particles is one of the main probes of a curved spacetime. 
Furthermore, null geodesics can be used to study gravitational waves at very large angular momenta, i.e. in the so-called eikonal limit, and a link between quasi-normal mode frequencies and unstable light rings can be drawn \cite{Cardoso:2008bp}.
Furthermore, all BHs formed from stellar collapse, that is, stationary, axially symmetric, and asymptotically flat solutions in $3+1$ dimensions with a non-extremal, topologically spherical Killing horizon must possess at least one circular photon orbit outside the horizon, thus
LRs are present in many black hole spacetimes, such as the Schwarzschild, Reissner-Nordstr\"om, Kerr, and Kerr-Newman spacetimes  
\cite{Cunha:2020azh}, as well as in spacetimes of other objects, that are dubbed ultracompact objects (UCOs) precisely because they feature LRs \cite{Iyer:1985,Cunha:2017qtt,Cardoso:2019rvt}.

The LRs of Schwarzschild black holes in a swirling universe (SBHSU) have previously been investigated in \cite{Moreira:2024sjq}.
Because of the swirling motion, the degenerate light ring of the Schwarzschild black hole splits into two LRs that are located symmetrically in planes above and below the equatorial plane.
Whereas the LRs counter-rotate with respect to each other, each of them co-rotates with respect to the swirling background universe.
Here we investigate how the additional angular momentum of the Kerr black hole affects the location, the rotational sense, and the number of LRs. 

A related objective is the calculation of the shadows of the Kerr black holes in a swirling universe.
While there has been work on photon spheres and shadows of rotating black holes for decades (see e.g.~\cite{Bardeen:1972fi,Claudel:2000yi,Teo:2003ltt,Hioki:2009na}), it is the observations of M87$^*$ and Sgr A$^*$ by the EHT collaboration \cite{EventHorizonTelescope:2019dse,EventHorizonTelescope:2022wkp} that make the study of shadows most relevant.
Recently, the shadow of a SBHSU has been investigated in \cite{Moreira:2024sjq},
revealing the interesting new feature, that the shadow became twisted, i.e., it inherited the odd-$\mathbb{Z}_2$ symmetry of the swirling universe with respect to the equatorial plane.
This symmetry is broken by the spin of the Kerr black hole. 
Here, we investigate how this affects the shadow.

We also address further properties of the KBHSU spacetime such as its conical singularities and its ergoregions. 
Our studies show that the value $ajM=0.25$, i.e. the combination of the swirling parameter $j$, the (seed) angular momentum parameter $a$, and the mass parameter $M$ of the seed Kerr black hole, is a critical value for the properties of the spacetime.

This manuscript is organised as follows: in section 2, we recall the metric of the KBHSU spacetime, derive the conical singularities, and determine the ergoregions.
In section 3, we address the motion of light-like particles, evaluate the LRs and determine their dependence on the parameters $j$ and $a$ of the spacetime.
We obtain the shadows in section 4, and conclude in section 5.

We have adopted geometrized units, i.e., $G = c = 1$. The swirling parameter has dimensions of $[j] = M^{-2}$. We will choose $M=1$ in the following, unless otherwise stated.

\section{The spacetime and its properties}

The metric tensor describing the KBHSU can be generated by applying an Ehlers transformation on a Kerr black hole seed solution within the Ernst formalism \cite{Astorino:2022aam}.
In Boyer-Lindquist coordinates, it reads \footnote{This is the metric as given in the auxiliary files provided with \cite{Astorino:2022aam}.}~:

\begin{equation}
\label{eq:metric}
    {\rm d}s^2 = \frac{1}{\mathcal{F}(r,\theta)} \left( {\rm d}\varphi - \omega {\rm d}t  \right)^2 + \mathcal{F}(r,\theta) \left[ -\rho^2 {\rm d}t^2 + \Sigma \sin^2\theta \left( \frac{{\rm d}r^2}{\Delta} + {\rm d}\theta^2 \right) \right],
\end{equation}

where the functions $\mathcal{F}$ and $\omega$ are expressed as a second-order series in the swirling parameter $j$ as follows~:

\begin{equation}
    \mathcal{F} = \frac{\mathcal{F}_0 + j \mathcal{F}_1 + j^2 \mathcal{F}_2}{R^2 \Sigma \sin^2\theta} \ \ \ \ \ , \ \ \ \ \ 
    \omega = \frac{\omega_0 + j \omega_1 + j^2 \omega_2}{\Sigma} \ ,
\end{equation}
with
\begin{eqnarray}
    \mathcal{F}_0 &=& R^4, \ \ \ \quad 
    \mathcal{F}_1 = 4 a M \Xi \cos\theta R^2, \ \ \ \quad 
    \mathcal{F}_2 = 4 a^2 M^2 \Xi^2 \cos^2\theta + \Sigma^2 \sin^4\theta  \nonumber
    \\
    \omega_0 &=& 2 a M r, \ \ \ \quad 
 \omega_1 = - 4\cos\theta \left( (r^3 -a^2 M)\Delta + a(r-M) \mathcal{W}\right)
\nonumber  \\
    \omega_2 &=& 2M \left[ 3ar^5 - a^5(r+2M) + 2a^3r^2(r+3M) -r^3 (\cos^2\theta -6)\mathcal{W} + a^2 \mathcal{W} \left( \cos^2\theta(3r-2M) + 6(r-M) \right) \right]
\end{eqnarray}
and abbreviations
\begin{eqnarray}
    \Delta &=& r^2-2Mr+a^2, \ \ \ \qquad \ \ \ 
    \rho^2 = \Delta \sin^2\theta, \ \ \ \qquad \ \ \ 
    \Sigma = \left( r^2+a^2 \right)^2 - \Delta a^2 \sin^2\theta, \nonumber 
    \\ 
    \mathcal{W} &=& \Delta a \cos^2\theta, \ \ \ \qquad 
    \Xi = r^2 \left( \cos^2\theta -3 \right) -a^2 \left( 1+\cos^2\theta \right), \ \ \ \qquad \ \ \ 
    R^2 = r^2 + a^2 \cos^2\theta. \nonumber
\end{eqnarray}

Here, the quantities $M$ and $a=J/M$ are inherited from the Kerr seed solution\footnote{Note that $J$ is the seed black hole's total angular momentum that has no link to the swirling parameter.}. We will refer to these parameters as ``the mass parameter'' and ``the angular momentum parameter'' in the following, respectively.
For $j=0$ the above metric describes the Kerr spacetime, while for $a=0$ the line element corresponds to that of a Schwarzschild black hole
in a swirling universe. 
When the quantities originating from the black hole vanish, i.e., for $a=M=0$, we recover the pure swirling background. 

The compact object still possesses two horizons 
which, in these coordinates, are obtained by $\Delta = 0$, i.e., $r_h^{(\pm)}=M\pm \sqrt{M^2 -a^2}$. 
Note that, although the locations of the horizons are independent of $j$ and mapped on the same surfaces as the Kerr seed solution, the geometry of the horizons when considering an isometric embedding  
deforms the horizon shape due to the background's frame dragging. 
However, despite the modifications in the event horizons' shape, the area remains unaffected \cite{Gjorgjieski:2024cko}.

\subsection{Conical singularity}
KBHSU spacetime is non-asymptotically flat and has a non-removable conical singularity. 
We can identify the existence of an axial angular deficit/excess by considering the ratio between the perimeter and radius of a small circle around the rotational axis at both the north and the south poles \cite{Astorino:2022aam,DiPinto:2024axv,DiPinto:2025yaa} 
\begin{equation}
    \delta_0=\lim_{\theta\to 0}\frac{1}{\theta}\int_{0}^{2\pi} \sqrt\frac{g_{\varphi\varphi}}{g_{\theta\theta}} \,d\varphi \qquad \text{and} \qquad 
    \delta_\pi=\lim_{\theta\to \pi}\frac{1}{\pi-\theta}\int_{0}^{2\pi} \sqrt\frac{g_{\varphi\varphi}}{g_{\theta\theta}} \,d\varphi.
\end{equation}
If $\delta_0=\delta_\pi=2\pi$, then there is no conical singularity. 
In the KBHSU spacetime, however, we have \cite{Astorino:2022aam}
\begin{equation}
    \delta_0=\frac{2\pi}{(1-4ajM)^2} \qquad \text{and} \qquad \delta_\pi=\frac{2\pi}{(1+4ajM)^2} ,
\end{equation}
for non-vanishing angular momentum parameter $a$ and finite $j$.
Thus the SBHSU does not suffer from conical singularities.
We note the divergence for the critical value $ajM=\pm 0.25$. To understand whether this divergence corresponds to a physical singularity we compute the Kretschmann scalar $K=R_{\mu\nu\sigma\gamma}R^{\mu\nu\sigma\gamma}$ of the solution when $j=\pm1/4aM$. The expressions are too lengthy and convoluted to be shown explicitly, but have the following form:
\begin{equation}
    \left.K\right|_{j=\pm1/4aM}=\frac{\mathcal{G}\left(r,\theta\right)}{\left(1\mp\cos\theta\right)^6}\,.
\end{equation}
Therefore, the solution is singular along the upper semiaxis, $\theta=0$, for $4ajM=1$, and along the lower semiaxis, $\theta=\pi$, for $4ajM=-1$.

\subsection{Ergoregions}

\begin{figure}[h!] 
    \centering

    \begin{subfigure}[b]{\textwidth}
        \centering
        \begin{subfigure}[b]{3.5cm}
            \includegraphics[width=3.5cm]{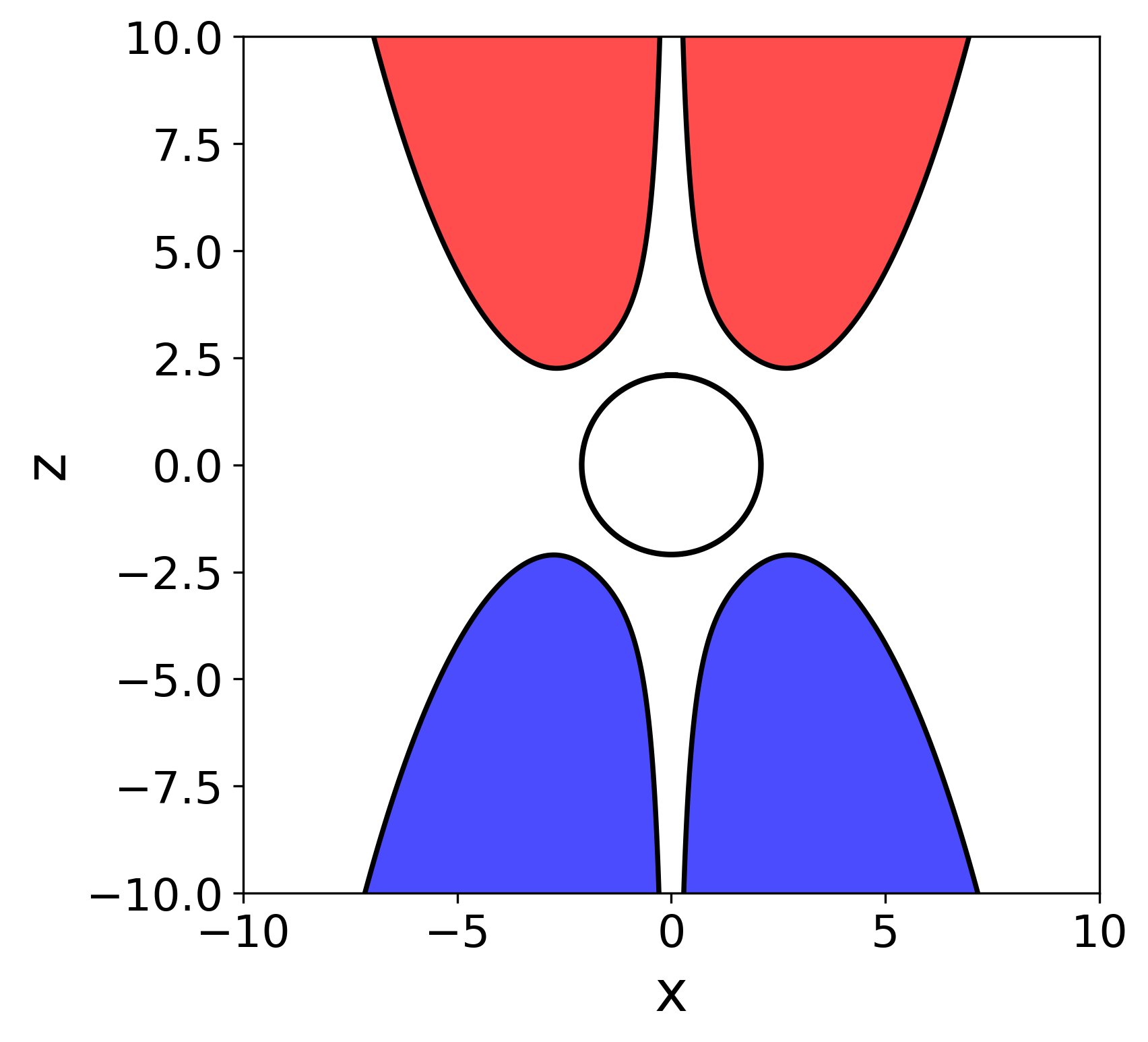}
            \caption*{$jM^2=0.1$, \protect\linebreak $ajM=0.01$ }
        \end{subfigure}
        \begin{subfigure}[b]{3.5cm}
            \includegraphics[width=3.5cm]{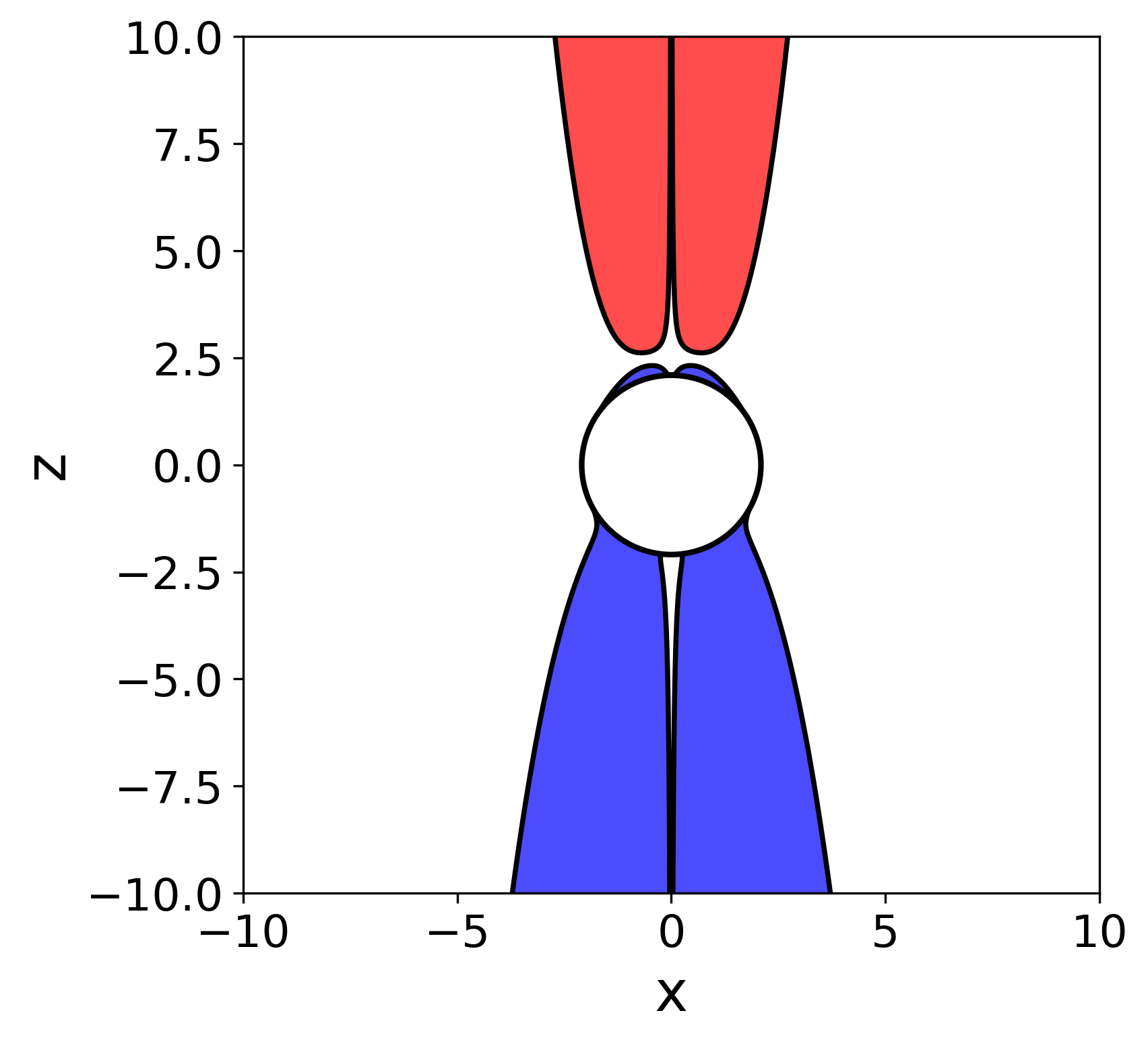}
            \caption*{$jM^2=1$, \protect\linebreak $ajM=0.1$}
        \end{subfigure}
        \begin{subfigure}[b]{3.5cm}
            \includegraphics[width=3.5cm]{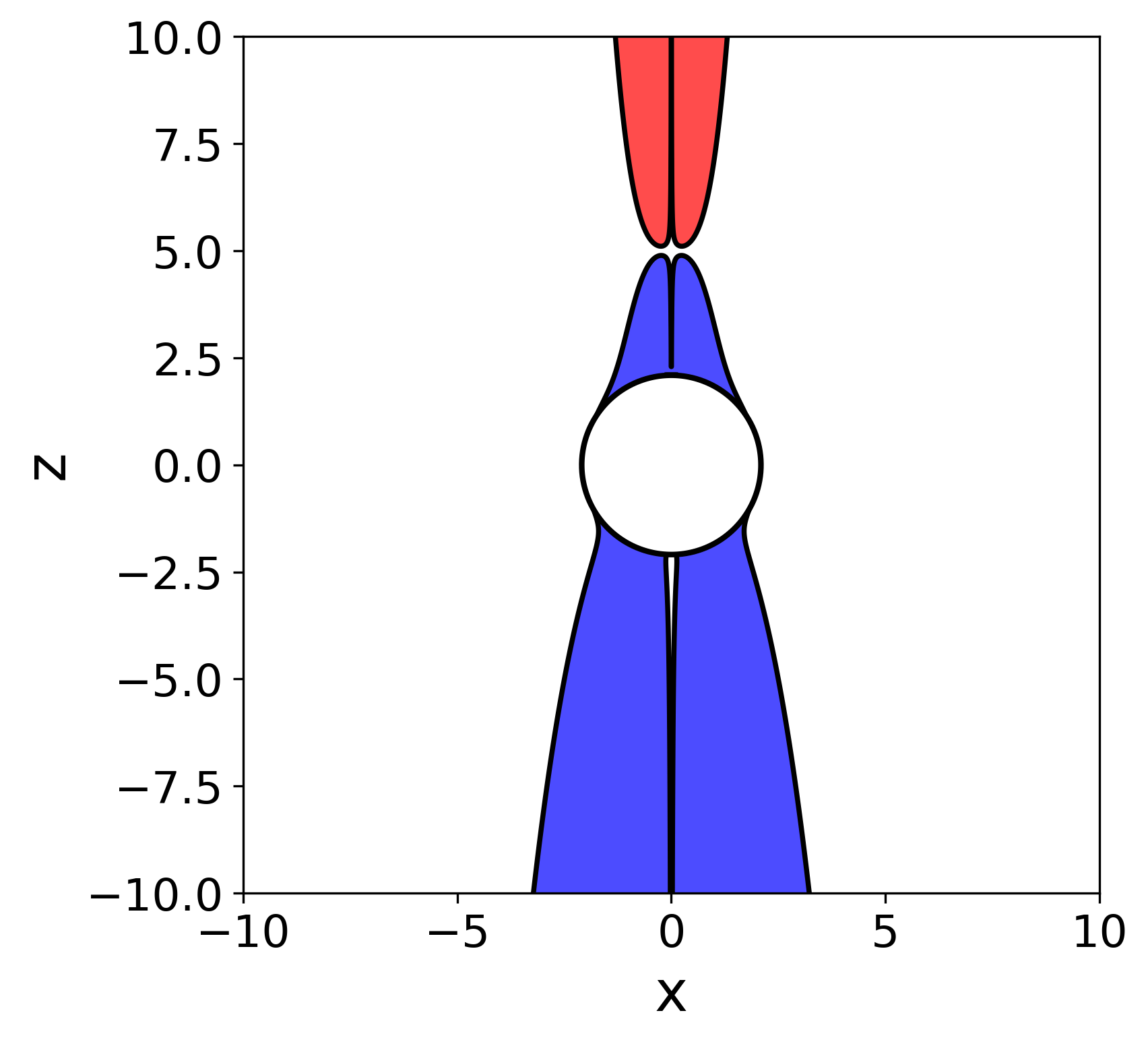}
            \caption*{$jM^2=2$, \protect\linebreak $ajM=0.2$}
        \end{subfigure}
        \begin{subfigure}[b]{3.5cm}
            \includegraphics[width=3.5cm]{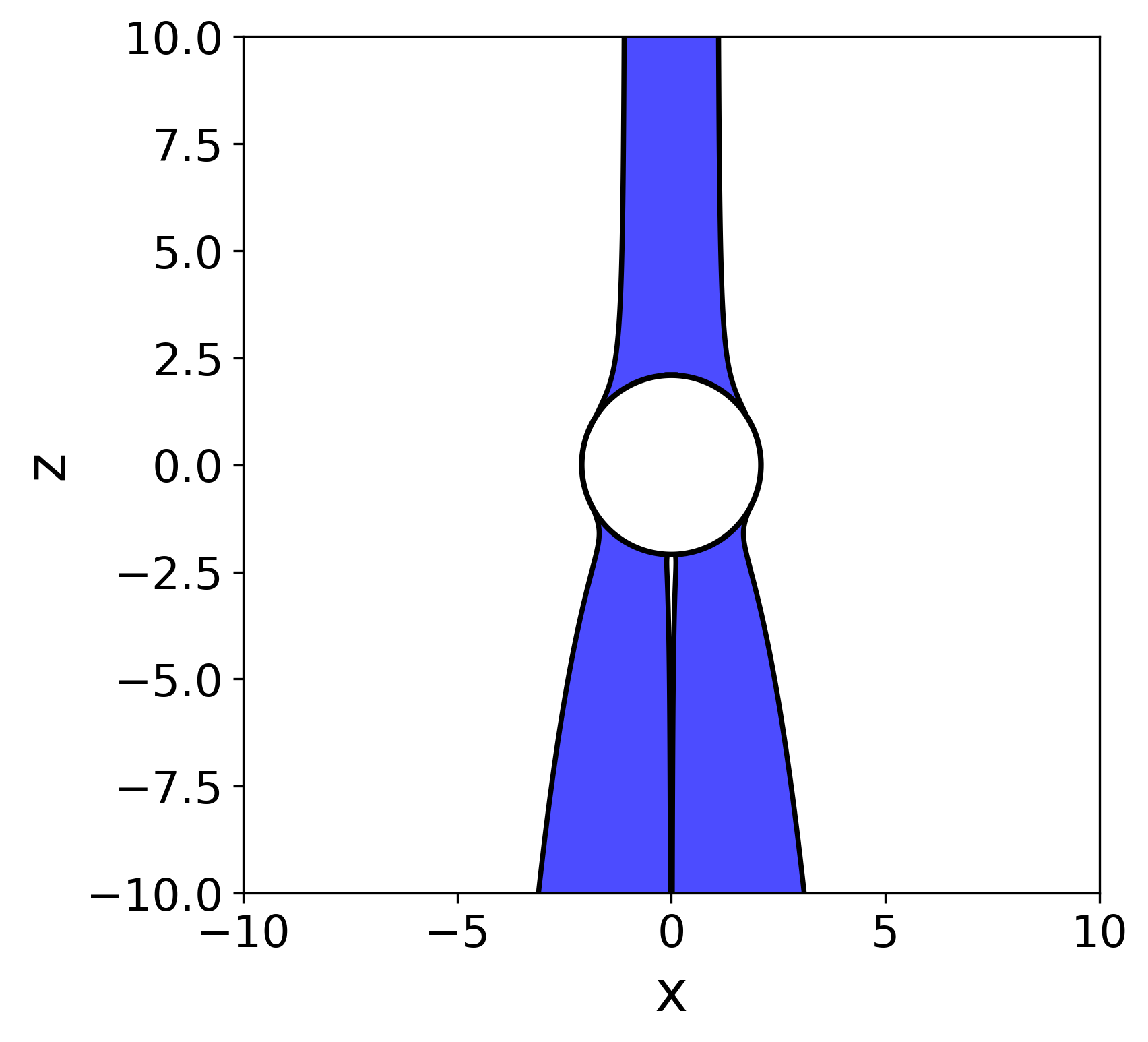}
            \caption*{$jM^2=2.5$, \protect\linebreak $ajM=0.25$}
        \end{subfigure}
        \begin{subfigure}[b]{3.5cm}
            \includegraphics[width=3.5cm]{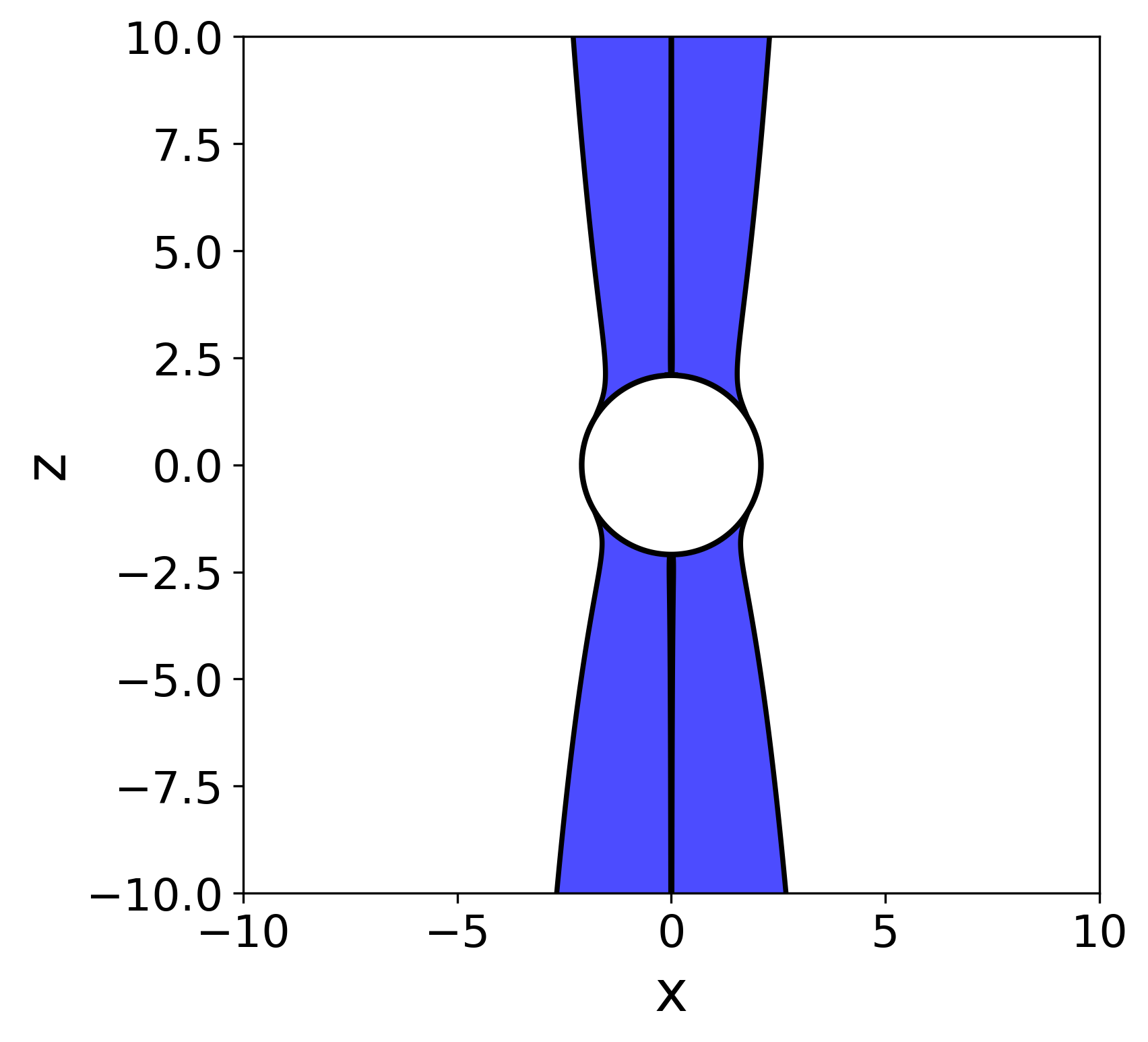}
            \caption*{$jM^2=10$,\protect\linebreak  $ajM=1$}
        \end{subfigure}
        \caption{$a=0.1M$}
        \label{fig:xz-ergoregion_a_01}
    \end{subfigure}
    
\hfill

    \begin{subfigure}[b]{\textwidth}
        \centering
        \begin{subfigure}[b]{3.5cm}
            \includegraphics[width=3.5cm]{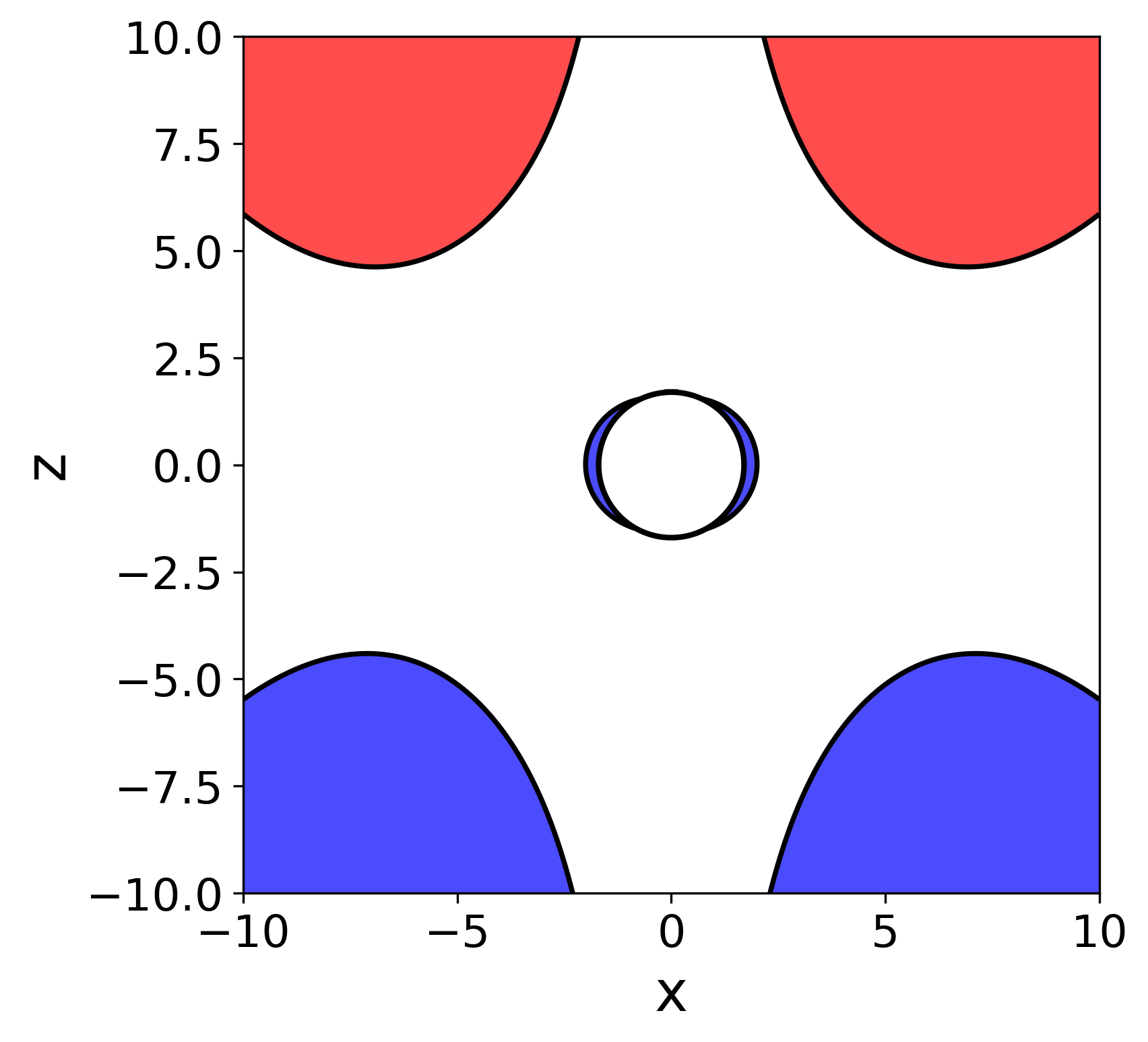}
            \caption*{$jM^2=0.0125$, \protect\linebreak $ajM=0.01$}
        \end{subfigure}
        \begin{subfigure}[b]{3.5cm}
            \includegraphics[width=3.5cm]{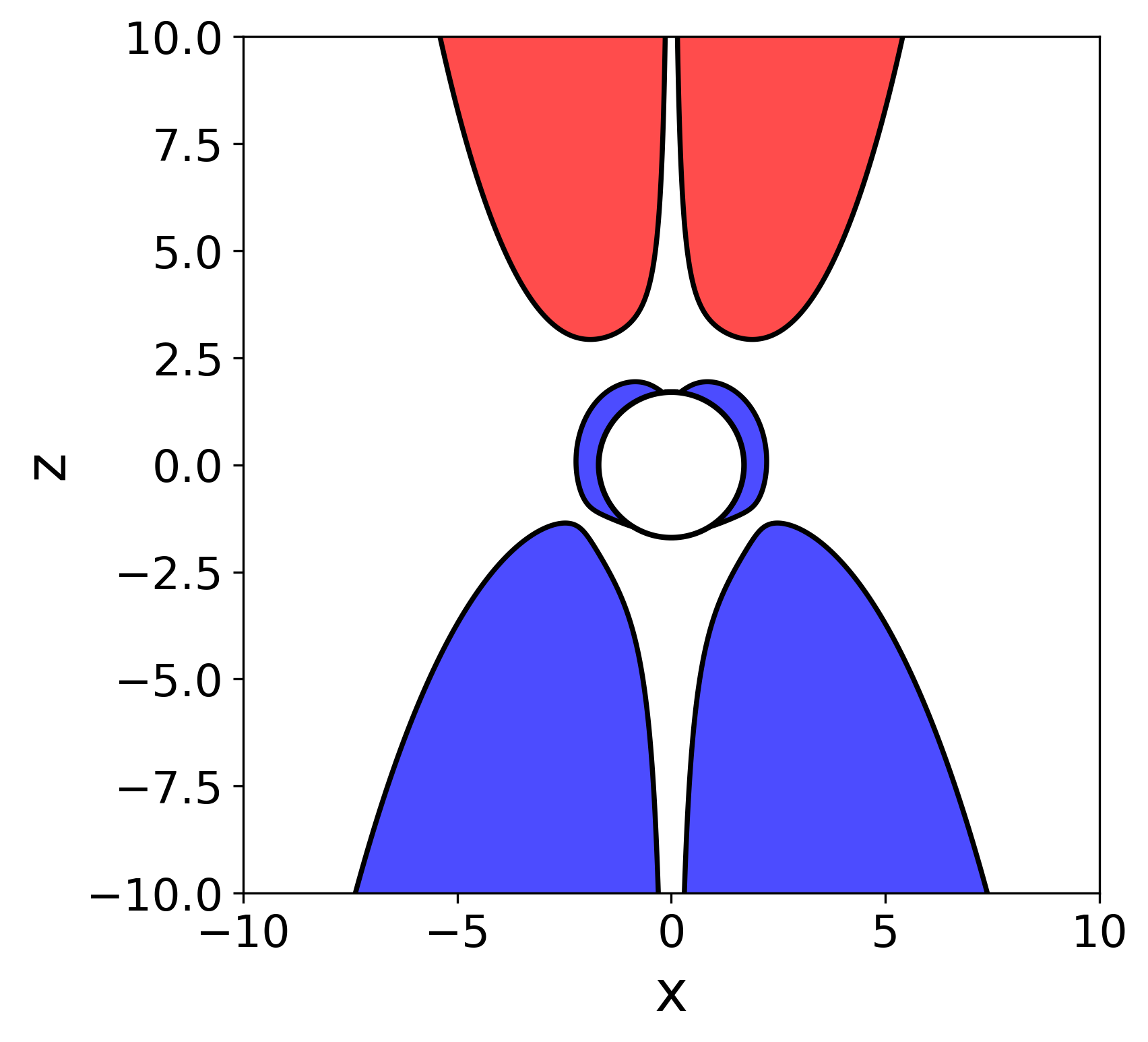}
            \caption*{$jM^2=0.125$, \protect\linebreak $ajM=0.1$}
        \end{subfigure}
        \begin{subfigure}[b]{3.5cm}
            \includegraphics[width=3.5cm]{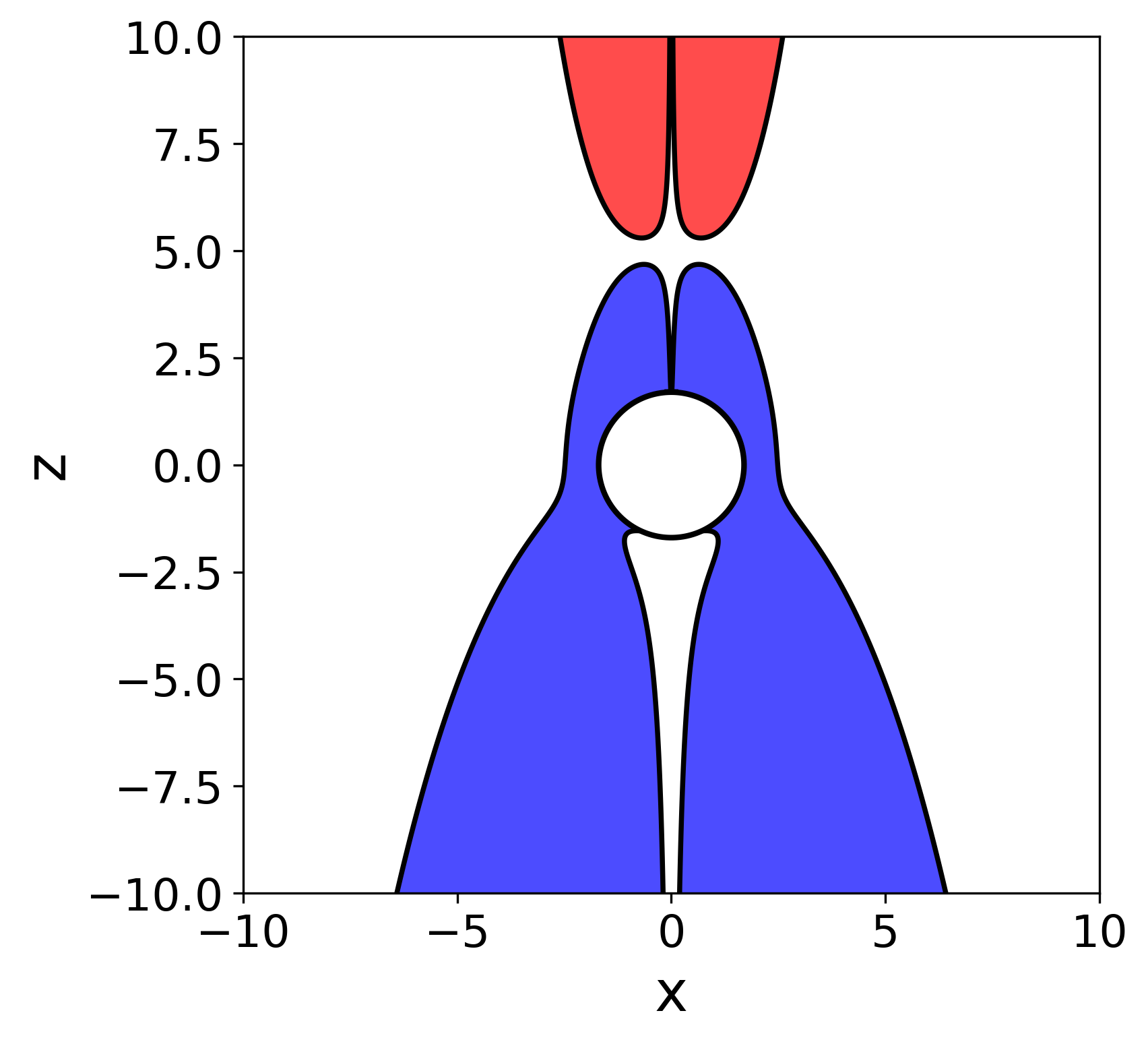}
            \caption*{ $jM^2=0.25$, \protect\linebreak $ajM=0.2$}
        \end{subfigure}
        \begin{subfigure}[b]{3.5cm}
            \includegraphics[width=3.5cm]{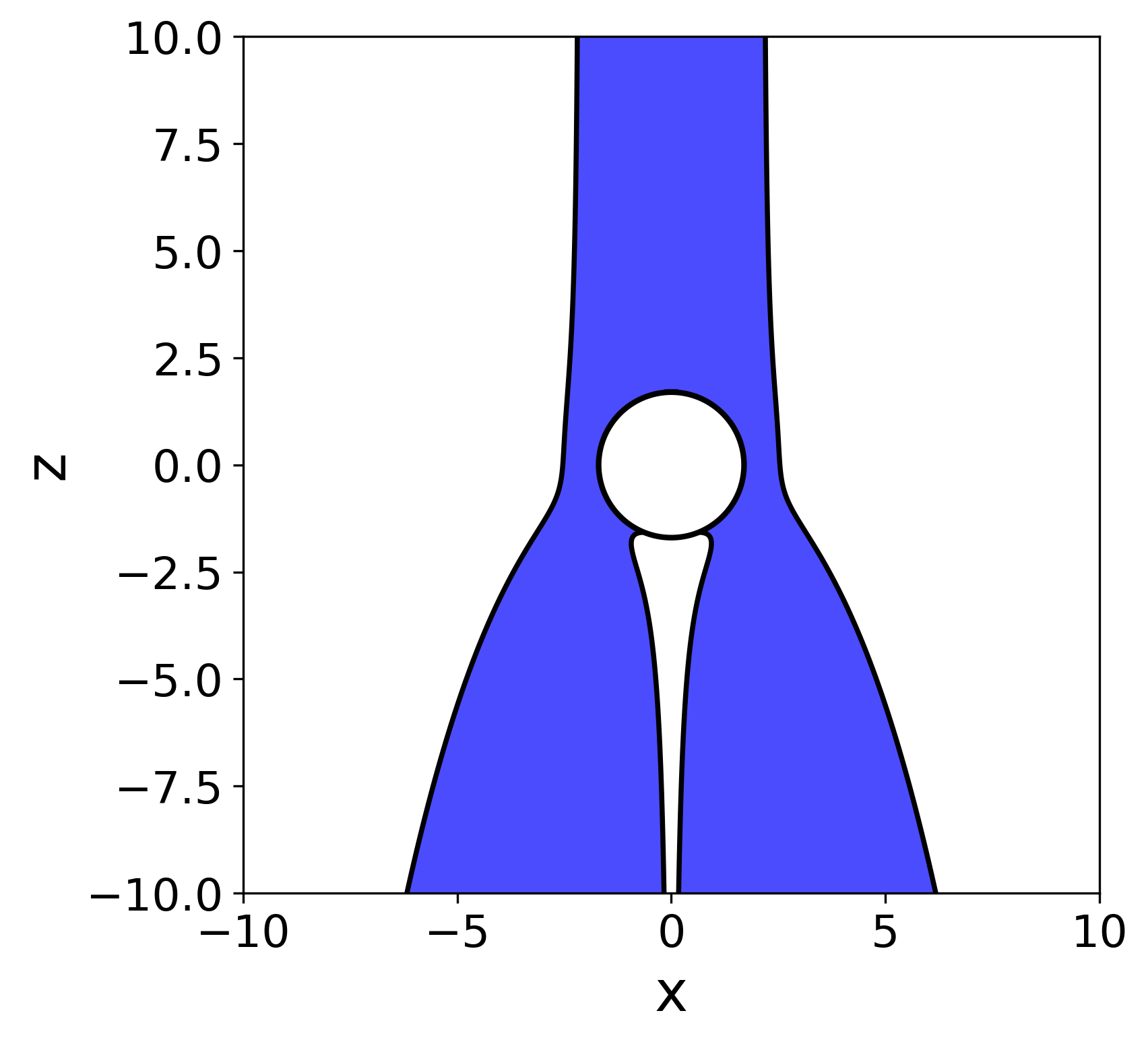}
            \caption*{ $jM^2=0.3125$, \protect\linebreak $ajM=0.25$}
        \end{subfigure}
        \begin{subfigure}[b]{3.5cm}
            \includegraphics[width=3.5cm]{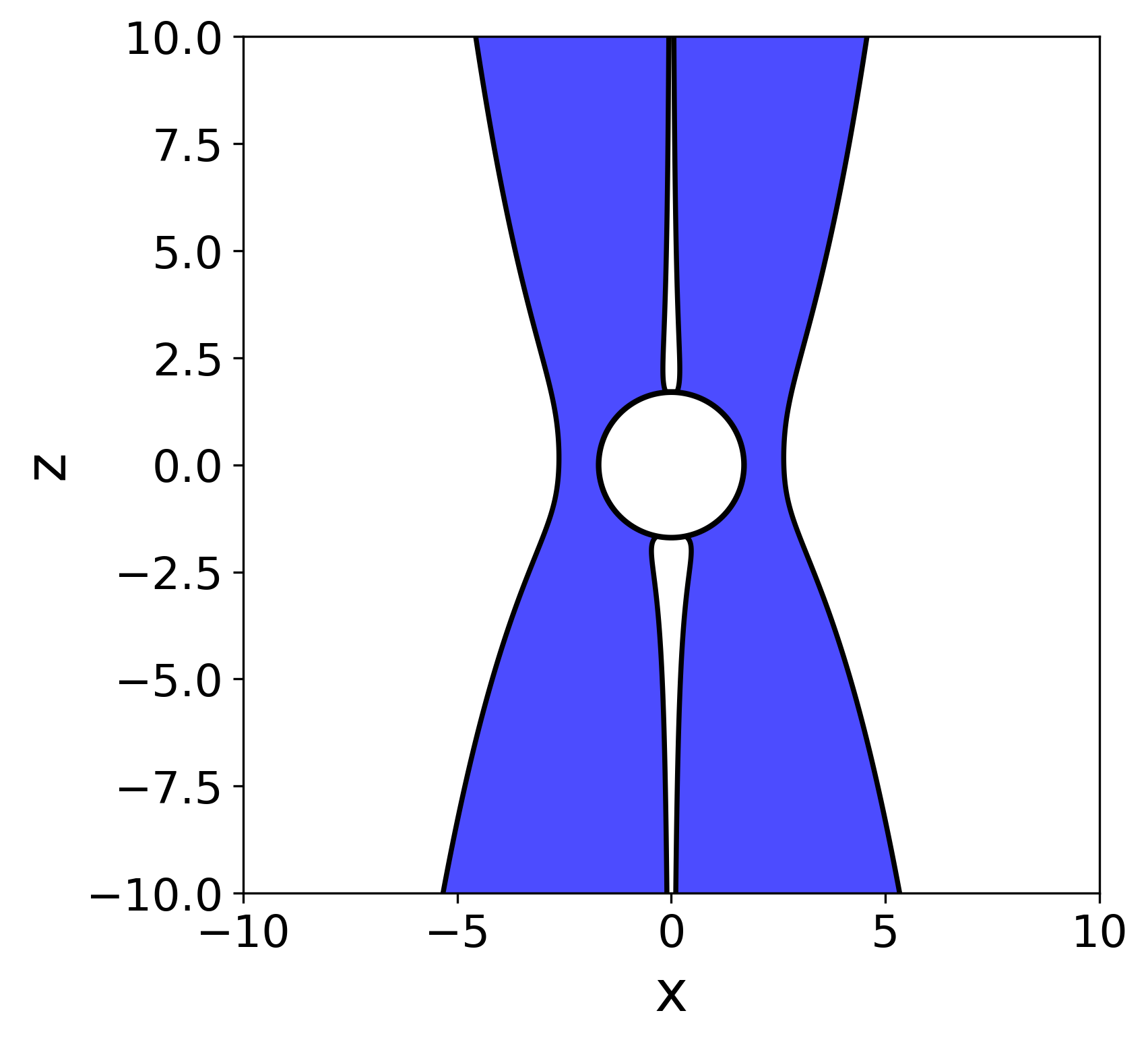}
            \caption*{ $jM^2=1.25$,\protect\linebreak $ajM=1$}
        \end{subfigure}
        \caption{$a=0.8M$}
        \label{fig:xz-ergoregion_a_09}
    \end{subfigure}        
    
\caption{Cross section ($x=r \sin\theta \cos\varphi$, $y = r \sin\theta \sin\varphi$, $z=r\cos\theta$, with $\varphi=0,\pi$) of the ergoregions (coloured) with different values of $jM^2$ and (a) $a=0.1M$  and (b) $a=0.8M$. 
The angular velocity of frame dragging $\Omega=\omega$ in these regions is positive when coloured blue and negative when coloured red. 
The region behind the event horizon $r<r_h^{(+)}$ has been excluded from these plots. 
\label{fig:xz-ergoregion}
}
\end{figure}

A typical characteristic observed in the swirling family of solutions is the appearance of non-compact ergoregions extending from the equatorial plane to infinity \cite{Astorino:2022aam,Gjorgjieski:2024cko,DiPinto:2024axv,DiPinto:2025yaa}. 
An ergoregion is defined by $g_{tt} > 0$ with boundary at $g_{tt}=0$. For the KBHSU spacetime this reads $\omega^2/{\cal F} - \rho^2 {\cal F} > 0$ with boundary at $\omega^2=\rho^2 {\cal F}^2$. 
The angular velocity $\Omega$ of the frame dragging is given by $\Omega=-g_{t\varphi}/g_{\varphi\varphi}=\omega$, which is not constant. 
We focus our analysis on the region outside the event horizon. 
The cross section of the ergoregions with the plane $y=0$ ($\varphi=0,\pi$) and their frame dragging directions are shown in Fig.~\ref{fig:xz-ergoregion} for two values of the angular momentum parameter, $a=0.1M$ and $a=0.8M$, and several values of $jM^2$, excluding the region behind the event horizon. 
For sufficiently small $j$, we observe three disconnected surfaces: one compact region inherited from the black hole which encompasses the horizon, and two non-compact regions which extend to spatial infinity above and below the equatorial plane.
The latter are already present
in the swirling universe \cite{Astorino:2022aam,Capobianco:2023kse}. 
The ergoregion in the upper hemisphere counter-rotates with respect to the black hole's spin, i.e., $\Omega < 0$ (coloured red), while the one in the lower hemisphere co-rotates with $\Omega > 0$ (coloured blue). 
Note that the ergoregion inherited from the Kerr spacetime around the event horizon is always present for a finite value of the angular momentum parameter $a$, but only becomes visible in the plots for large enough $a$ (see Fig.~\ref{fig:xz-ergoregion_a_09}).

Similar to the SBHSU \cite{Moreira:2024sjq}, we find that by increasing $j$, the boundaries of both the upper and lower ergoregions move closer to the horizon. 
However, whilst in the $a=0$ case they never interfere with the horizon, for $a\neq 0$ and sufficiently large $j$, the two disconnected patches below and above the equatorial plane merge with the ergoregion surrounding the horizon. 
For a positive swirling parameter and increasing $j$, the lower ergoregion with $\Omega >0$ joins the black hole's ergosphere first. Once $j$ reaches the critical value such that $ajM=0.25$, the upper patch joins the merged region to form an infinitely extended ergoregion that surrounds the horizon. 
This single ergoregion now has angular velocity $\Omega > 0$ everywhere and possesses a cylindrical shape. 
For very small $a$ and large $j$ the ergoregion shrinks to very small width.

In summary, the structure of the ergoregions for small $j$ appears similar to that of the SBHSU case \cite{Moreira:2024sjq}, but differs significantly when $j$ increases. 
In the SBHSU, there is always symmetry about the equatorial plane.
However, when $a$ increases, the spin-spin interaction starts to play a crucial role in the geometry of the ergoregions. 
The symmetry about the equatorial plane is broken, and the ergoregions eventually merge, forming a single region that extends all the way to infinity.

\noindent


\section{Photon orbits}

In this section, we discuss how light-like particles move on geodesics in the KBHSU spacetime. 
We employ the Hamiltonian formulation with
\begin{equation}
    \mathscr{H}=\frac{1}{2}g^{\mu\nu}p_\mu p_\nu=0,
\end{equation}
where $p_\mu\equiv g_{\mu\nu}\dot{x}^\nu$ is the photon's 4-momentum. 
Then the respective Hamilton's equations read   
\begin{equation}
    \dot{x}^\mu=\frac{\partial \mathscr{H}}{\partial p_\mu}, \qquad \dot{p}^\mu=-\frac{\partial \mathscr{H}}{\partial x_\mu}.
    \label{Ham_eqs}
\end{equation}
The spacetime admits two constants of motion owing to the geometry's stationarity and axisymmetry, which give rise to the Killing vector fields $\partial_t$ and $\partial_\varphi$, respectively. 
Such fields entail 
two quantities that are conserved along the geodesics, related to the particle's four-momentum:
\begin{equation}\label{cyclic variables}
    p_t=g_{tt}\dot t + g_{t\varphi}\dot\varphi=-E \qquad \text{and} \qquad p_\varphi=g_{\varphi\varphi}\dot \varphi + g_{t\varphi}\dot t = L  \ ,
\end{equation}
where $E$ and $L$ are usually referred to as the energy and angular momentum of the particle, respectively. 
Solving the system (\ref{cyclic variables}) for the velocities and inserting the metric (\ref{eq:metric}), one finds
\begin{equation}
    \dot t = \frac{E-\omega L}{\mathcal{F}\rho^2} \ \ ,  \ \
    \dot \varphi = \mathcal{F}L+\frac{\omega(E-\omega L)}{\mathcal{F} \rho^2} \ .
    \end{equation}
Additionally, the normalization condition becomes
    \begin{equation}
     \Sigma \sin^2\theta\left(\dot r^2 + \Delta\dot\theta^2\right) =\Delta\left(-  L^2+\frac{(E-\omega L)^2}{\mathcal{F}^2 
    \rho^2}\right) \ .
\end{equation}
In contrast to the Kerr spacetime, the $r$- and $\theta$-motion is not separable in the KBHSU spacetime, i.e., no equivalent of the Carter constant exists. 
The spacetime is algebraically general, that is, Petrov type I \cite{Astorino:2022aam}.

\subsection{Light rings}

Light can travel on special circular paths, light rings (LRs), in the close vicinity of compact objects.
A Schwarzschild black hole with mass parameter $M$ has an unstable and planar LR at $r_{p}=3M$. 
Due to the SO(3) symmetry of the event horizon, $r_{p}=3M$ then corresponds to the radius of the photon sphere surrounding the Schwarzschild black hole. Furthermore, a Kerr black hole possesses two LRs that are also located on the equatorial plane and are both unstable. 
The prograde ring has a radius that decreases from $3M$ to $M$ as the angular momentum parameter $a$ increases from zero to $M$, the so-called \textit{extremal case}, while the radius of the retrograde ring increases from $3M$ to $4M$.

In general, the location of the LRs can be determined by finding the critical points of the effective potential $V$, which 
for a stationary and axially symmetric solution depends exclusively on the metric tensor and conserved charges. 
The effective potential reads
\begin{equation}\label{LR_potential}
    V=-\frac{1}{D}(E^2g_{\varphi\varphi}+2EL g_{t\varphi}+L^2g_{tt})  \ \ \ , \ \ \ D=g_{t\varphi}^2-g_{tt}g_{\varphi\varphi} \ . 
    \end{equation}

Following \cite{Cunha:2016bjh,Cunha:2017qtt},  we write $V$ in terms of a pair of more manageable potentials $H_{\pm}$ which 
depend only on the spacetime geometry
\begin{equation}
    V=-\frac{L^2}{D}g_{\varphi\varphi}(\sigma-H_+)(\sigma-H_-) \ \ , \ \ 
    H_{\pm}(r,\theta)=\frac{-g_{t\varphi} \pm \sqrt{D}}{g_{\varphi\varphi}} \ ,
\end{equation}
where $\sigma\equiv E/L$ is the inverse impact parameter. 
In this formalism, LRs are found as solutions of $V=0=\nabla V$. 
This therefore implies that a critical point of $H_\pm(r,\theta)$ must correspond to the coordinates of a LR, i.e., a pair $(r_{\text{LR}},\theta_{\text{LR}})$ such that $\nabla H_\pm=0$.

For the KBHSU, with the metric (\ref{eq:metric}) the potentials take a particularly simple form:
\begin{equation}
\label{eq:hpot}
H_{\pm}(r,\theta)=\omega \pm \mathcal{F} \rho.
\end{equation}

\begin{figure}[htbp]
    \centering
    \begin{subfigure}[b]{\textwidth}
        \centering
        \begin{subfigure}[b]{4.9cm}
            \includegraphics[width=4.9cm]{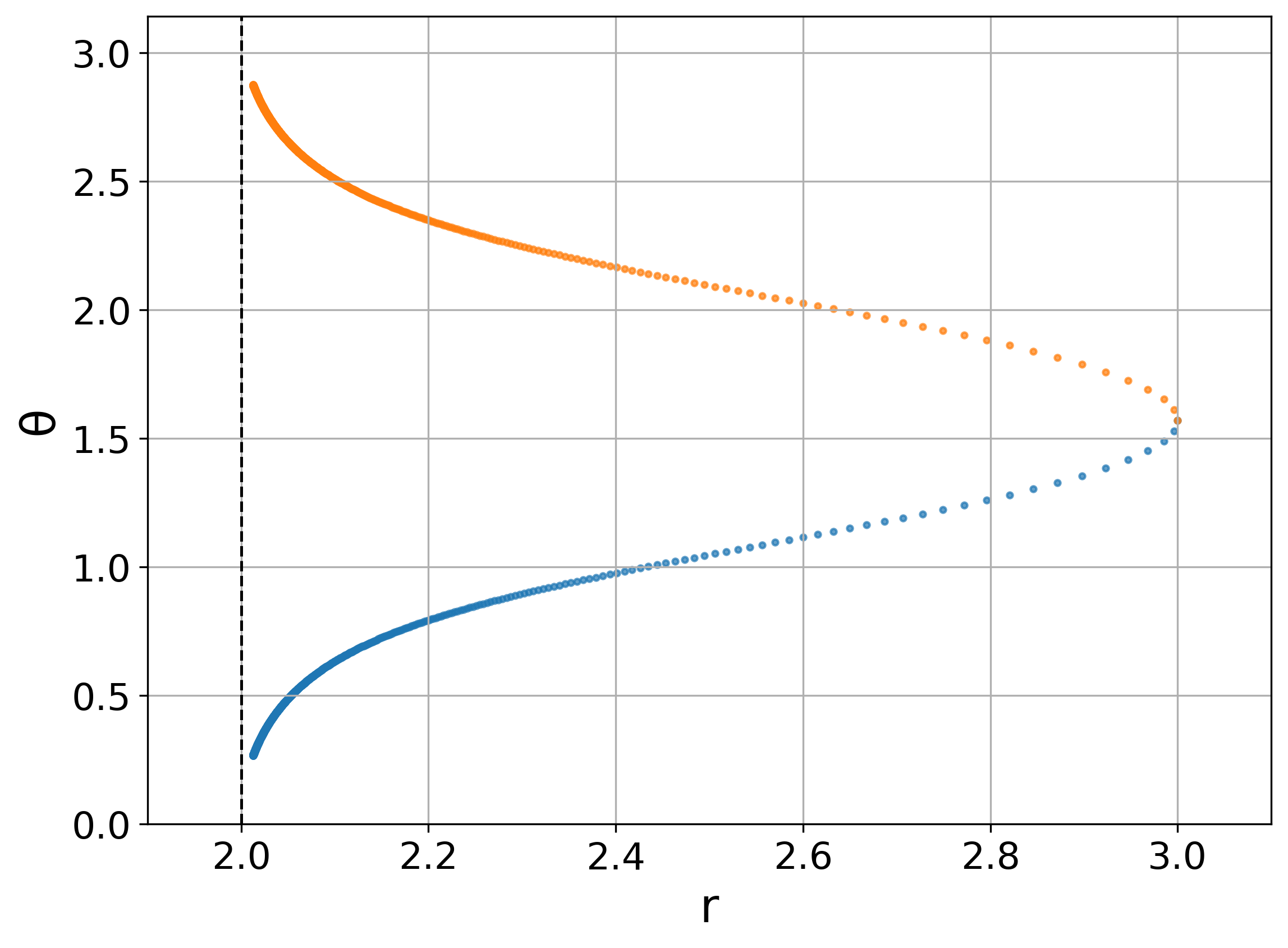}
        \end{subfigure}
        \begin{subfigure}[b]{4.9cm}
            \includegraphics[width=4.9cm]{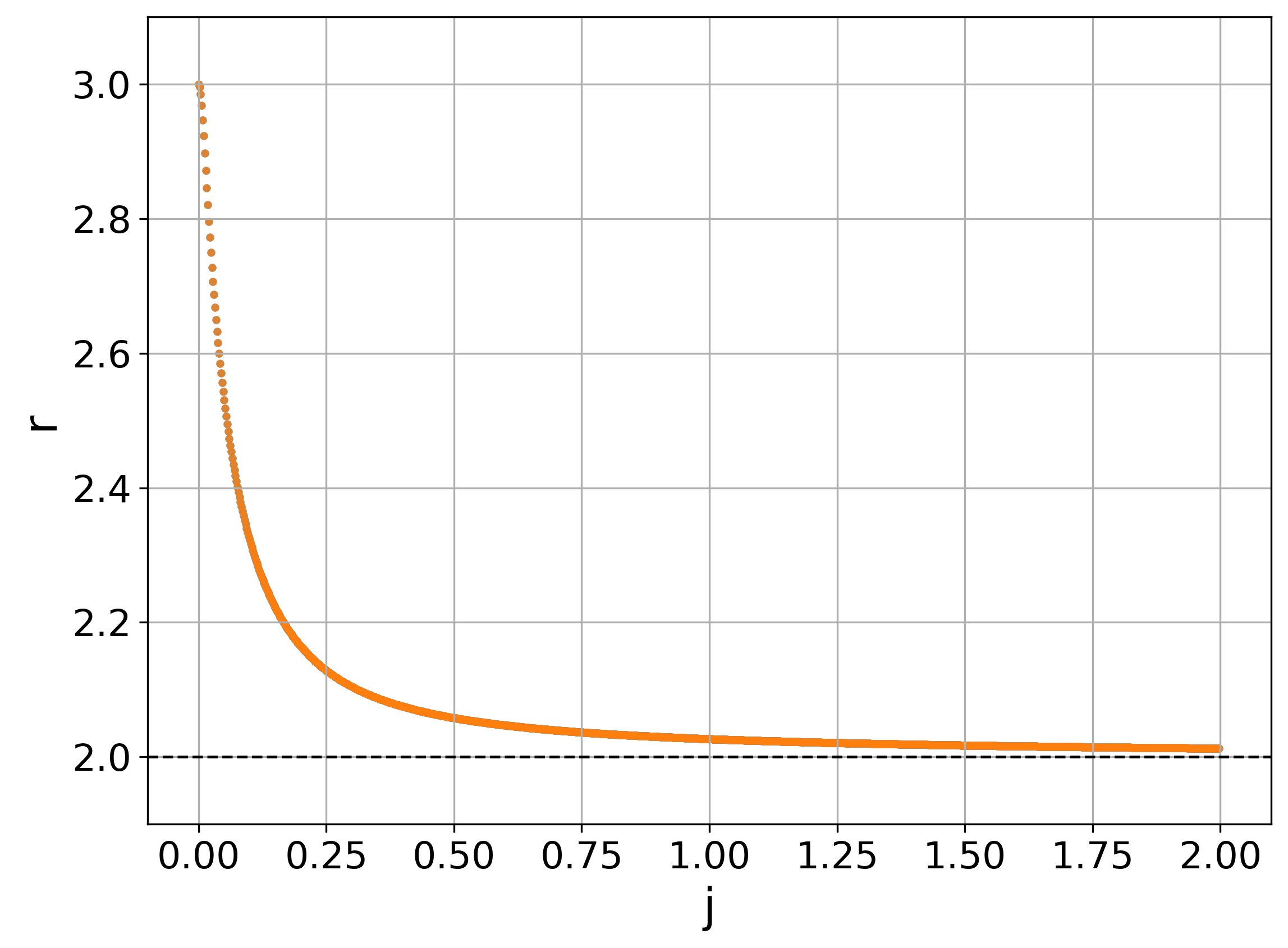}
        \end{subfigure}
        \begin{subfigure}[b]{4.9cm}
            \includegraphics[width=4.9cm]{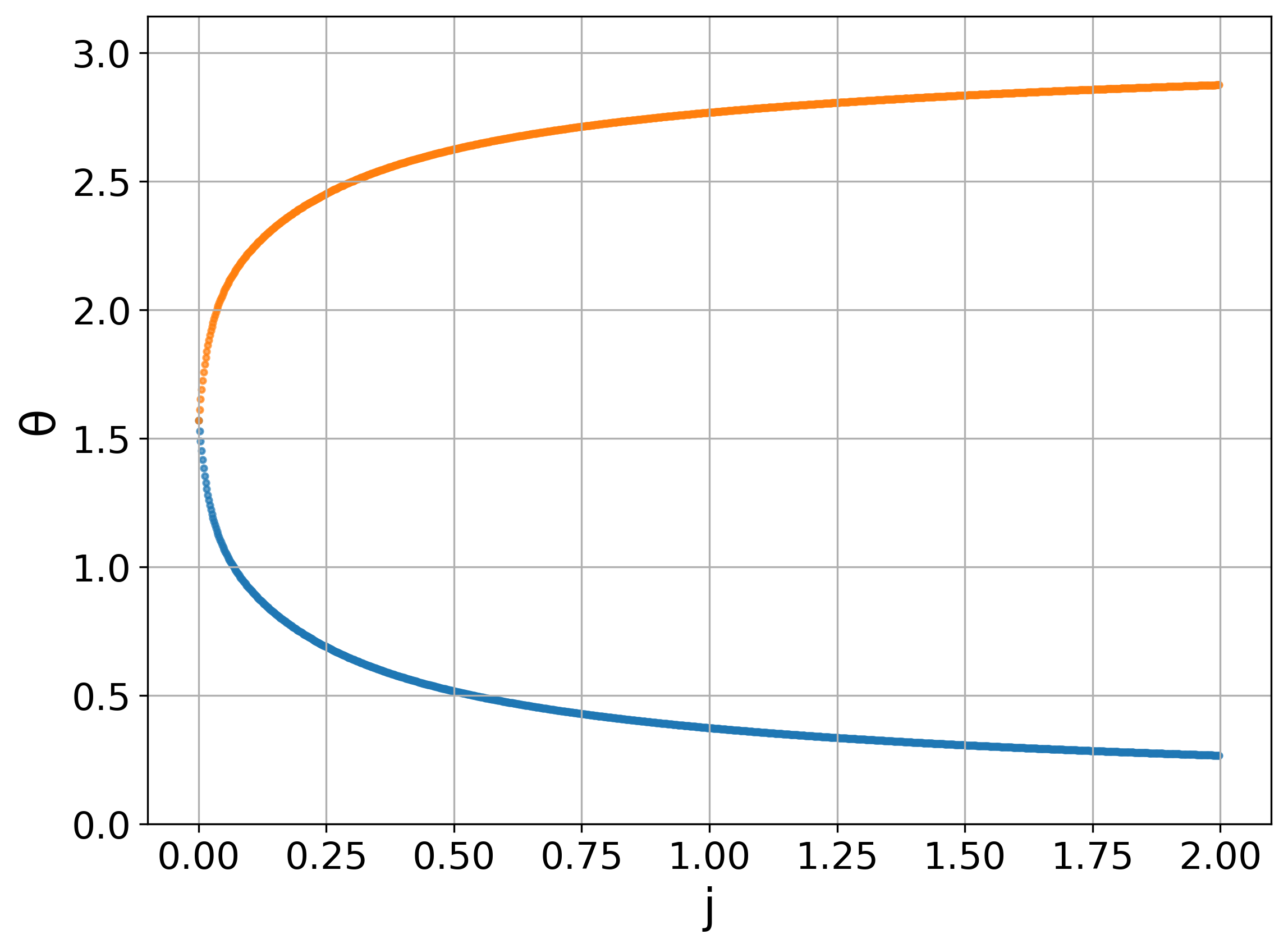}
        \end{subfigure}
        \caption{$a=0$}
        \label{LR_Locations}
    \end{subfigure}
    
    \begin{subfigure}[b]{\textwidth}
        \centering
        \begin{subfigure}[b]{4.9cm}
            \includegraphics[width=4.9cm]{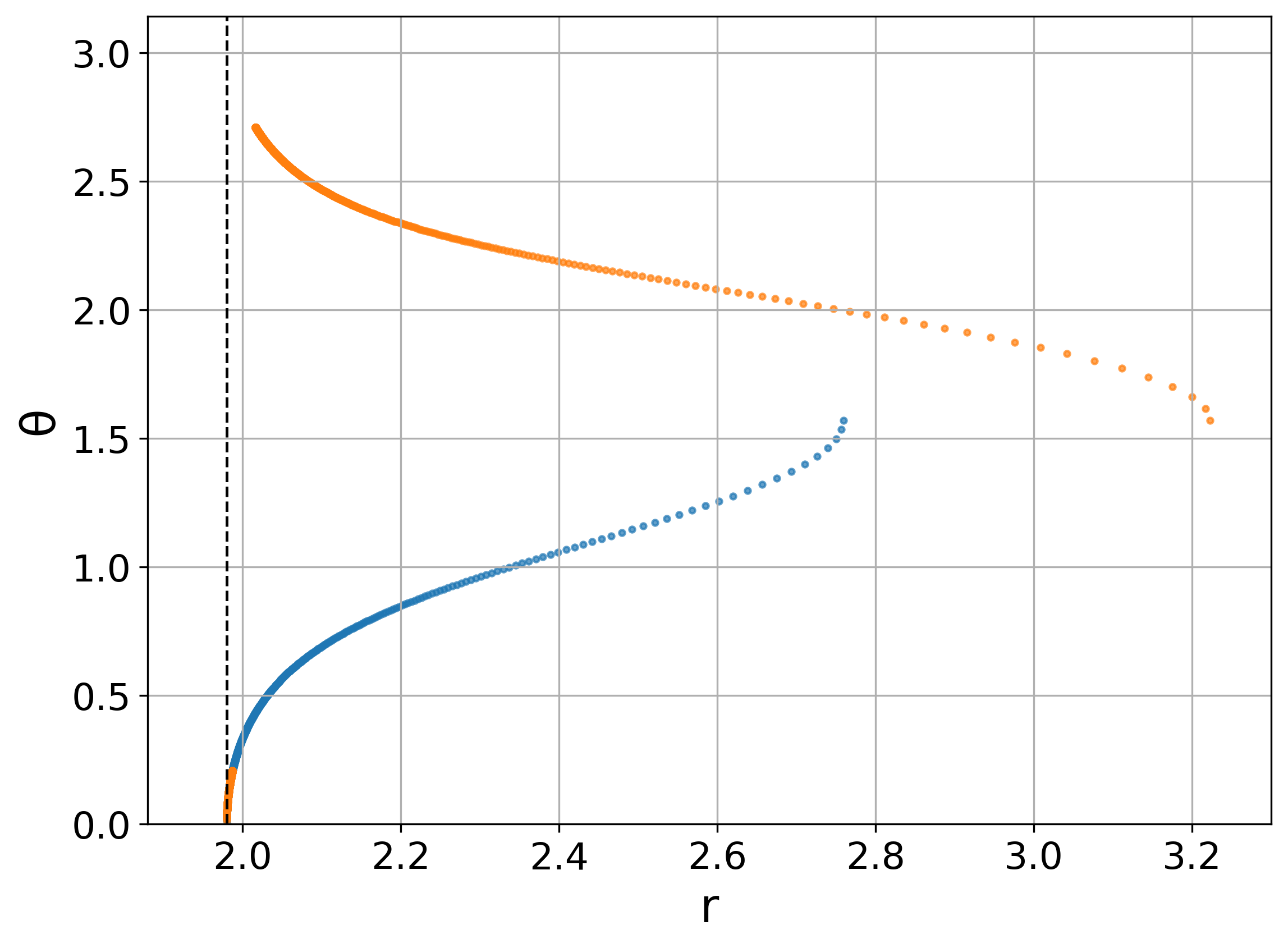}
        \end{subfigure}
        \begin{subfigure}[b]{4.9cm}
            \includegraphics[width=4.9cm]{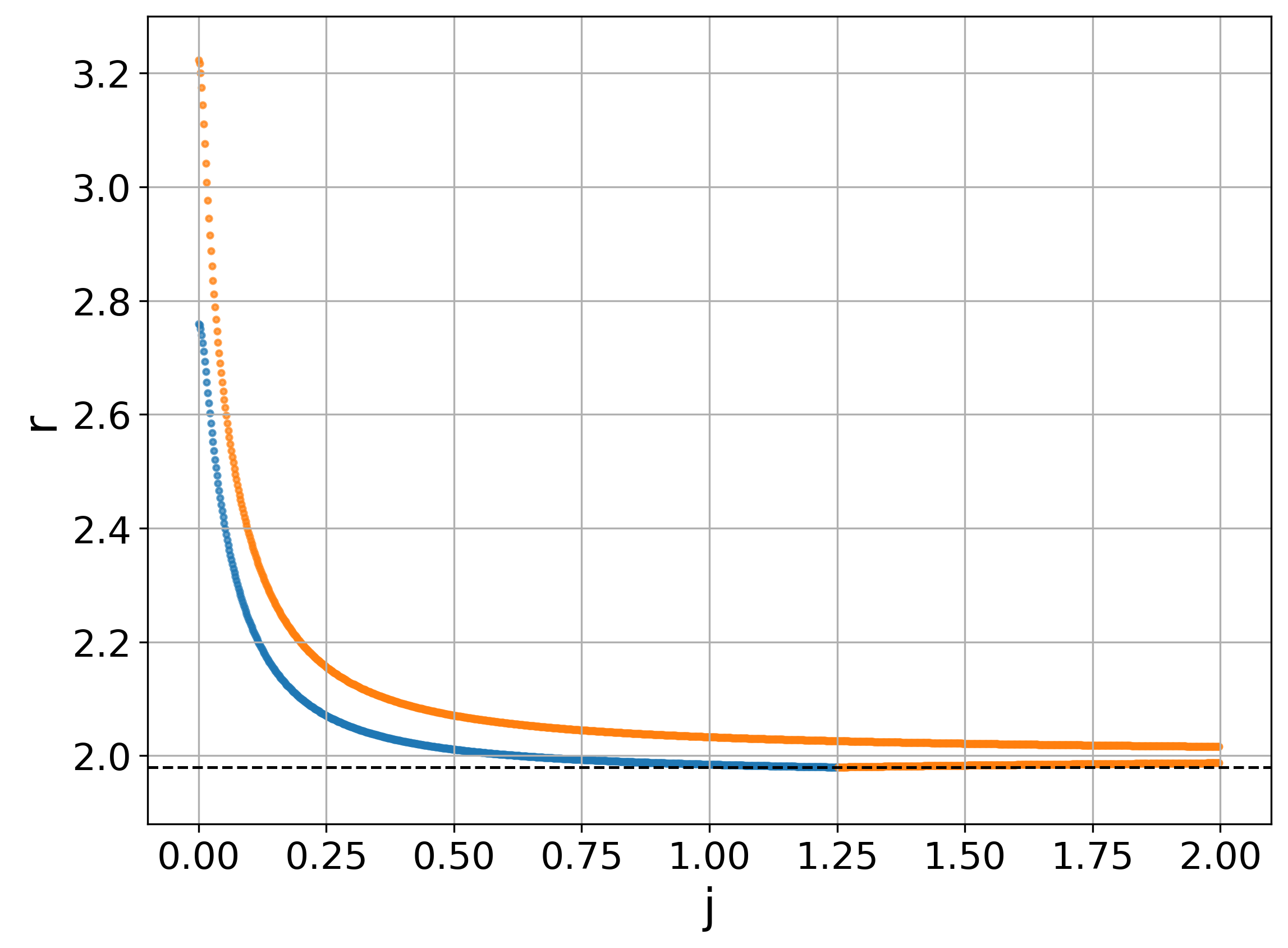}
        \end{subfigure}
        \begin{subfigure}[b]{4.9cm}
            \includegraphics[width=4.9cm]{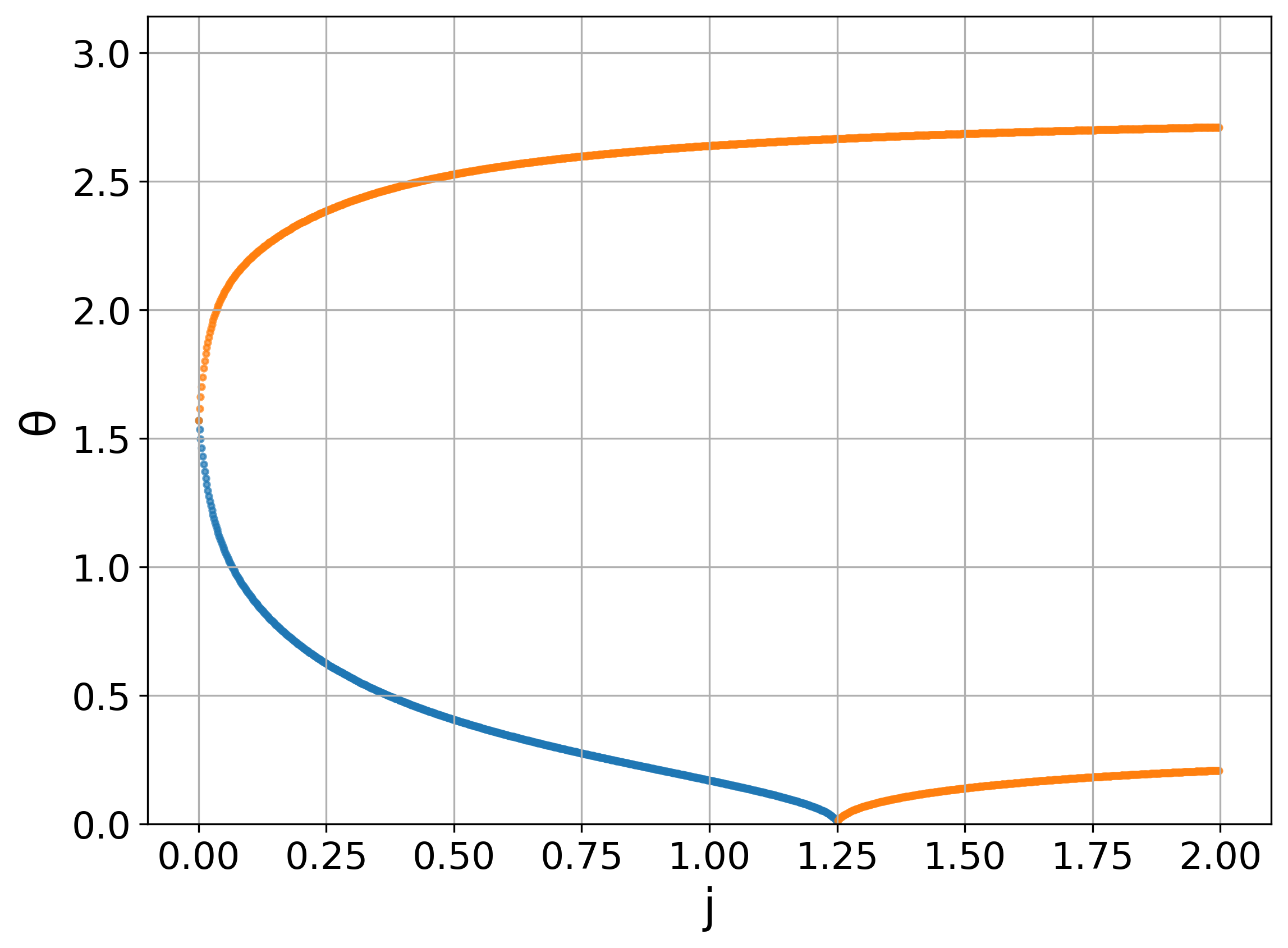}
        \end{subfigure}
        \caption{$a=0.2M$}
        \label{LR_Locations_b}
    \end{subfigure}

    \begin{subfigure}[b]{\textwidth}
        \centering
        \begin{subfigure}[b]{4.9cm}
            \includegraphics[width=4.9cm]{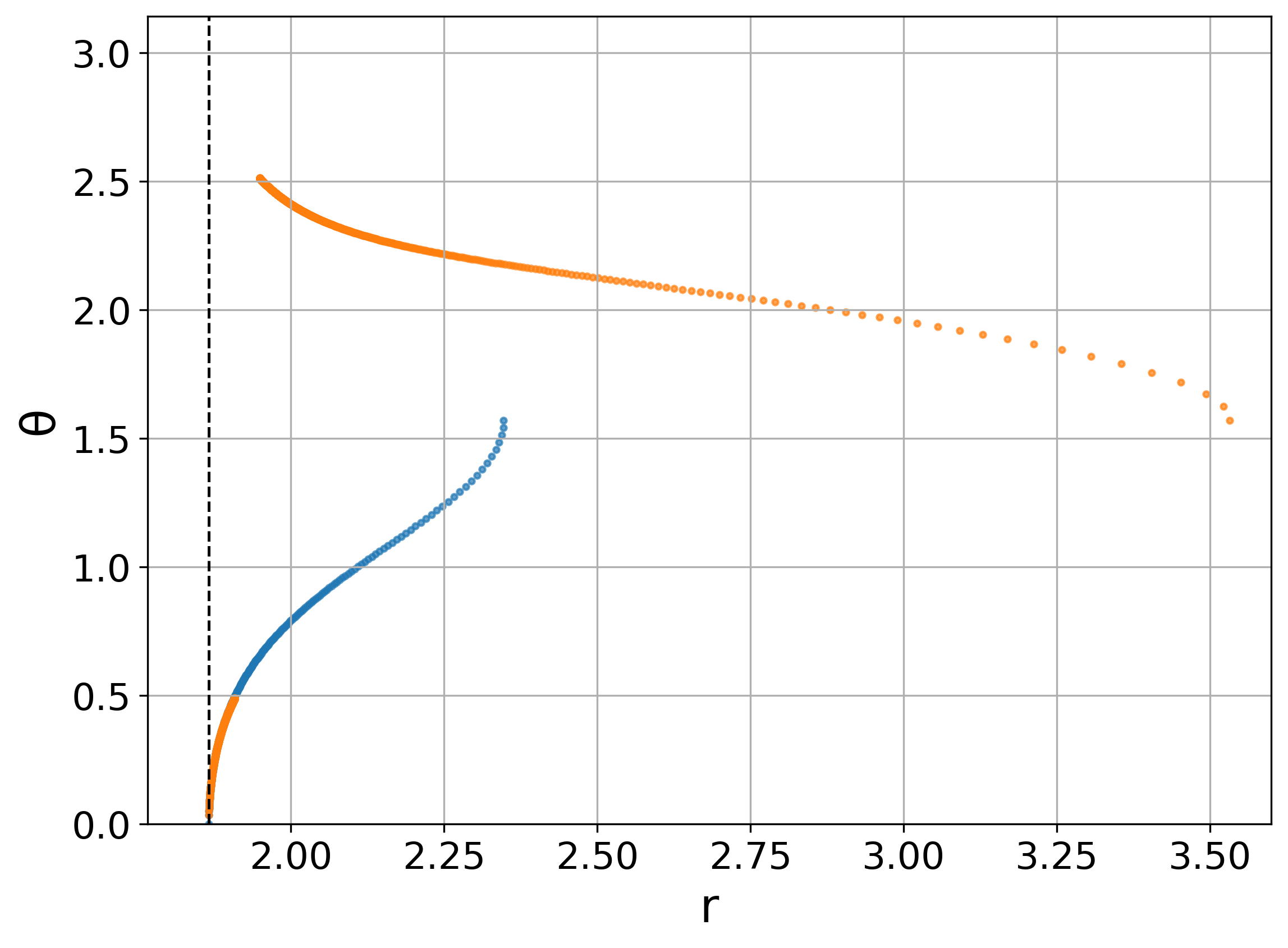}
        \end{subfigure}
        \begin{subfigure}[b]{4.9cm}
            \includegraphics[width=4.9cm]{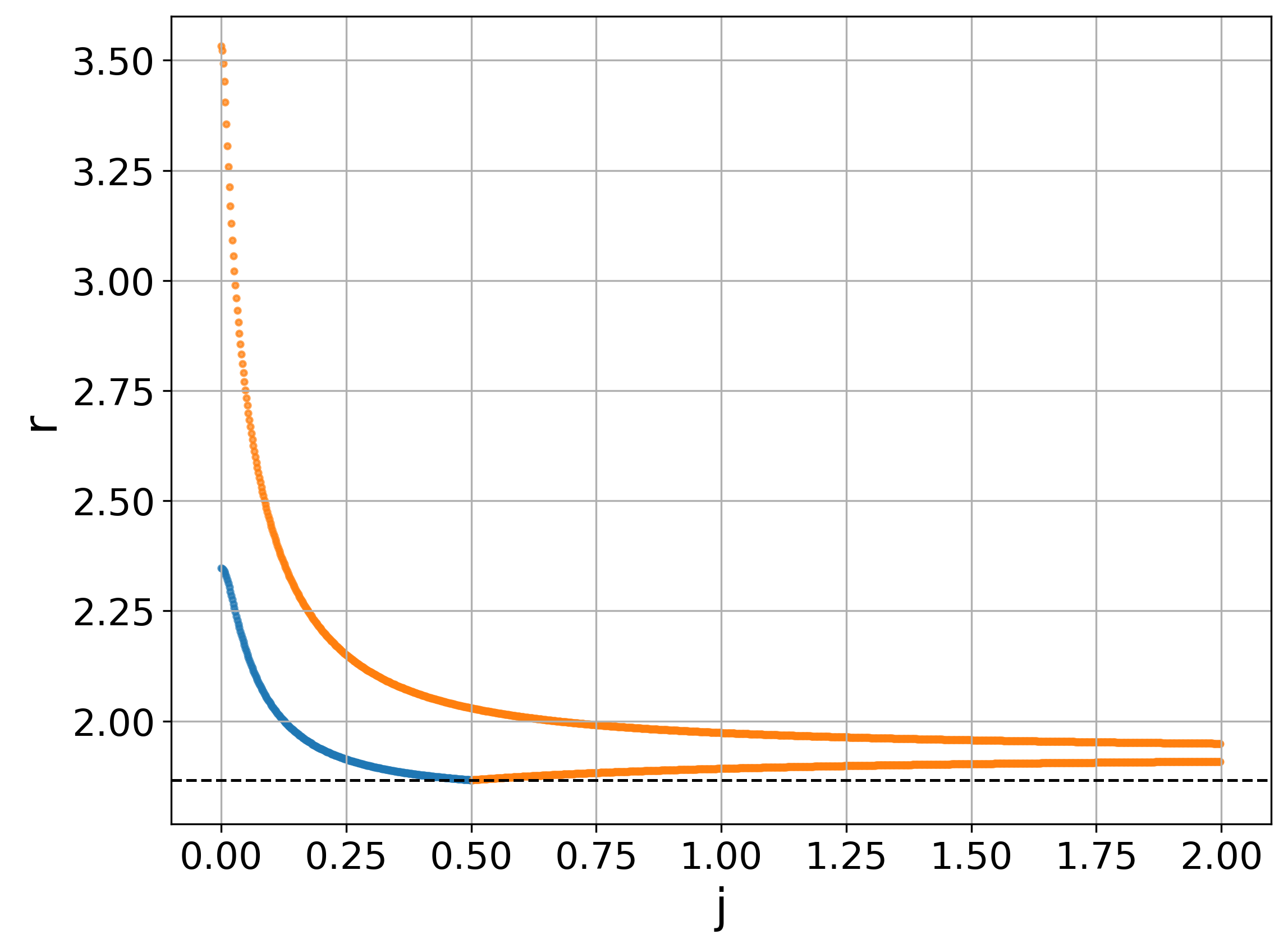}
        \end{subfigure}
        \begin{subfigure}[b]{4.9cm}
            \includegraphics[width=4.9cm]{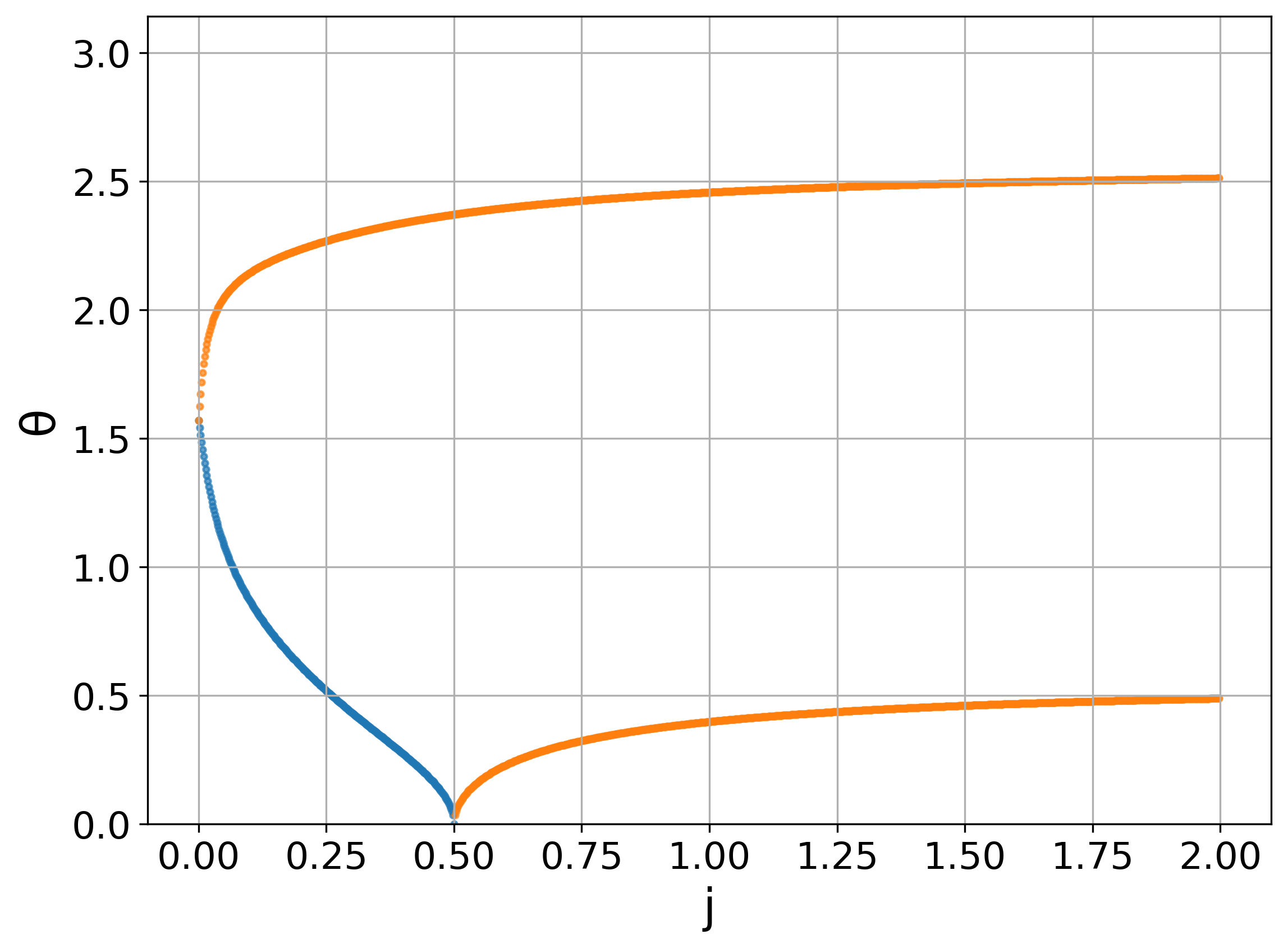}
        \end{subfigure}
        \caption{$a=0.5M$}
        \label{LR_Locations_c}
    \end{subfigure}

    \begin{subfigure}[b]{\textwidth}
        \centering
        \begin{subfigure}[b]{4.99cm}
            \includegraphics[width=4.9cm]{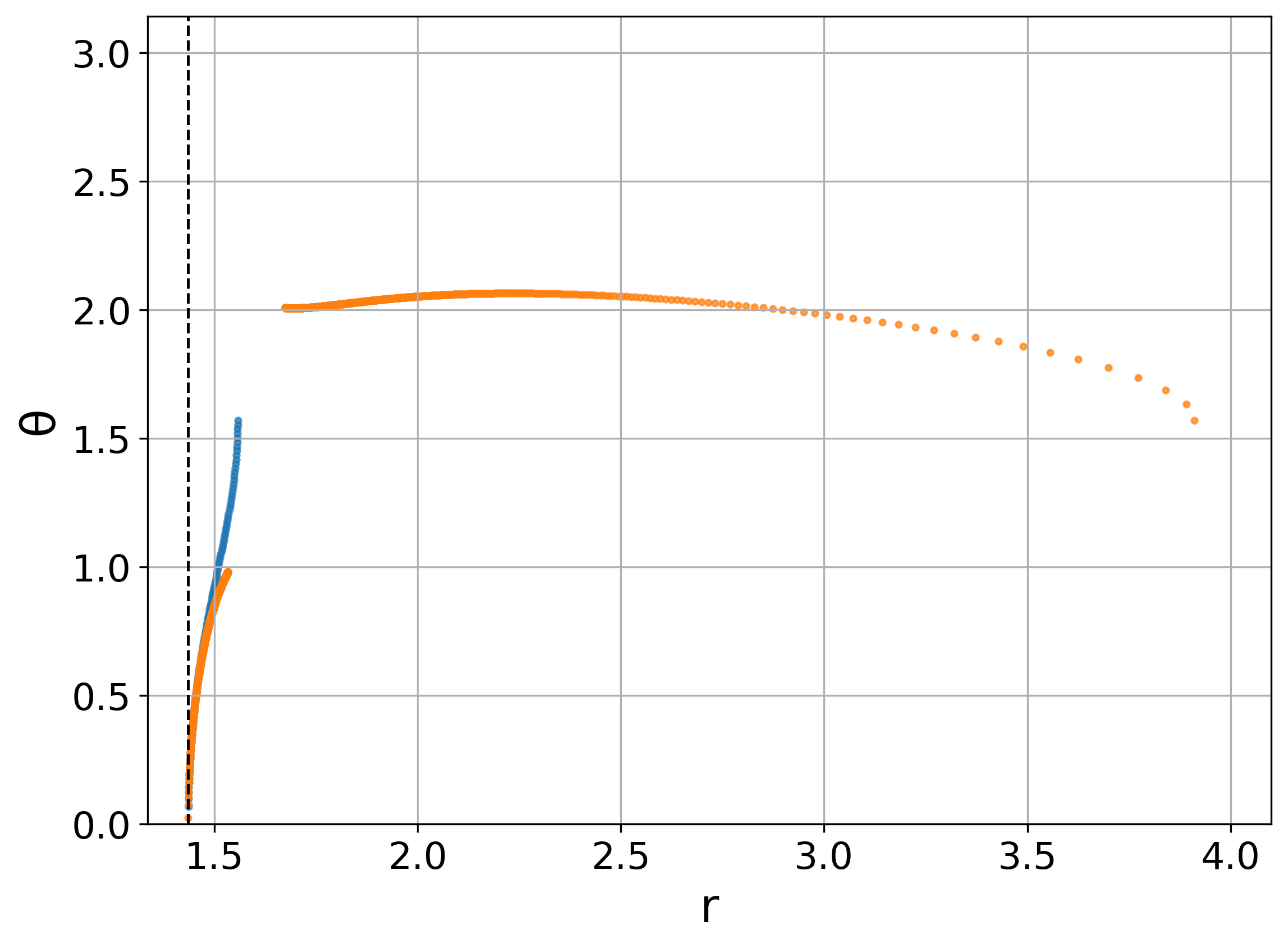}
        \end{subfigure}
        \begin{subfigure}[b]{4.9cm}
            \includegraphics[width=4.9cm]{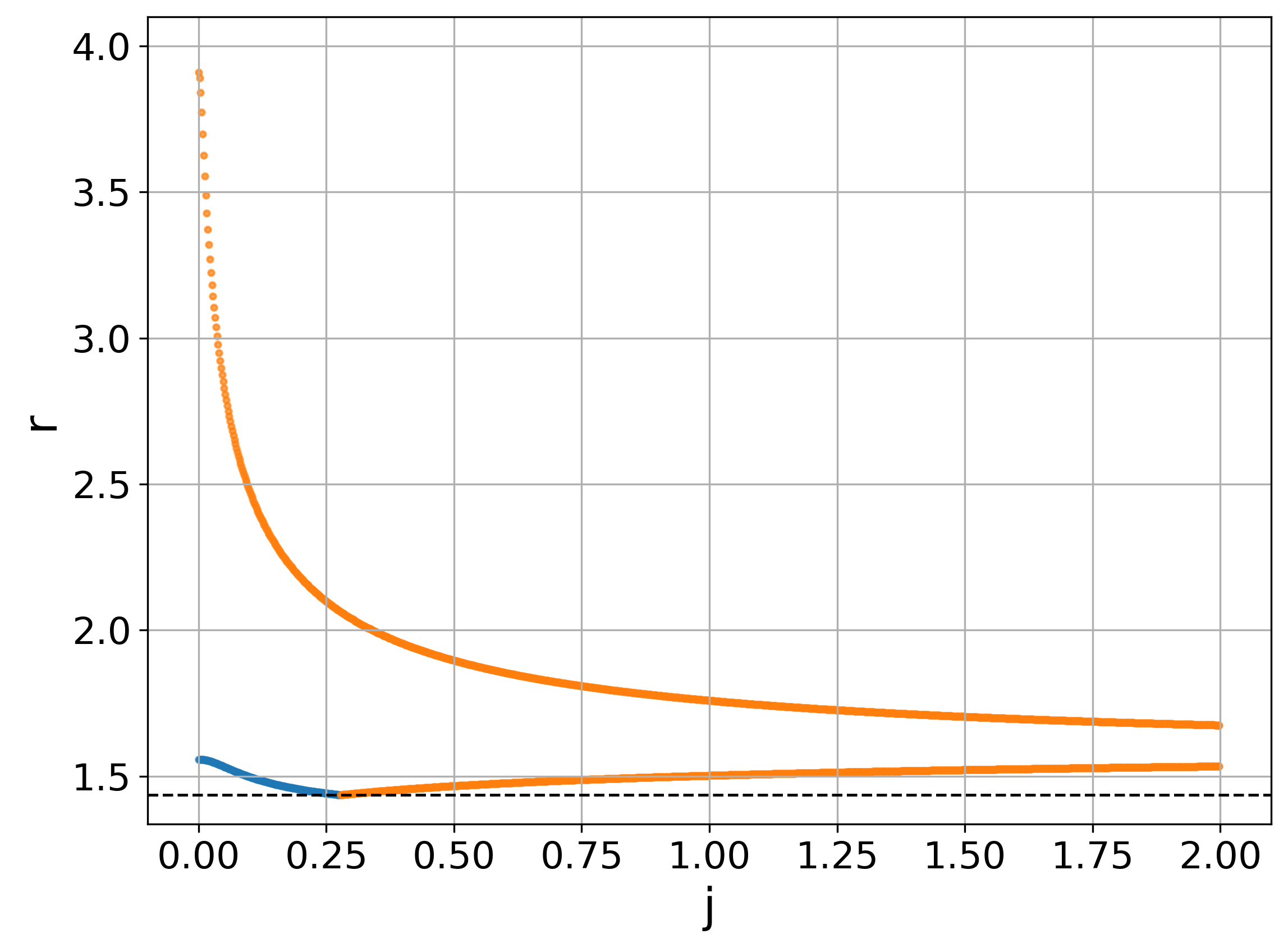}
        \end{subfigure}
        \begin{subfigure}[b]{4.9cm}
            \includegraphics[width=4.9cm]{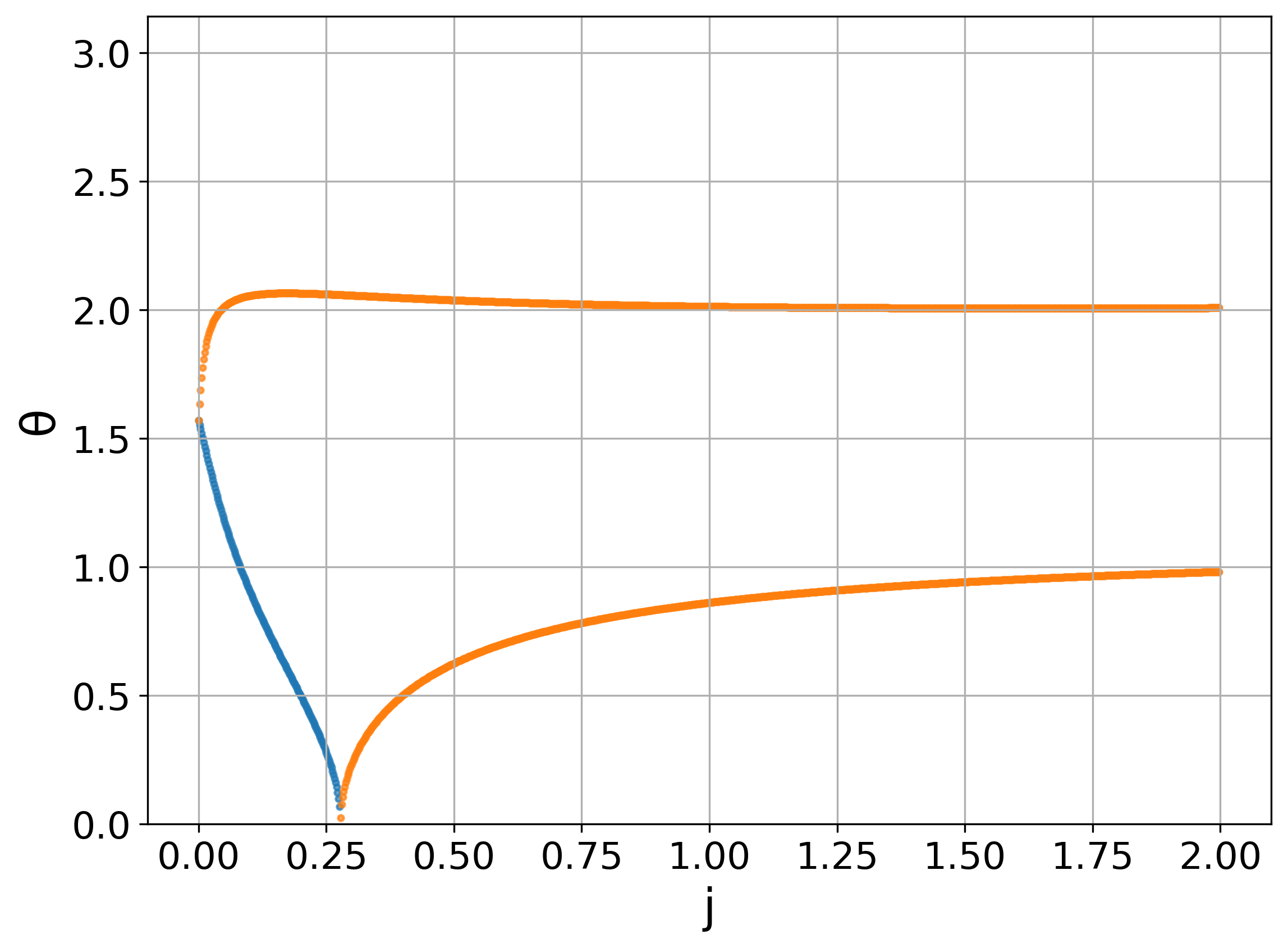}
        \end{subfigure}
        \caption{$a=0.9M$}
        \label{LR_Locations_d}
    \end{subfigure}
    
    \begin{subfigure}[b]{\textwidth}
        \centering
        \begin{subfigure}[b]{4.9cm}
            \includegraphics[width=4.9cm]{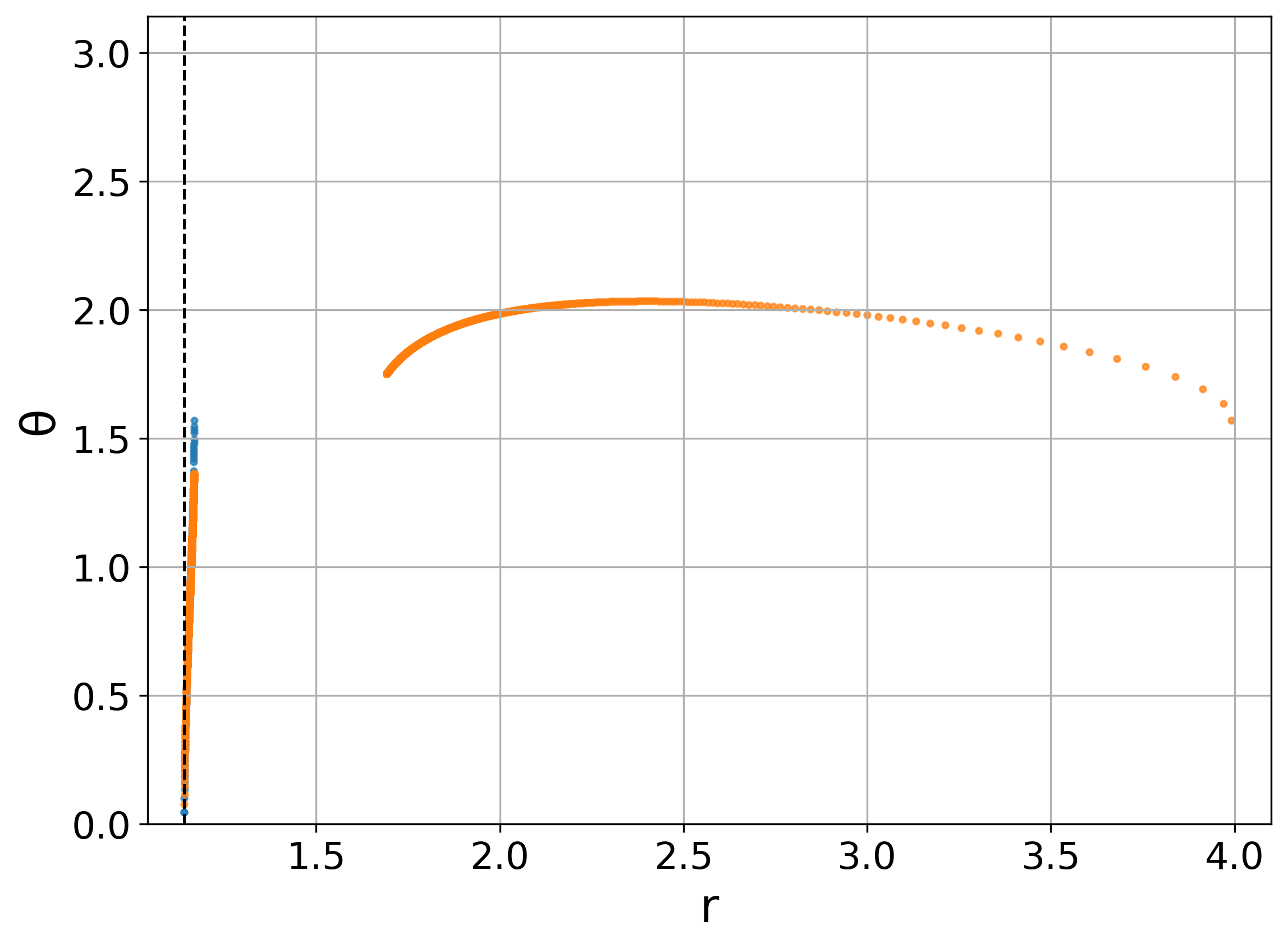}
        \end{subfigure}
        \begin{subfigure}[b]{4.9cm}
            \includegraphics[width=4.9cm]{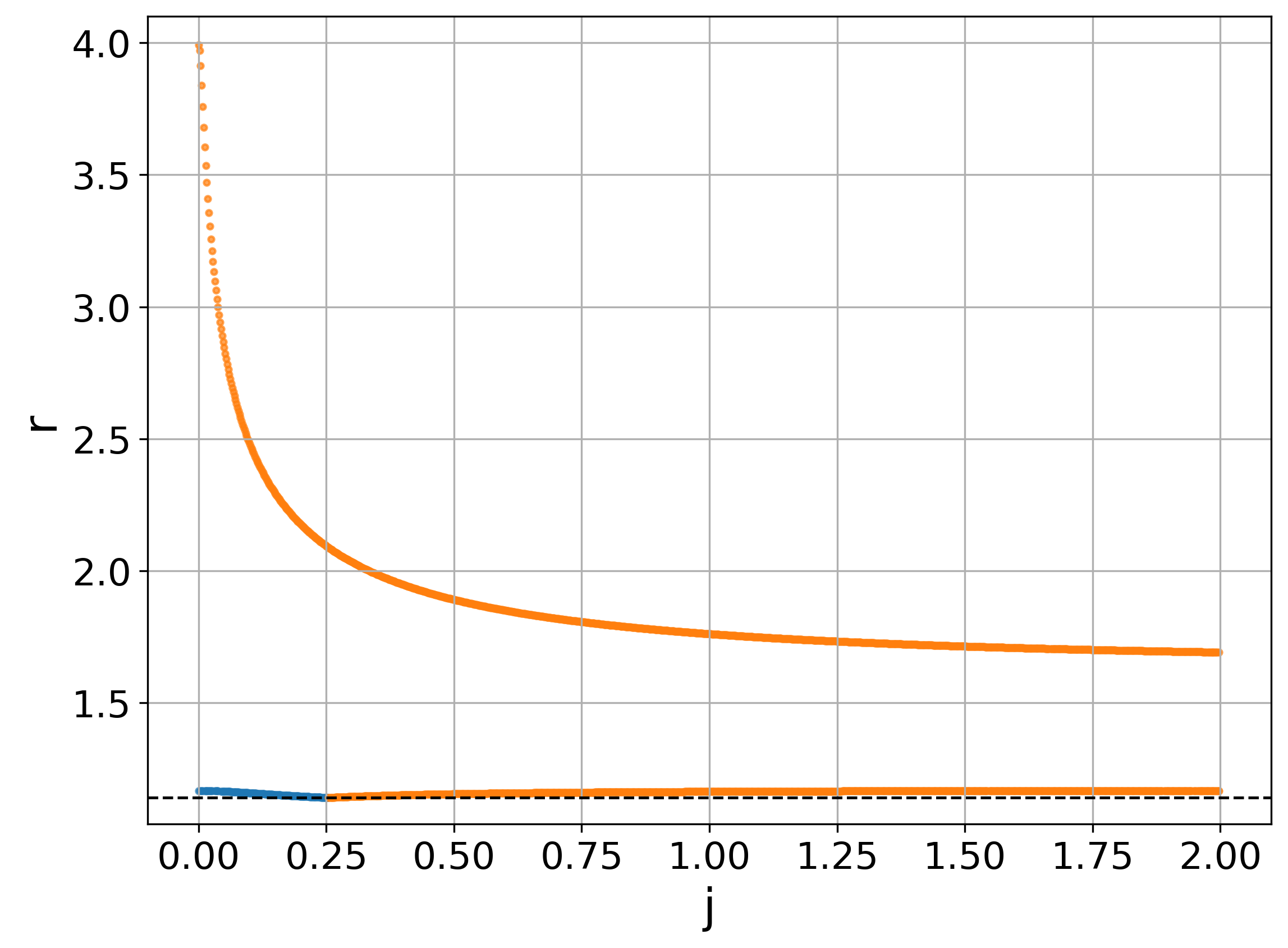}
        \end{subfigure}
        \begin{subfigure}[b]{4.9cm}
            \includegraphics[width=4.9cm]{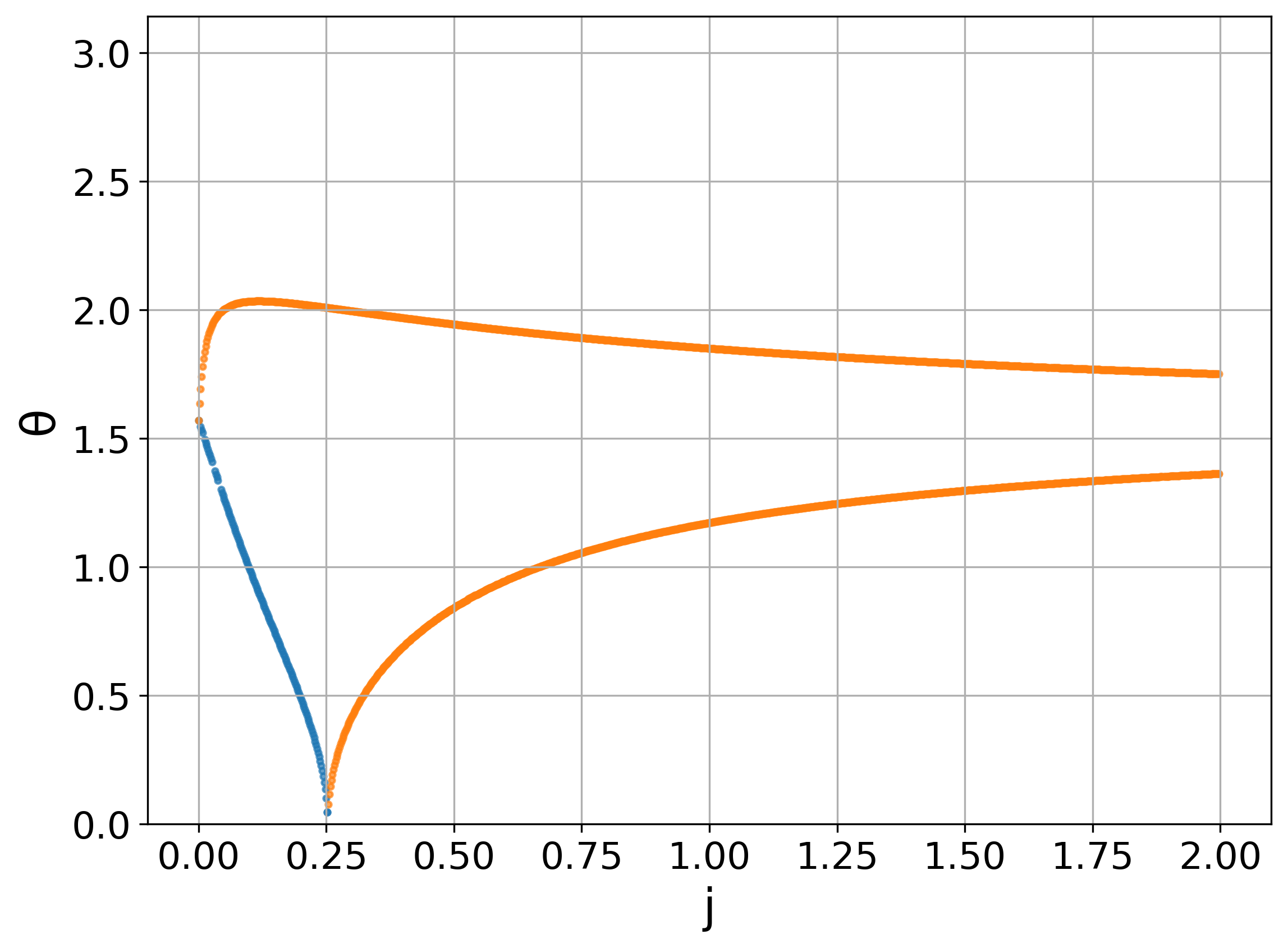}
        \end{subfigure}
        \caption{$a=0.99M$}
        \label{LR_Locations_e}
    \end{subfigure}
    \caption{Locations of the Light Rings for the KBHSU solution at multiples of $jM^2=0.002$ with $M=1$. LRs corresponding to $H_+$ are coloured blue and LRs corresponding to $H_-$ are in orange. In the left-hand column we plot the coordinates of the LRs in the $(r, \theta)$-plane, in the middle column we give the radial coordinate against $j$, and in the right-hand column we show the polar coordinate against $j$. The thick black dashed line represents the radius of the outer horizon.}
\end{figure}

In the SBHSU (where $a=0$), for $j=0$ there is one degenerate LR on the equatorial plane at $r=3M$ since in that case there is no distinction between co- and counter-rotation with respect to the swirling of the spacetime.
However, as $j$ increases, the degeneracy is broken and the LRs branch off: one into the upper hemisphere and one into the lower as seen in Fig.~\ref{LR_Locations}. 
This pair of LRs is pushed off the equatorial plane and at the same time pulled towards the horizon. Note, in particular, that both LRs have the same radius.
As $j\rightarrow\infty$, the LRs eventually reach the horizon $r=2M$, with one LR at the North pole $\theta=0$ and the other at the South pole $\theta=\pi$. 
In fact, for the SBHSU there is a symmetry of the effective potential such that if $(r,\theta)$ is a critical point of $H_+$ (shown in blue), then $(r,\pi-\theta)$ is also a critical point of $H_-$ (shown in orange) \cite{Moreira:2024sjq}. 

For the KBHSU where $a\neq0$, this symmetry no longer holds, though for small $a$, we observe similar features as those appearing in the SBHSU case: the LRs are again pushed off the equatorial plane and pulled towards the horizon. 
The branches now, however, start from the two different LR locations for vanishing $j$, since co- and counter-rotating rings no longer agree for Kerr black holes (see the left hand graphs in Figs.~\ref{LR_Locations_b}-\ref{LR_Locations_e}).

Due to the spin-spin interaction of the KBHSU, we observe that as $j$ increases to the critical value where $ajM=0.25$, the LR corresponding to the $H_+$ potential reaches the North pole $\theta=0$, and then disappears, passing over to $H_-$ and drawing away from the axis. We observe this effect for all the values of $a$ that we have studied, as demonstrated in Figs.~\ref{LR_Locations_b}-\ref{LR_Locations_e}.


\begin{figure}[t!] 
    \centering

    \begin{subfigure}[b]{\textwidth}
        \centering
        \begin{subfigure}[b]{4.1cm}
            \includegraphics[width=4.1cm]{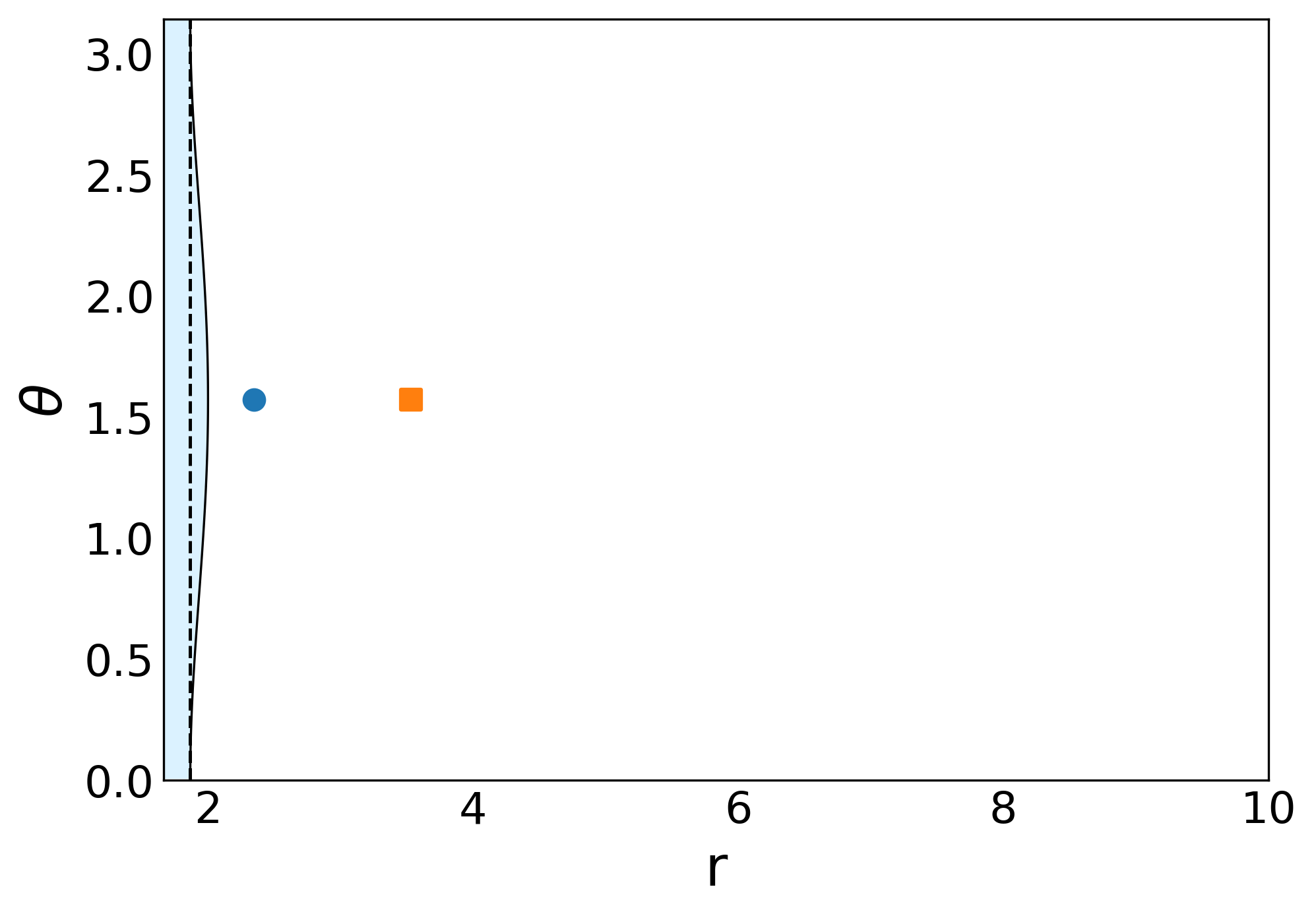}
            \caption*{$j=0$}
        \end{subfigure}
        \begin{subfigure}[b]{4.1cm}
            \includegraphics[width=4.1cm]{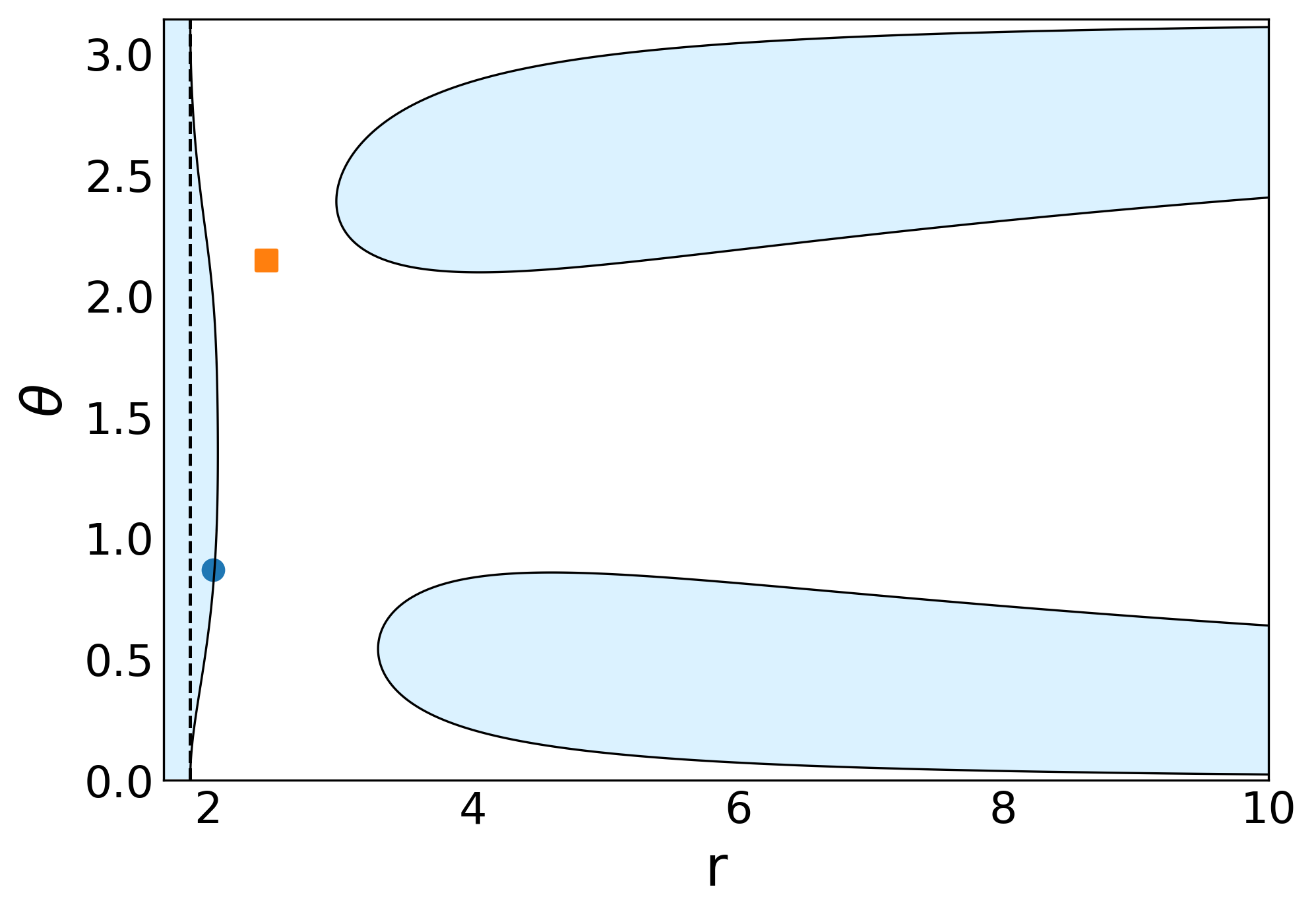}
            \caption*{$jM^2=0.1$}
        \end{subfigure}
        \begin{subfigure}[b]{4.1cm}
            \includegraphics[width=4.1cm]{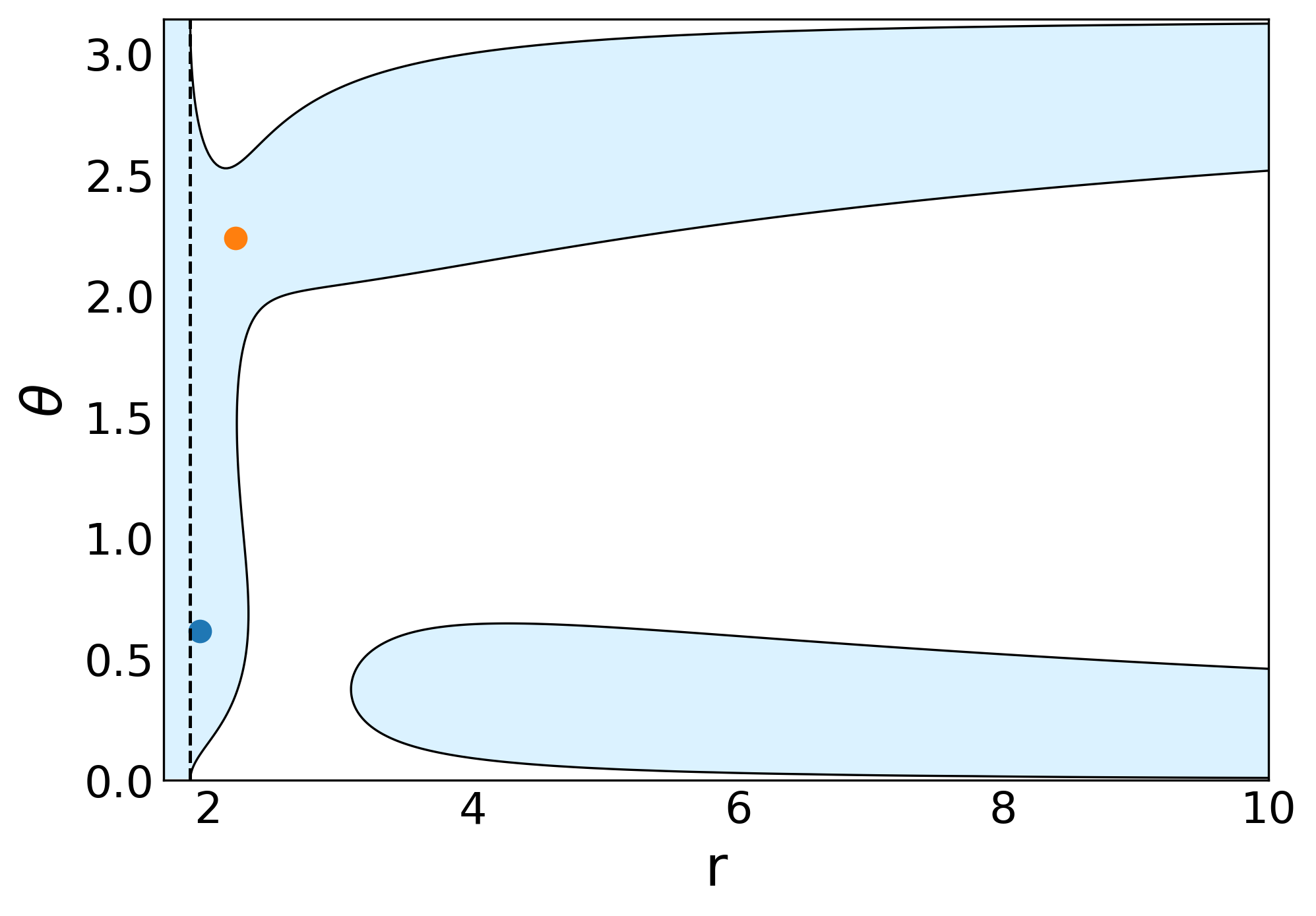}
            \caption*{$jM^2=0.2$}
        \end{subfigure}
        \begin{subfigure}[b]{4.1cm}
            \includegraphics[width=4.1cm]{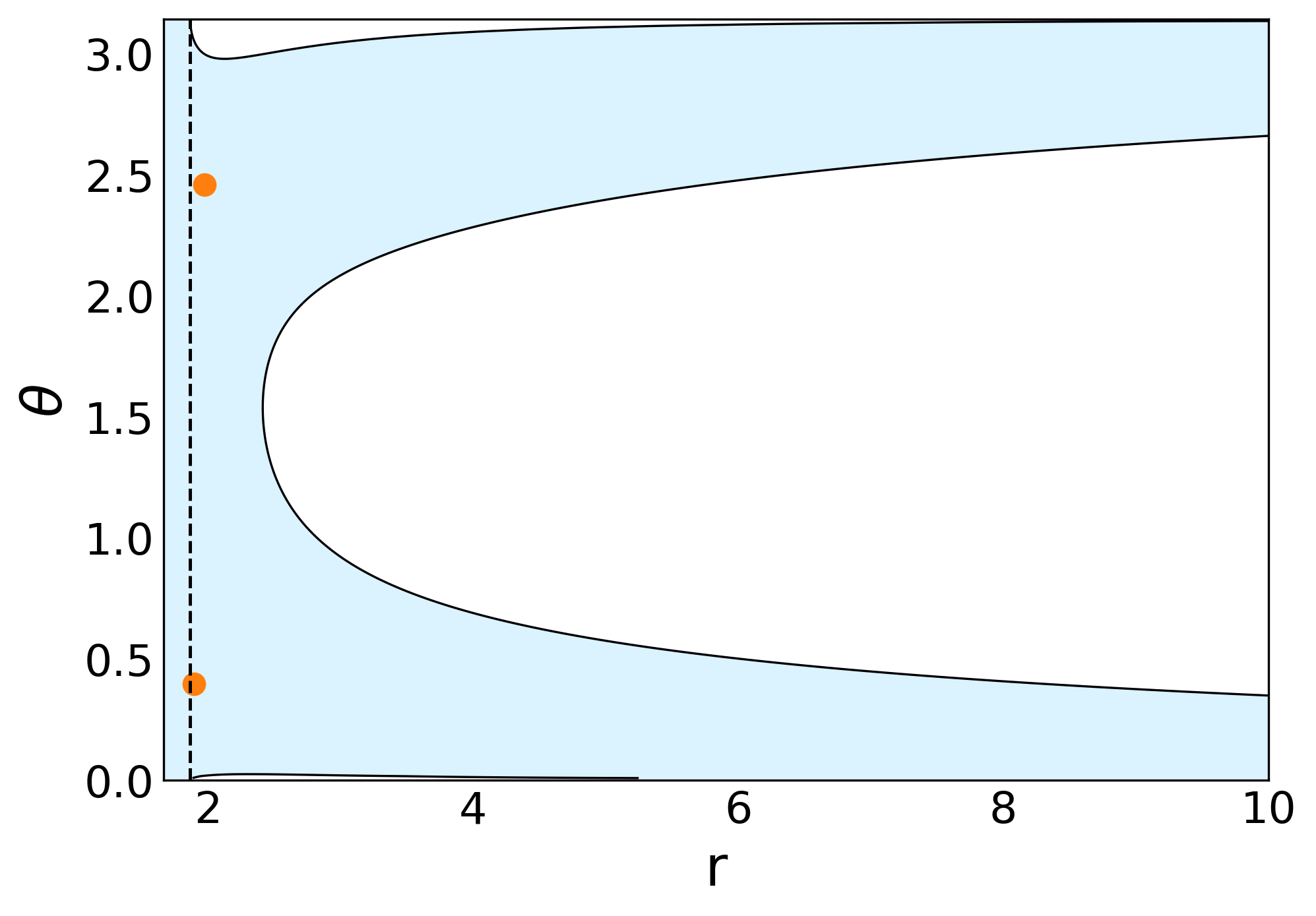}
            \caption*{$jM^2=1.0$}
        \end{subfigure}
        
        \caption{$a=0.5M$}
        \label{fig: LR_ergoregion_a_05}
    \end{subfigure}
    
\hfill

    \begin{subfigure}[b]{\textwidth}
        \centering
        \begin{subfigure}[b]{4.1cm}
            \includegraphics[width=4.1cm]{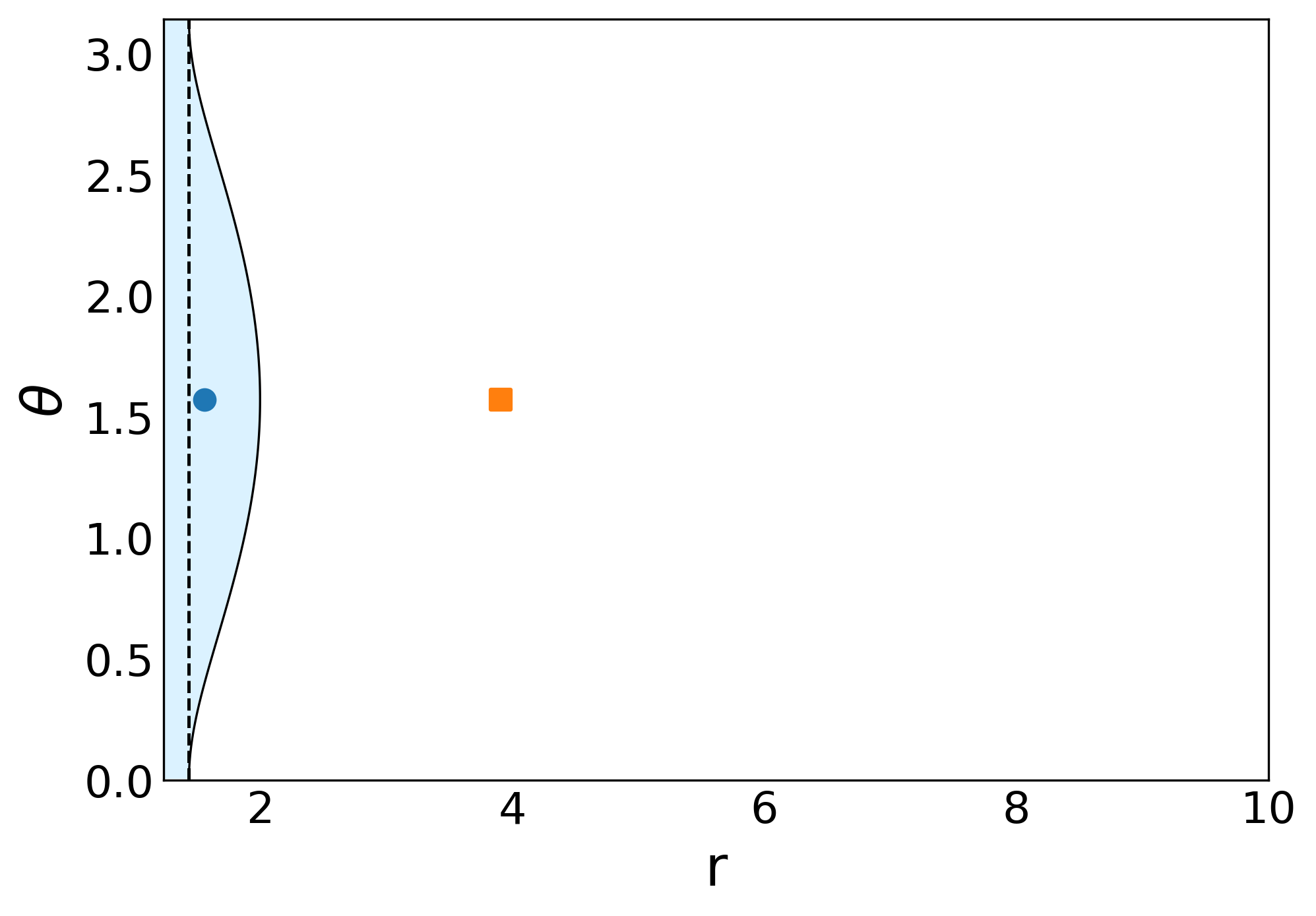}
            \caption*{$j=0$}
        \end{subfigure}
        \begin{subfigure}[b]{4.1cm}
            \includegraphics[width=4.1cm]{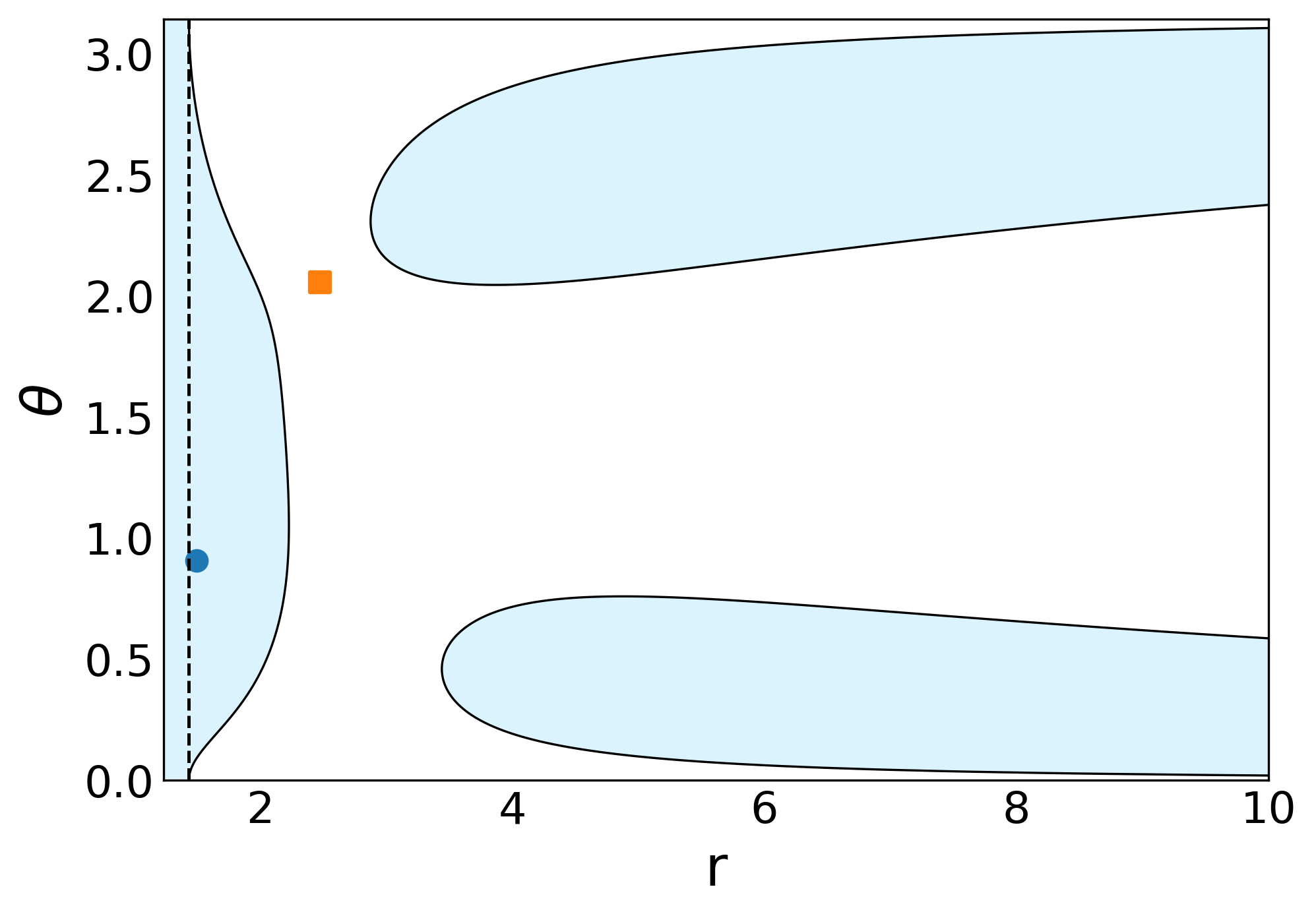}
            \caption*{$jM^2=0.1$}
        \end{subfigure}
        \begin{subfigure}[b]{4.1cm}
            \includegraphics[width=4.1cm]{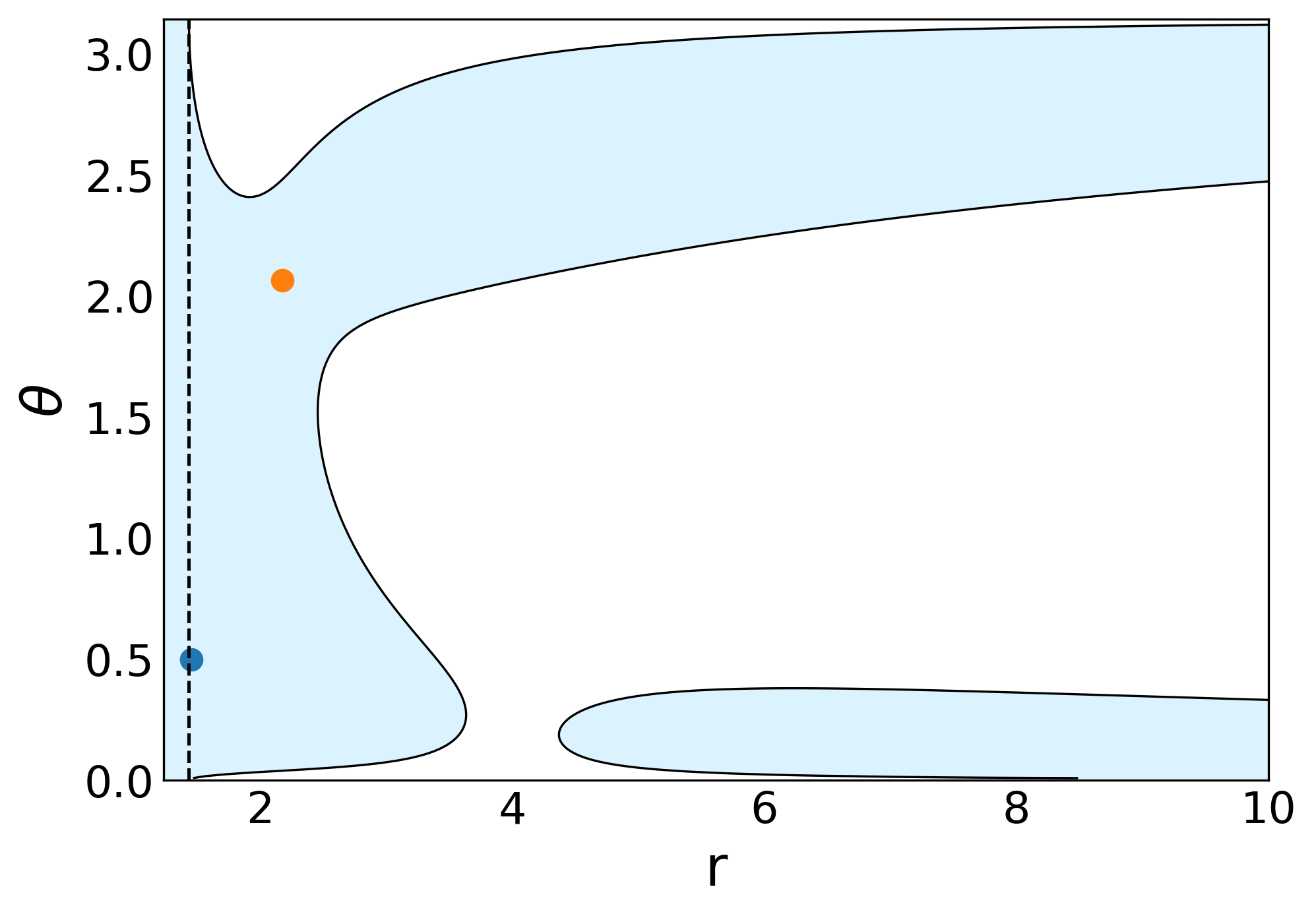}
            \caption*{$jM^2=0.2$}
        \end{subfigure}
        \begin{subfigure}[b]{4.1cm}
            \includegraphics[width=4.1cm]{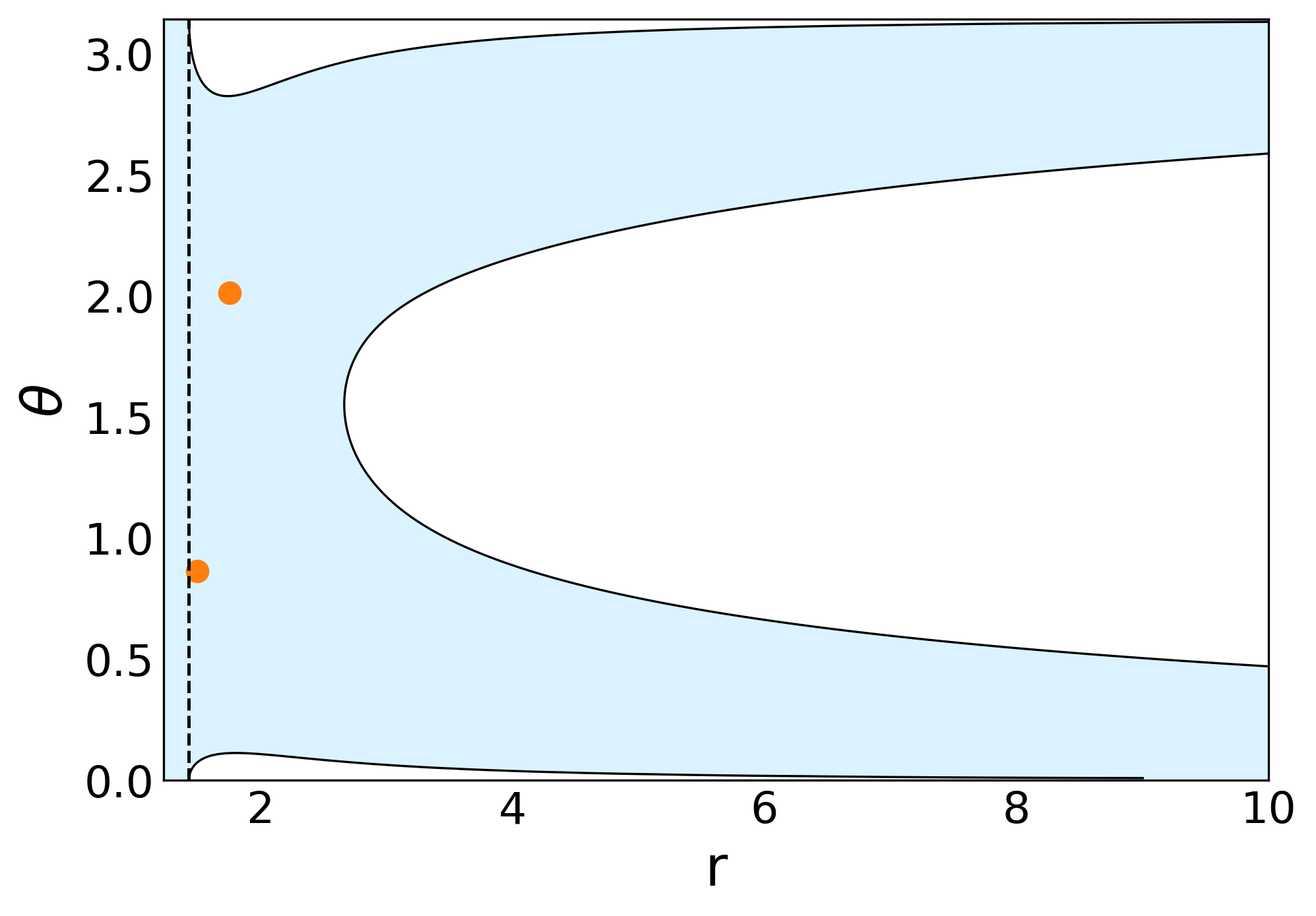}
            \caption*{$jM^2=1.0$}
        \end{subfigure}
        \caption{$a=0.9M$}
        \label{fig: LR_ergoregion_a_09}
    \end{subfigure}        
    
    \caption{Location of photon rings in the $(r,\theta)$-plane for $a=0.5M$ (upper row) and $a=0.9M$ (lower row). The ergoregions are highlighted in light blue and the vertical dashed line represents the radius of the event horizon. The direction of rotation of the light rings is indicated by their markers~: prograde orbits are represented by dots, while retrograde ones are represented by squares. The colors indicate that the LR were obtained from the $H_{\pm}$ potentials. Note that prograde and retrograde here are with respect to the direction of the black hole rotation.}
    \label{fig:LR_rotation}
\end{figure}

In Fig.~\ref{fig:LR_rotation}, we show the ergoregions for $a=0.5M$ and $a=0.9M$ and different values of $jM^2$ in the $(r, \theta)$-plane. We include the locations of the LRs for the respective parameter sets as well as the ergoregions (highlighted in light blue). Here, we indicate whether the direction of rotation of the LR is prograde (circles) or retrograde (squares) with respect to the angular momentum of the BH. 
To find the rotational direction of the photon rings, we consider the azimuthal rotation
given by the quantity $\Phi$ which is defined as:
\begin{equation}
    \Phi=\frac{\rm d\varphi}{\rm d t}=\frac{\dot\varphi}{\dot t}=\frac{\mathcal{F}^2\rho^2L}{E-\omega L}+\omega.
\end{equation}
In an asymptotically flat spacetime this would correspond to the rotational velocity with respect to a static asymptotic observer.
At a light ring, we have the condition $V=0$, i.e. $\mathcal{F}^2\rho^2L^2=(E-\omega L)^2$, i.e., $E-\omega L=\pm \mathcal{F}\rho L$, hence we find
\begin{equation}
    \Phi_\pm=\omega\pm\mathcal{F}\rho.
\end{equation}
Notice that this is identical to the potential $H_\pm$ (\ref{eq:hpot}), thus the rotation of the LR associated with the $H_+$ depends on the sign of $\Phi_+$ and the one with the $H_-$ depends on the sign of $\Phi_-$. 
For the pure Kerr case, we always have two LRs on the equatorial plane: one retrograde orbit and one prograde orbit which lies at a radius smaller than that of the retrograde orbit. This can be explained by the Lense-Thirring effect \cite{Teo:2003ltt}. 
For sufficiently large BH angular momentum $a$, the prograde orbit lies inside the ergoregion (Fig.~\ref{fig: LR_ergoregion_a_09}).  
As $j$ increases and the ergoregions expand, both LRs are pulled off the equatorial plane and eventually lie within the ergoregions, thus co-rotating with the swirling background.


\subsection{Light points}

An interesting phenomenon can be seen taking place in Fig. \ref{fig:LR_rotation}: the outermost light ring (depicted in orange) changes the sign of its angular velocity. For small values of $jM^2$ it is counter-rotating with the black hole, but when the swirling parameter is increased this light ring is captured by the ergoregion and it starts co-rotating with the black hole. Hence, its angular velocity must vanish at some point, indicating the presence of a \emph{light point} \cite{Grandclement:2016eng}. Such peculiar orbits arise when a light ring sits at the boundary of the ergoregion, there the frame dragging can only be compensated by a particle moving at the speed of light. It is precisely this that happens with a photon at a light point - it is counter-rotating at the speed of light but its speed is balanced by the frame-dragging such that it appears to be at rest when seen by an asymptotic observer.

It is possible to show that the light point occurs precisely at the point where the two disconnected ergoregions merge. To prove this, it is important to note that the light point is a saddle point of $g_{tt}$, hence it satisfies:
\begin{equation}
    g_{tt}=0\quad\wedge\quad \nabla g_{tt}=0\,. \label{eq:SaddleErgo}
\end{equation}
The first condition imposes some constraints on the effective potentials, as:
\begin{equation}
    H_\pm\overset{g_{tt}=0}{=}\frac{-g_{t\varphi}\pm \left|g_{t\varphi}\right|}{g_{\varphi\varphi}}\,,
\end{equation}
so if $g_{t\varphi}>0$ one has $H_+=0\,,H_-=-2\left|g_{t\varphi}\right|/g_{\varphi\varphi}$, if $g_{t\varphi}<0$ the role of the potentials is reversed. For concreteness, it is assumed that $g_{t\varphi}>0$. From the definition of the effective potentials it also follows that they are related to $g_{tt}$ by:
\begin{equation}
    g_{tt}=g_{\varphi\varphi}H_+ H_-\,.
\end{equation}
Hence, its gradient can be written as:
\begin{equation}
    \nabla g_{tt}=H_+ \nabla\left(g_{\varphi\varphi}H_-\right)+g_{\varphi\varphi}H_-\nabla H_+\,.
\end{equation}
As seen before, the ergoregion condition implies that $H_+=0$, moreover $H_-<0$ and $g_{\varphi\varphi}$ is also positive everywhere outside the axis. And the axis has no common points with the ergosphere so it is impossible to have $g_{tt}=0$ and $g_{\varphi\varphi}=0$ simultaneously. Therefore, to satisfy $\nabla g_{tt}=0$ one must have $\nabla H_+ =0$, so in summary
\begin{equation}
    g_{tt}=0\: \wedge \: \nabla g_{tt}=0 \Rightarrow \begin{cases}
        H_+=0 \: \wedge \: \nabla H_+ =0\,, \quad g_{t\varphi}>0\,,\\        
        H_-=0 \: \wedge \: \nabla H_- =0\,, \quad g_{t\varphi}<0\,,
    \end{cases}
\end{equation}
But the conditions on the right hand side are precisely the LR conditions, indicating the presence of a light point ($H_\pm=0$) precisely where the ergorregion has a saddle point. Thus, this not only proves that when a light ring coincides with a saddle point of the boundary of the ergoregion it is a light point, but rather that whenever the boundary of the ergoregion has a saddle point, that point corresponds to a light point.

\begin{figure}
    \centering
    \includegraphics[width=0.7\linewidth]{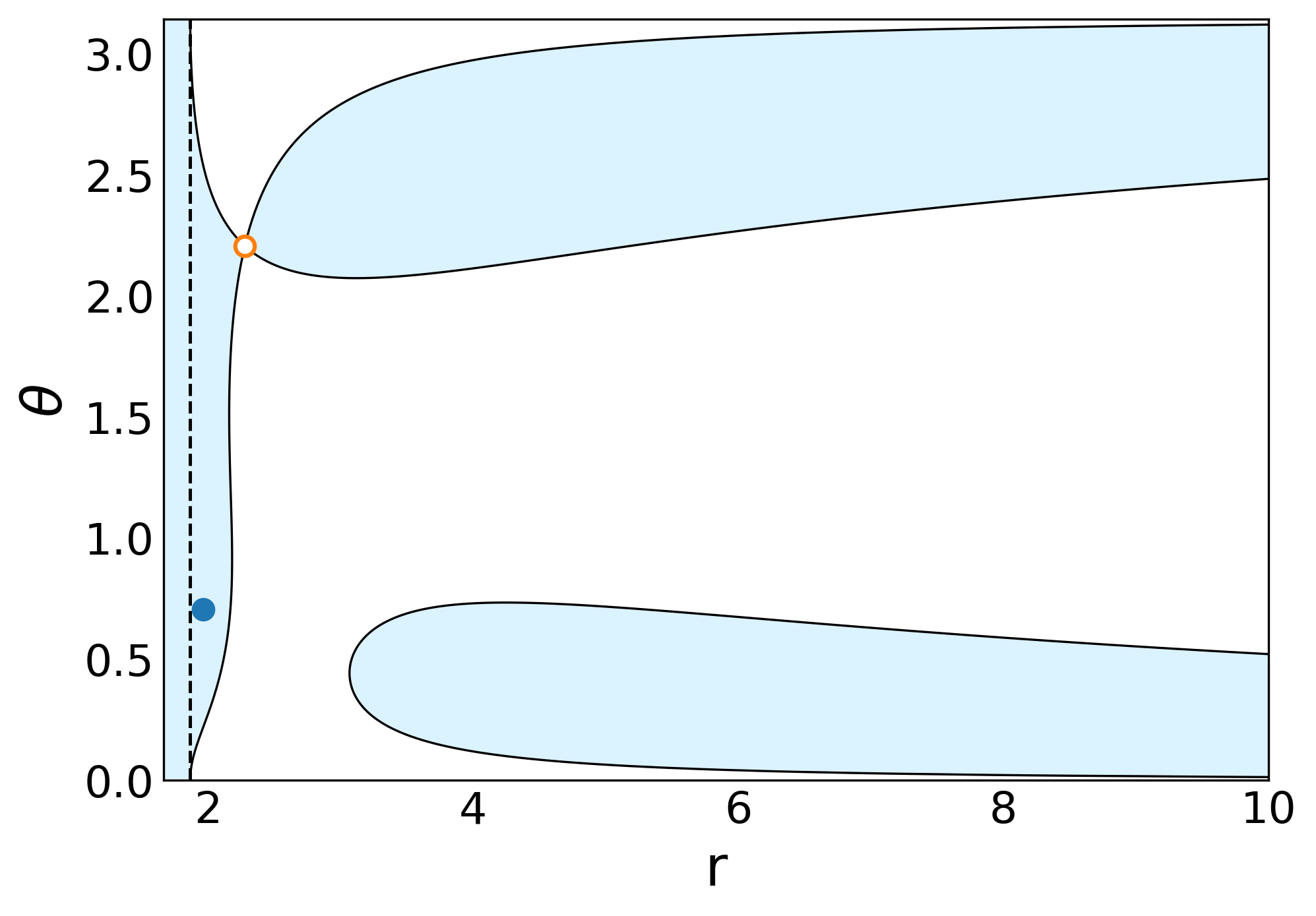}
    \caption{Ergoregion of the KBHSU for $a/M=1/2$ and $jM^2\simeq 0.158579$, highlighted in blue. Here, the appearance of a saddle point is made clear, occurring when two previously disconnected regions merge. The light ring (blue circle) and light point (orange circle) of this solution are also shown. To highlight the absence of rotation the light point is represented by an open circle. 
    }
    \label{fig:LightPoint}
\end{figure}

The explicit form of the equations in Eq.~(\ref{eq:SaddleErgo}) is very convoluted and it was not possible to obtain a closed-form solution for the parameters that yield such a configuration or even the location of the light point. Nonetheless, by solving it numerically we found that for $a/M=1/2$ and $jM^2\simeq 0.158579$ the ergoregion develops precisely the type of saddle point described above, and the light point is indeed present there. This is presented in Fig.~\ref{fig:LightPoint}. To indicate the absence of angular velocity the light point is represented by an open circle.

To the best of our knowledge, the appearance of light points had only been observed for spinning boson stars on the onset of the formation of an ergotorus \cite{Grandclement:2016eng}, where the conditions $g_{tt}=0\: \wedge \: \nabla g_{tt}=0$ are also satisfied but correspond to a local maximum, instead of a saddle point. Light rings that appear at the onset of the ergoregion's appearance are also stable, related to the fact that they correspond to local extrema of $g_{tt}$. Therefore, this represents the first instance in the literature of a light point in a black hole spacetime, which is unstable and possesses a distinct formation mechanism as compared to previous cases (merging of disconnected sections of ergoregions).

\subsection{Topological charge}

The fact that the light rings are critical points of some potential defined on the 2-dimensional space spanned by the non-Killing coordinates for the definition of a topological charge associated with the spacetime outside the possible event horizons. This topological charge yields strict bounds on the possible number of light rings on the spacetime.

For stationary, axially symmetric asymptotically flat spacetimes this topological charge has been used to establish that all spacetimes containing $N$ non-extremal event horizons always possess at least $N$ light rings per rotation sense~\cite{Cunha:2020azh,Cunha:2024ajc}. However, the asymptotic swirling behavior of the metric (\ref{eq:metric}) places it outside the scope of those theorems. Nonetheless, such theorems have also been used to extract information on the number of light rings in other non-asymptotically flat, including asymptotically swirling spacetimes~\cite{Moreira:2024sjq, Junior:2021dyw, Junior:2021svb}.

For the following analysis, it is more convenient to work with the metric in Bach-Weyl coordinates $\{\rho,z\}$\cite{Moreira:2024sjq}. The relation between the two coordinate systems is given by:
\begin{eqnarray}
    \rho &=& \sqrt{\Delta} \sin\theta, \nonumber \\
    z &=& (r-M) \cos\theta , \label{eq:KerrBW_transf}
\end{eqnarray}
such that the metric now reads
\begin{equation}
\label{eq:metricBW}
    {\rm d}s^2 = \frac{1}{\mathcal{F}(r,\theta)} \left( {\rm d}\varphi - \omega {\rm d}t  \right)^2 + \mathcal{F}(r,\theta) \left[ -\rho^2 {\rm d}t^2 + \frac{\Sigma \sin^2\theta}{\Delta+(M^2-a^2)\sin^2\theta} \left({\rm d}\rho^2 + {\rm d}z^2 \right) \right],
\end{equation}
where the $r$ and $\theta$ functions of $\rho$ and $z$ are implicitly given by the transformation in Eq.~(\ref{eq:KerrBW_transf}). The horizon corresponds now to the rod $\rho=0\,,-\sqrt{M^2-a^2}<z<\sqrt{M^2-a^2}$, while the remaining points with $\rho=0$ correspond to the rotation axis.

Let $\mathbf{v}^\pm$ be the vector field defined from the potential $H_\pm$ as:
\begin{equation}
    \mathbf{v}^\pm = \left(\frac{1}{\sqrt{g_{\rho\rho}}}\partial_\rho H_\pm,\frac{1}{\sqrt{g_{zz}}} \partial_z H_\pm\right)=\left(v_\rho^\pm,v_z^\pm\right)\,.
\end{equation}
This vector field can also be parametrized by its magnitude and an angle $\Omega$ such that:
\begin{equation}
    v^\pm_\rho=v^\pm \cos\Omega\,,\quad v^\pm_z=v^\pm\sin\Omega\,,
\end{equation}
where $v^\pm=|\mathbf{v}^\pm|=\sqrt{(v^\pm_\rho)^2+(v^\pm_z)^2}$. Consider a closed curve, $\mathcal{C}$, in the $\{\rho,z\}$ plane. After a revolution along this curve, the angle $\Omega$ must be the same as in the beginning modulo $2\pi$, that is:
\begin{equation}
    \oint_\mathcal{C}\mathrm{d}\Omega=2\pi \omega_{\mathcal{C}}\, ,
\end{equation}
where $\omega \in \mathbb{Z}$ is the winding number of the vector field along the curve. The contributions to this winding number come from the critical points enclosed by $\mathcal{C}$, with an extremum (maximum or minimum) contributing with $+1$ and a saddle point contributing with $-1$. Notice that, since each critical point corresponds to the location of a light ring the winding number along some curve $\mathcal{C}$ can be used to extract information on the number of light rings it encloses. 

In order to compute the total number of light rings on the spacetime outside the black hole, a contour $\mathcal{C}$ is considered such that in the appropriate limit, it encloses all the exterior spacetime. In Bach-Weyl coordinates, such a contour can be made of the union of the following six segments:
\begin{align}
    \mathcal{I}_1=&\left\{\rho=M\xi\, ,-M\xi<z<M\xi\right\}\,,\\
    \mathcal{I}_2=&\left\{\frac{M}{\xi}<\rho<M\xi\, ,z=M\xi\right\}\,,\\
    \mathcal{I}_3=&\left\{\rho=\frac{M}{\xi}\, ,\sqrt{M^2-a^2}<z<M\xi\right\}\,,\\
    \mathcal{I}_4=&\left\{\rho=\frac{M}{\xi}\, ,-\sqrt{M^2-a^2}<z<\sqrt{M^2-a^2}\right\}\,,\\
    \mathcal{I}_5=&\left\{\rho=\frac{M}{\xi}\, ,-M\xi<z<-\sqrt{M^2-a^2}\right\}\,,\\
    \mathcal{I}_6=&\left\{\frac{M}{\xi}<\rho<M\xi\, z=-\xi\right\}\,.
\end{align}
That is $\mathcal{C}=\bigcup_i \mathcal{I}_i$. Notice that $\xi\in\mathbb{R}$ is a dimensionless parameter. In the limit $\xi\rightarrow\infty$, the contour $\mathcal{C}$ encloses all the spacetime outside the horizon, such that $\mathcal{I}_1,\,\mathcal{I}_2$ and $\mathcal{I}_6$ capture the behavior of the potential at spatial infinity, $\mathcal{I}_4$ the horizon contribution, and the behavior near the rotational axis is captured by $\mathcal{I}_3$ together with $\mathcal{I}_5$. Under this decomposition, the total topological charge can also be decomposed in the contribution along each segment:
\begin{equation}
    w_\mathcal{C}=\oint_\mathcal{C}\mathrm{d}\Omega=\sum_i\omega_{\mathcal{I}_i}\,,
\end{equation}
where
\begin{align}
    \nonumber
    \omega_{\mathcal{I}_1}&=\frac{1}{2\pi}\intop_{-M\xi}^{M\xi}\left.\frac{\mathrm{d}\Omega}{dz}\right|_{\rho=M\xi}\mathrm{dz}\:,&
    \omega_{\mathcal{I}_2}&=\frac{1}{2\pi}\intop_{M\xi}^{M/\xi}\left.\frac{\mathrm{d}\Omega}{d\rho}\right|_{z=M\xi}\mathrm{d\rho}\:,\\    
    \omega_{\mathcal{I}_3}&=\frac{1}{2\pi}\intop_{M\xi}^{\sqrt{M^2-a^2}}\left.\frac{\mathrm{d}\Omega}{dz}\right|_{\rho=M/\xi}\mathrm{dz}\:,&
    \omega_{\mathcal{I}_4}&=\frac{1}{2\pi}\intop_{\sqrt{M^2-a^2}}^{-\sqrt{M^2-a^2}}\left.\frac{\mathrm{d}\Omega}{dz}\right|_{\rho=M/\xi}\mathrm{dz}\:,\quad\\
    \omega_{\mathcal{I}_5}&=\frac{1}{2\pi}\intop_{-\sqrt{M^2-a^2}}^{-M\xi}\left.\frac{\mathrm{d}\Omega}{dz}\right|_{\rho=M/\xi}\mathrm{dz}\:,&
    \omega_{\mathcal{I}_6}&=\frac{1}{2\pi}\intop_{M/\xi}^{M\xi}\left.\frac{\mathrm{d}\Omega}{d\rho}\right|_{z=-M\xi}\mathrm{d\rho} .\nonumber
\end{align}

Now one can study the behavior of the vector field $\mathbf{v^\pm}$ along each of the segments by making the appropriate perturbative expansions. Before continuing this analysis, it is important to notice that the effective potentials can be transformed into one another via the transformation $j\rightarrow -j$ and $\theta\rightarrow \pi-\theta$ (or equivalently $z\rightarrow -z$ in Bach-Weyl coordinates), i.e. $H_+^a(r,\theta)=-H_-^{-a}(r,\pi-\theta)$. Therefore, it is sufficient to analyze the properties of one of the potentials, for instance $H_+$. Moreover, the behavior of the vector field close to $\rho=0$ is strictly fixed by the regularity of the axis and the non-extremality of the horizon (which will be assumed from now on). Such a study has been done in~\cite{Cunha:2024ajc}. There, it was concluded that each timelike rod (topologically spherical horizon) contributes with $-1$ to the total topological charge, so:
\begin{equation}
    \lim_{\xi\rightarrow\infty} \left(\omega_{\mathcal{I}_3}+\omega_{\mathcal{I}_4}+\omega_{\mathcal{I}_5}\right) = -1\,.
\end{equation}
Now we consider the asymptotic behavior of the vector field on the remaining segments:

Let us consider the behavior along $\mathcal{I}_1$, for $\rho=M\xi\rightarrow\infty$ one obtains:
\begin{equation}
    \left(v^+_\rho,v^+_z\right)\sim\left(3|j|,0\right)\,.
\end{equation}
Since $\mathbf{v^+}$ is constant up to zeroth order in $\mathcal{O}(\xi^{-1})$ there is no contribution to the winding number along this segment, i.e. $\omega_{\mathcal{I}_1}=0$.

Next, we consider the segment $\mathcal{I}_2$, where $z\rightarrow\infty$, then:
\begin{equation}
    \left(v^+_\rho,v^+_z\right)\sim\frac{1}{\sqrt{\left(1-4ajM\right)^2+\rho^4j^2}}\left(3j^2\rho^2-\frac{\left(1-4ajM\right)^2}{\rho^2},4j\left(1-4ajM\right)\right)\,.
\end{equation}
Notice that if $v^\pm_z>0\Rightarrow\Omega\in]0,\pi[$ and $v^\pm_z<0\Rightarrow\Omega\in]-\pi,0[$, then, it is possible to express the angle $\Omega\left(\rho,z\right)$ depending on the two possibilities for the value of $1-4ajM$:
\begin{equation}
    \Omega\left(\rho,z\right)=\mathrm{sign}(j)\mathrm{sign}\left(1-4ajM\right)\arccos\left(\frac{v^\pm_\rho}{|\mathbf{v}^\pm|}\right)\,.
\end{equation}
Then one can compute $\omega_{\mathcal{I}_2}$ to be:
\begin{equation}
    2\pi\omega_\mathcal{{I}_2}=\left[\arccos\left(\frac{v^+_\rho}{|\mathbf{v}^+|}\right)\right]_{\rho=M\xi}^{\rho=M/\xi} = \mathrm{sign}(j)\mathrm{sign}\left(1-4ajM\right)\pi + \mathcal{O}\left(\xi^{-1}\right)\,.
\end{equation}

Only the behavior along $\mathcal{I}_6$ remains to be considered, this corresponds to the limit $z\rightarrow-\infty$, where:
\begin{equation}
    \left(v^+_\rho,v^+_z\right)\sim\frac{1}{\sqrt{\left(1+4ajM\right)^2+\rho^4j^2}}\left(3j^2\rho^2-\frac{\left(1+4ajM\right)^2}{\rho^2},4j\left(1+4ajM\right)\right)\,.
\end{equation}
By the same procedure as before one finds:
\begin{equation}
    2\pi\omega_{\mathcal{I}_6} = \left[\arccos\left( \frac{v^+_\rho}{|\mathbf{v}^+|} \right)\right]^{\rho=M\xi}_{\rho=M/\xi} = -\mathrm{sign}(j)\mathrm{sign}\left(1+4ajM\right)\pi + \mathcal{O}\left(\xi^{-1}\right)\,.
\end{equation}
It is useful to compute the contribution from the three segments $\mathcal{I}_1\,,\mathcal{I}_2$ and $\mathcal{I}_6$ together, which yield:
\begin{equation}
    \lim_{\xi\rightarrow\infty} \left( \omega_{\mathcal{I}_1}+\omega_{\mathcal{I}_2}+\omega_{\mathcal{I}_6}\right) = \frac{\mathrm{sign}\left(j\right)}{2}\left[\mathrm{sign}\left(1-4ajM\right)-  \mathrm{sign}\left(1+4ajM\right) \right]\,.
\end{equation}
There are three possibilities for the term in square brackets:
\begin{equation}
    \mathrm{sign}\left(1-4ajM\right)-  \mathrm{sign}\left(1+4ajM\right) =
    \begin{cases}
    \ \   2\,,\quad a j M<-1/4\\
    \ \  0\,,\quad -1/4<a j M<1/4\\
        -2\,,\quad ajM>1/4
    \end{cases}\, .
\end{equation}

So one has three possibilities for the total topological charge of the spacetime depending on the value of the parameter $ajM$, which, considering $j>0$ are:
\begin{equation}
    \omega^+=\lim_{\xi\rightarrow \infty}\omega^+_C=
    \begin{cases}
      \ \  0\,,\quad a j M<-1/4\\
        -1\,,\quad -1/4<a j M<1/4\\
        -2\,,\quad ajM>1/4
    \end{cases}\,.
\end{equation}
For $j<0$ the possibilities are the same but correspond to different regions of the parameter space. 

\begin{figure}
    \centering
    \includegraphics[width=0.4\linewidth]{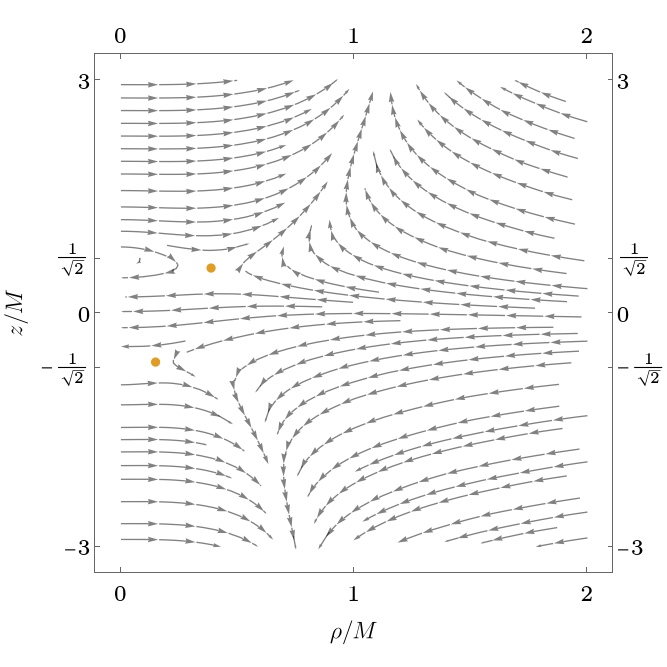}
    \includegraphics[width=0.4\linewidth]{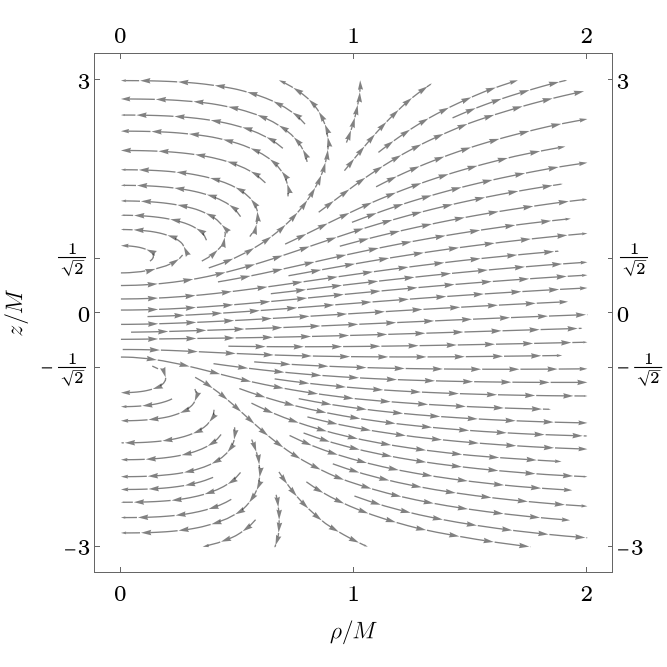}\\
    \includegraphics[width=0.4\linewidth]{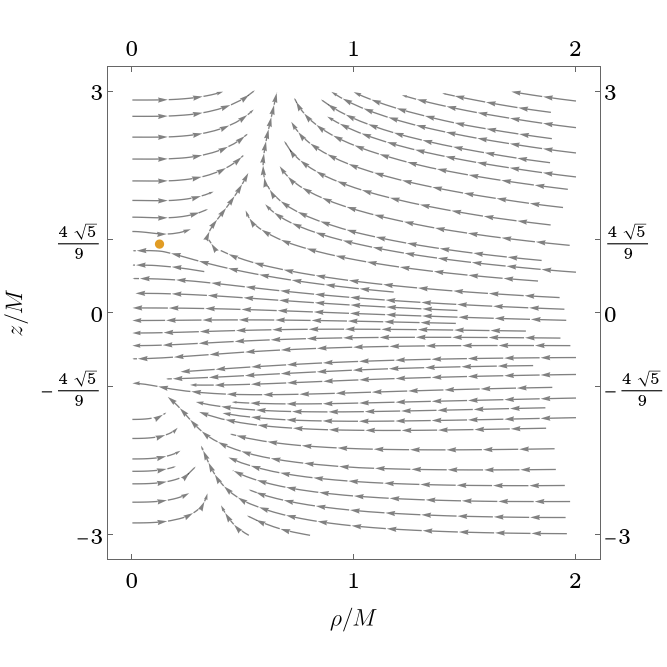}
    \includegraphics[width=0.4\linewidth]{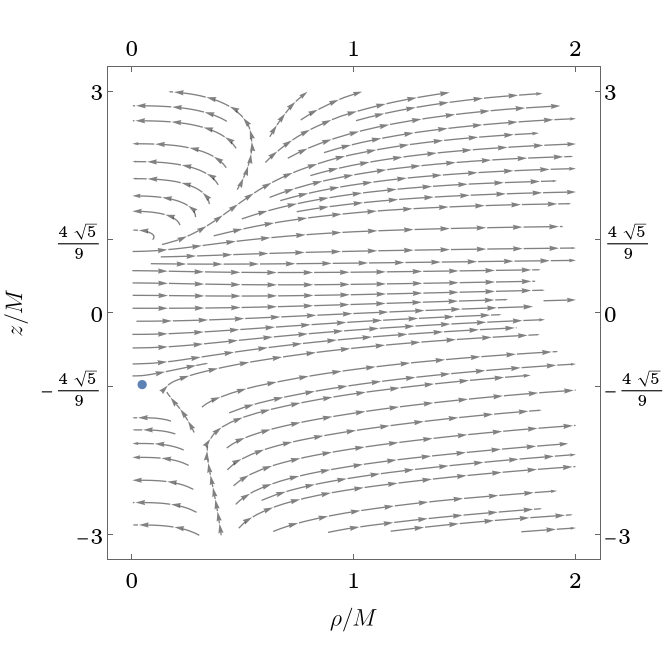}\\
    \includegraphics[width=0.4\linewidth]{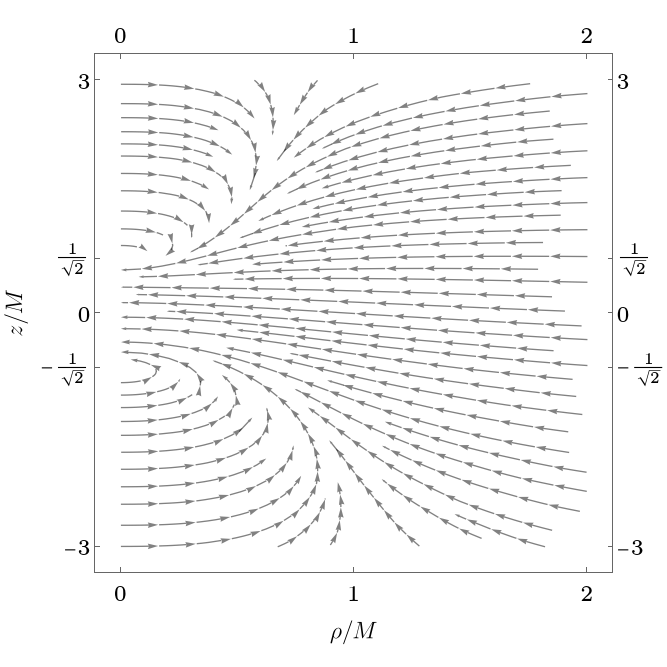}
    \includegraphics[width=0.4\linewidth]{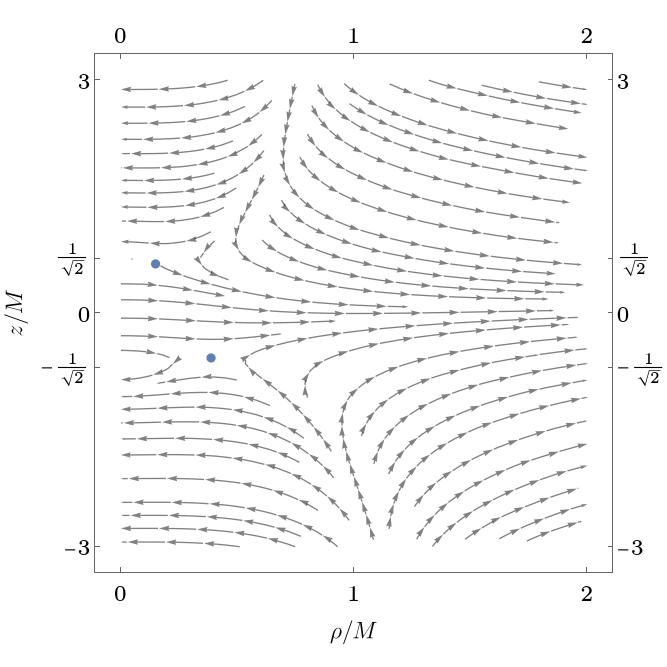}    
    \caption{Vector fields $\mathbf{v}^\pm$ for $jM^2=1$ and several values of the black hole spin parameter. On the left (right) column the vector $\mathbf{v}^-$ ($\mathbf{v}^+$) is plotted, with $a/M=\{-\sqrt{2}/2\,,\,-1/9,\,\sqrt{2}/2\}$ from top to bottom. It is possible to verify that all configurations possess in total 2 light rings, but for the top and bottom values of $a$ one potential contributes with both light rings, while for the middle row each potential contributes with one. }
    \label{fig:VectorFields}
\end{figure}

At first glance this result seems to contradict the expectation that the topological charge is conserved under deformations of the spacetime, as one can change only one parameter and obtain three different values for the total topological charge for each potential. This occurs because the deformation is not smooth, in fact at the limiting values of the regions, i.e. $4ajM=\pm1$ the metric is singular, as discussed previously. Another remark is that the light rings associated with the potential $H_-$ can be obtained by the substitution $a\rightarrow-a$ and $\theta\rightarrow\pi-\theta$, thus the topological charge associated with $\mathbf{v}^-$ is:
\begin{equation}
    \omega^-=\lim_{\xi\rightarrow0}\omega^+_C=
    \begin{cases}
        -2\,,\quad a j M<-1/4\\
        -1\,,\quad -1/4<a j M<1/4\\
    \ \    0\,,\quad ajM>1/4
    \end{cases}\,.
\end{equation}
Interestingly, the sum of the topological charges is independent of the value of $4ajM$, $\omega^++\omega^-=-2$.
The topological charge gives only the surplus of unstable light rings with respect to the stable ones, however, computing explicitly the location of the LRs for the considered spacetime, one never finds stable LRs. So there are only unstable LRs whose number is the modulus of the topological charge, which in this case is always $2$, despite the fact that the contribution from each potential may vary.
In Fig.~\ref{fig:VectorFields} the vector potentials are plotted for different values of $ajM$ such that all possibilities are covered.

\section{Shadows}
\begin{figure}[h!]
    \centering
    \begin{subfigure}[t]{0.2\textwidth}
        \centering
        \includegraphics[width=\textwidth]{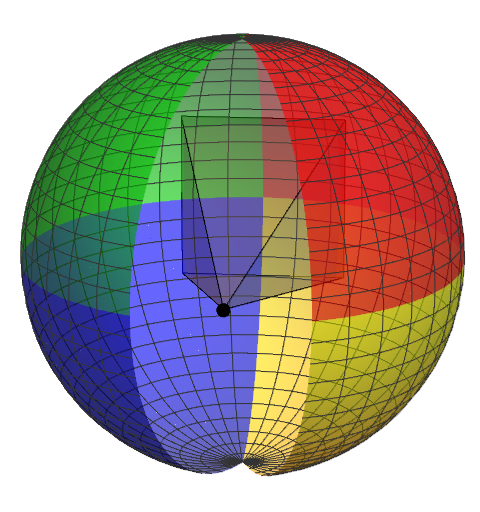}
        \caption{Schematic representation}
        \label{code_schematic}
    \end{subfigure}
    \begin{subfigure}[t]{0.2\textwidth}
        \centering
        \includegraphics[width=\textwidth]{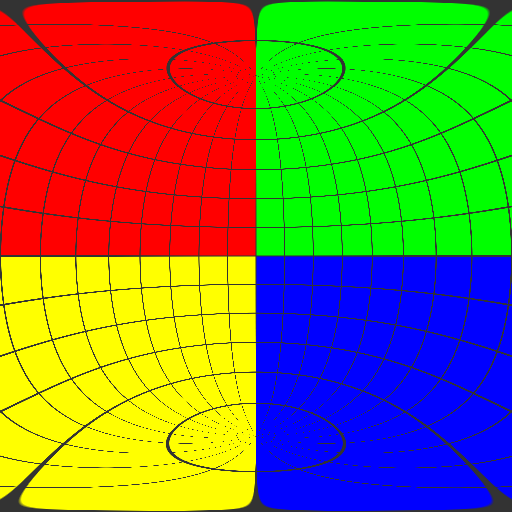}
        \caption{Minkowski}
        \label{lensing_Mink}
    \end{subfigure}
    \begin{subfigure}[t]{0.2\textwidth}
        \centering
        \includegraphics[width=\textwidth]{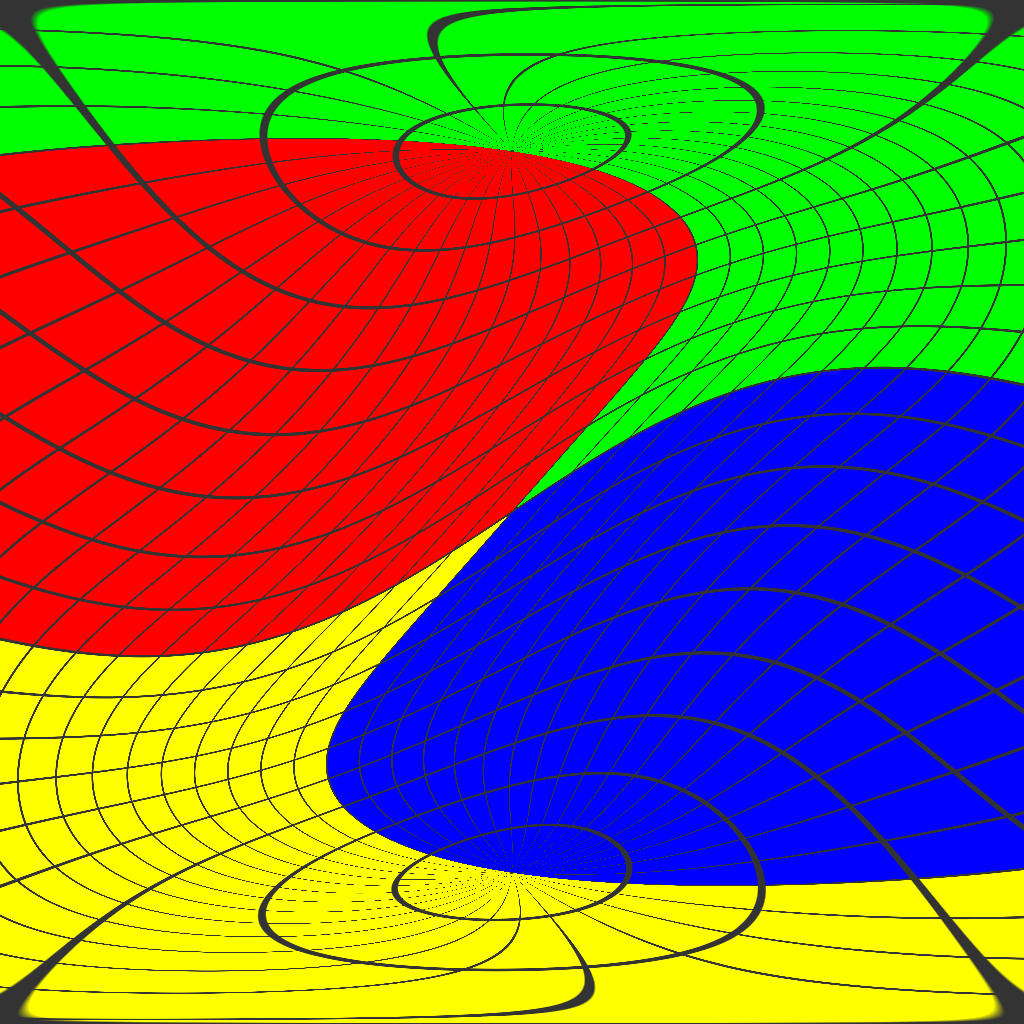}
        \caption{Swirling universe}
        \label{lensing_swirling}
    \end{subfigure}

    \caption{(a) 
    A schematic representation of the numerical setup. 
    The small black sphere represents the observer, while the sphere with four coloured sections represents the celestial sphere (or the observer's sky). 
    The observer's field of view is represented by a pentahedron in dark gray, resulting in the image seen by the observer. 
    In (b), we show the view from a Minkowski spacetime, while in (c), the view is from a swirling universe with j = 0.0005.}
    \label{PYHOLE_schematic}
\end{figure}

In this section, we will focus on the BH shadow and the lensing effect generated by the presence of the BH. 
The methods to study the shadow's silhouette 
depend on the integrability of the equations of motion. 
When the geodesic motion for photons is separable, such as in the pure-Kerr case, the BH shadow can be obtained analytically \cite{Bardeen:1972fi}; for non-separable equations, a full numerical approach is required. 
Here, we follow the numerical 
method applied previously for the SBHSU solution \cite{Moreira:2024sjq}. 
Thus we apply a backwards ray-tracing method to integrate the equations of motion \cite{Cunha:2016bjh}.
The geodesic equations for photons are then evolved backward in time starting from the \textit{observer}'s position. 
The photon's trajectory can either end by falling into the black hole, or once it reaches the \textit{celestial sphere}. 
This is equivalent to considering light isotropically coming from every point in the celestial sphere and reaching the observer. 
A schematic representation is shown in Fig.~\ref{PYHOLE_schematic}.

The lensing images were produced using the PyHole package implemented in Python \cite{Cunha:2016bjh}. 
The observer is placed off-center at $(r_o,\theta_o)$ inside the celestial sphere, which is set to have a larger radius. 
The routine then integrates the equations of motion (\ref{Ham_eqs}) using the Hamiltonian formalism with initial conditions set from the observer's position as follows~:

\begin{eqnarray}
    p_r &=& \sqrt{g_{rr}} \cos{\alpha} \cos{\beta} \ , \\ \ p_\theta &=& \sqrt{g_{\theta\theta}} \sin{\alpha} \ , \\ \ p_\varphi &=& L = \sqrt{g_{\varphi\varphi}} \cos{\alpha} \sin{\beta} \ ,\\
    -p_t &=& E = \frac{1}{\sqrt{g_{\varphi\varphi}}} \left( \sqrt{D} - g_{t\varphi} \cos{\alpha} \sin{\beta} \right) ,
\end{eqnarray}                                                                                              where $D$ was already defined in (\ref{LR_potential}).
To obtain an image, we scan over the observing angles $(\alpha,\beta)$ fixed by the observer's field of view. 
In our images, the celestial sphere is divided into four sections, each represented with a different colour. 
Overall, the black hole shadows present a clear asymmetrical pattern with the asymmetry more pronounced for an observer on the equatorial plane. 

\begin{figure}[h!]
    \centering
    \begin{subfigure}[t]{0.2\textwidth}
        \centering
        \includegraphics[width=\textwidth]{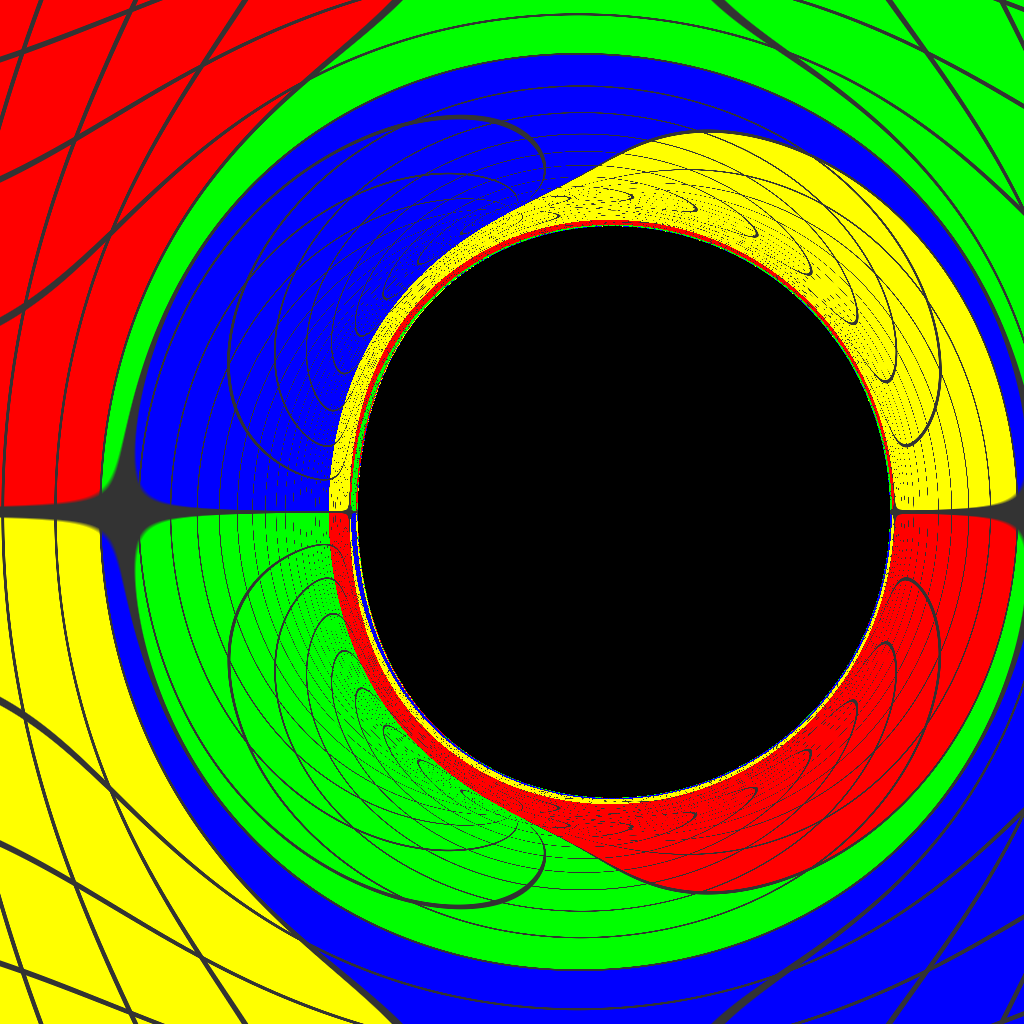}
        \caption{$j=0$}
    \end{subfigure}
    \begin{subfigure}[t]{0.2\textwidth}
        \centering
        \includegraphics[width=\textwidth]{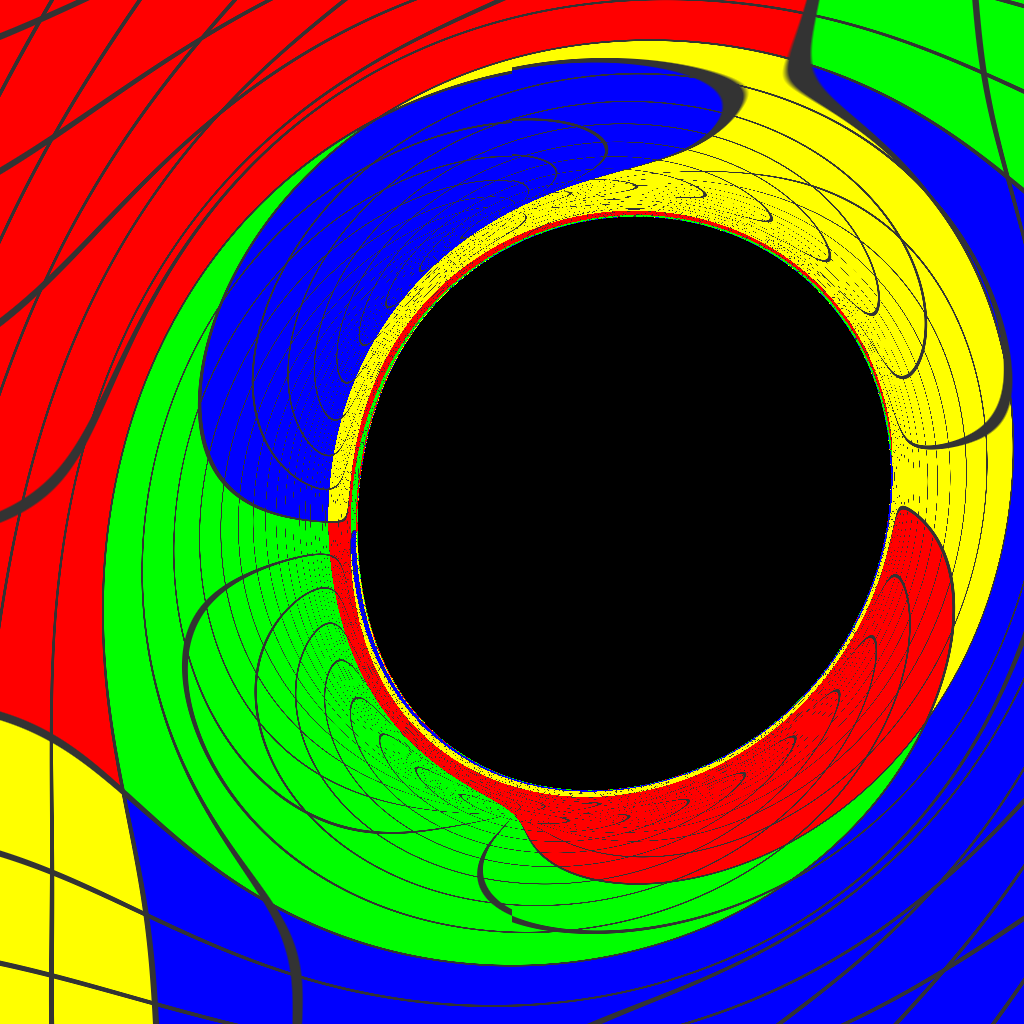}
        \caption{$j=0.0002$}
    \end{subfigure}
    \begin{subfigure}[t]{0.2\textwidth}
        \centering
        \includegraphics[width=\textwidth]{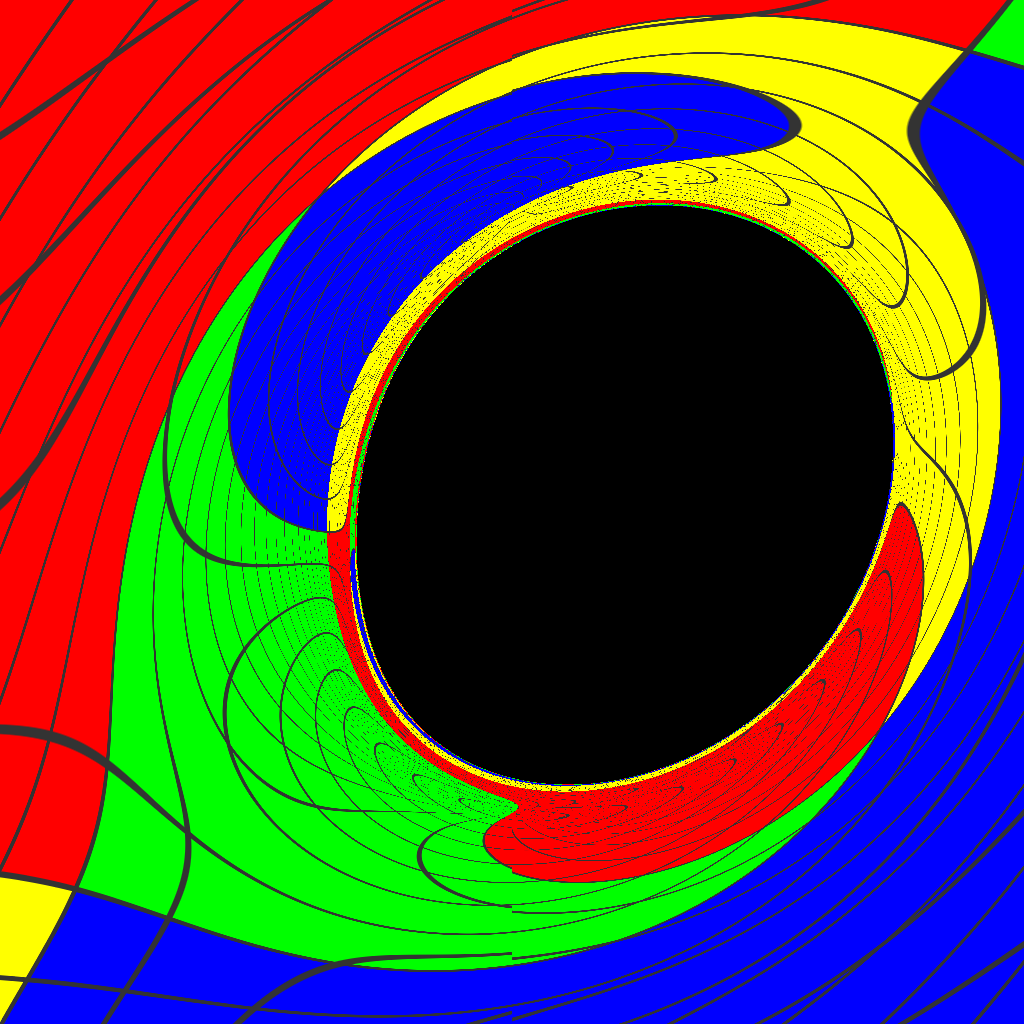}
        \caption{$j=0.0004$} 
    \end{subfigure} \\
    \begin{subfigure}[t]{0.2\textwidth}
        \centering
        \includegraphics[width=\textwidth]{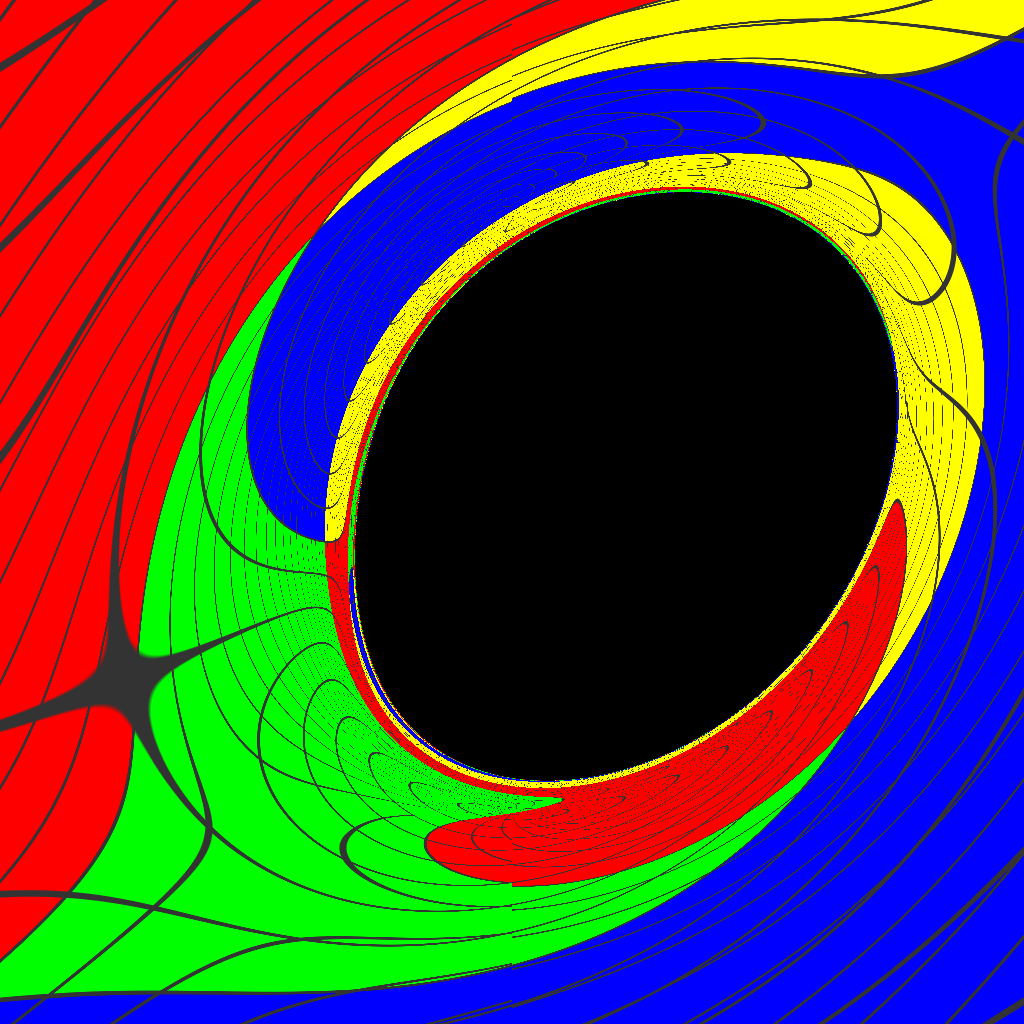}
        \caption{$j=0.0006$}
    \end{subfigure}
    \begin{subfigure}[t]{0.2\textwidth}
        \centering
        \includegraphics[width=\textwidth]{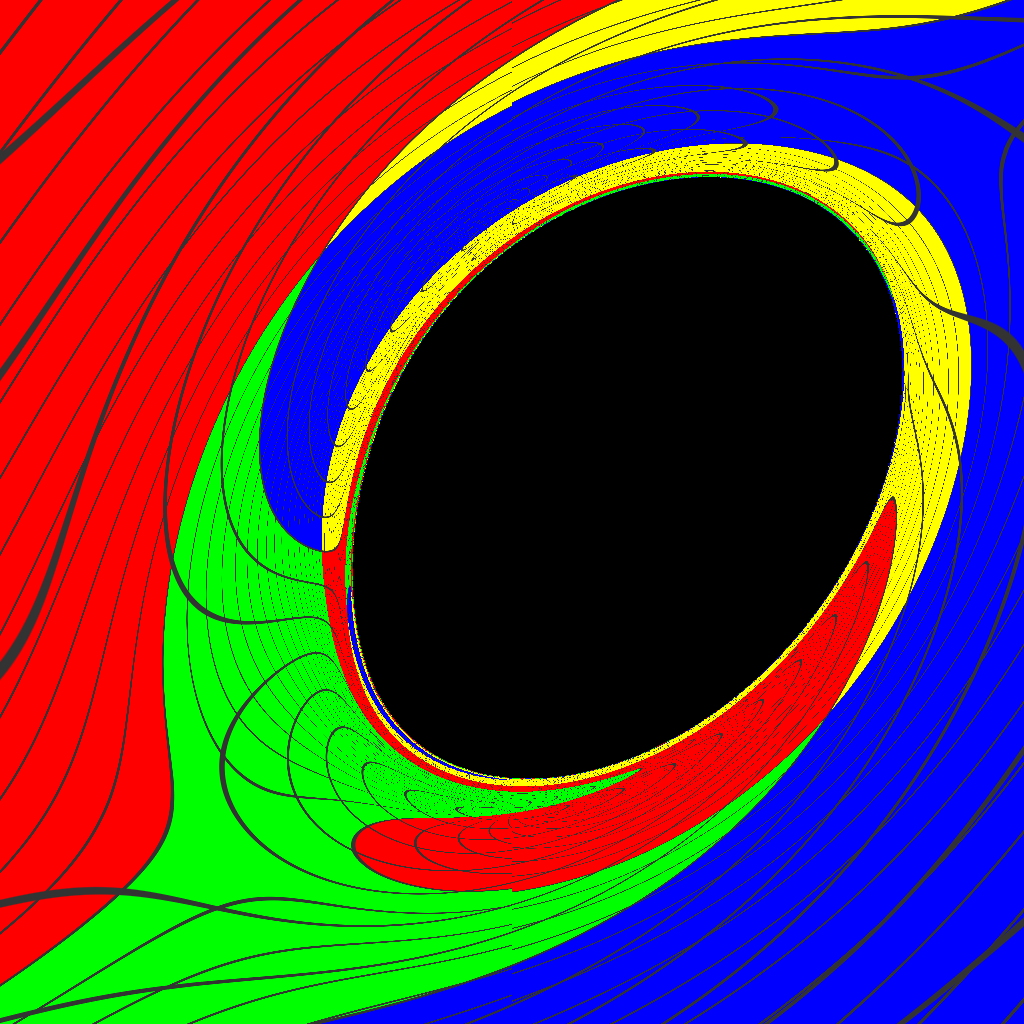}
        \caption{$j=0.0008$}
    \end{subfigure}
    \begin{subfigure}[t]{0.2\textwidth}
        \centering
        \includegraphics[width=\textwidth]{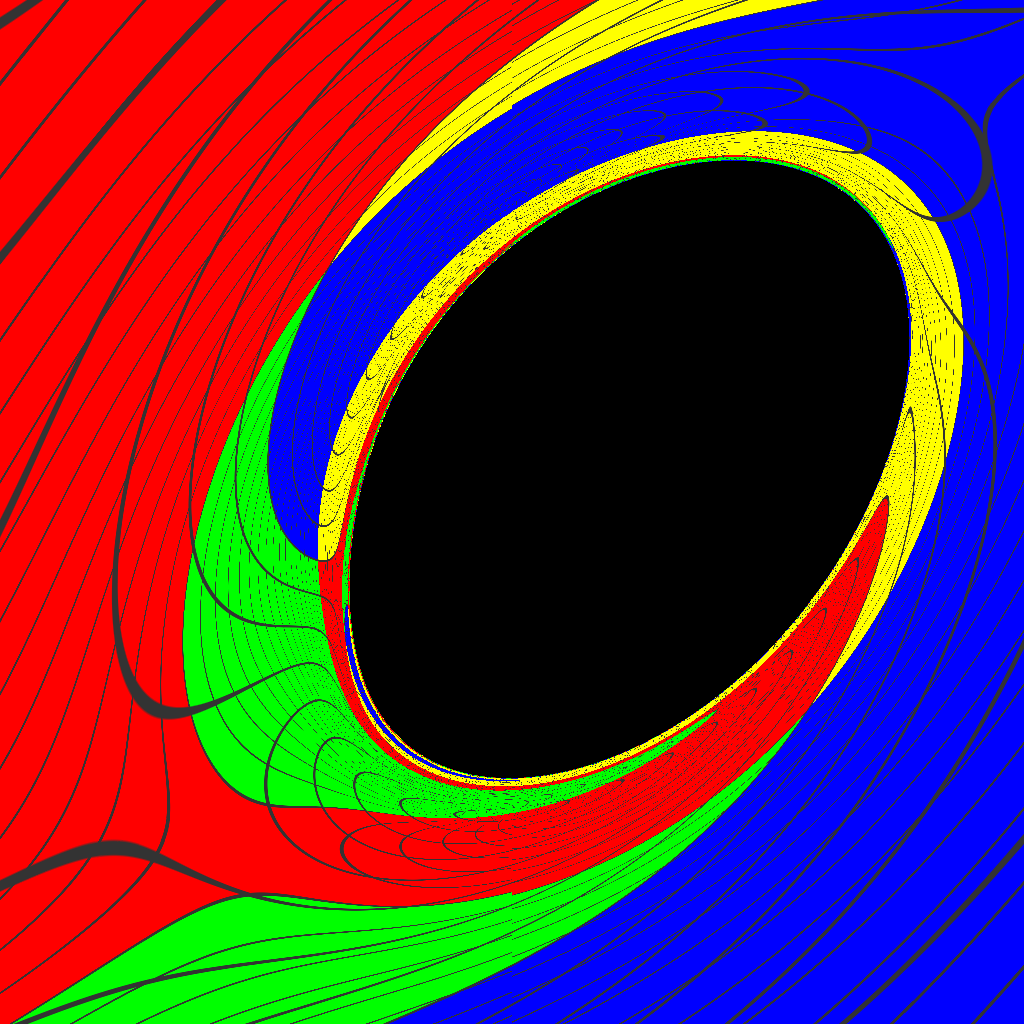}
        \caption{$j=0.001$}
    \end{subfigure}
    \caption{We show the lensing images for a rapidly rotating KBHSU with $a = 0.9M$ and different values of the swirling parameter $j$. 
    The observer is set on the equatorial plane at $r_o = 15$.}
    \label{Lensing_a_0.9_equatorial}
\end{figure}

\begin{figure}[h!]
    \centering
    \begin{subfigure}[t]{0.2\textwidth}
        \centering
        \includegraphics[width=\textwidth]{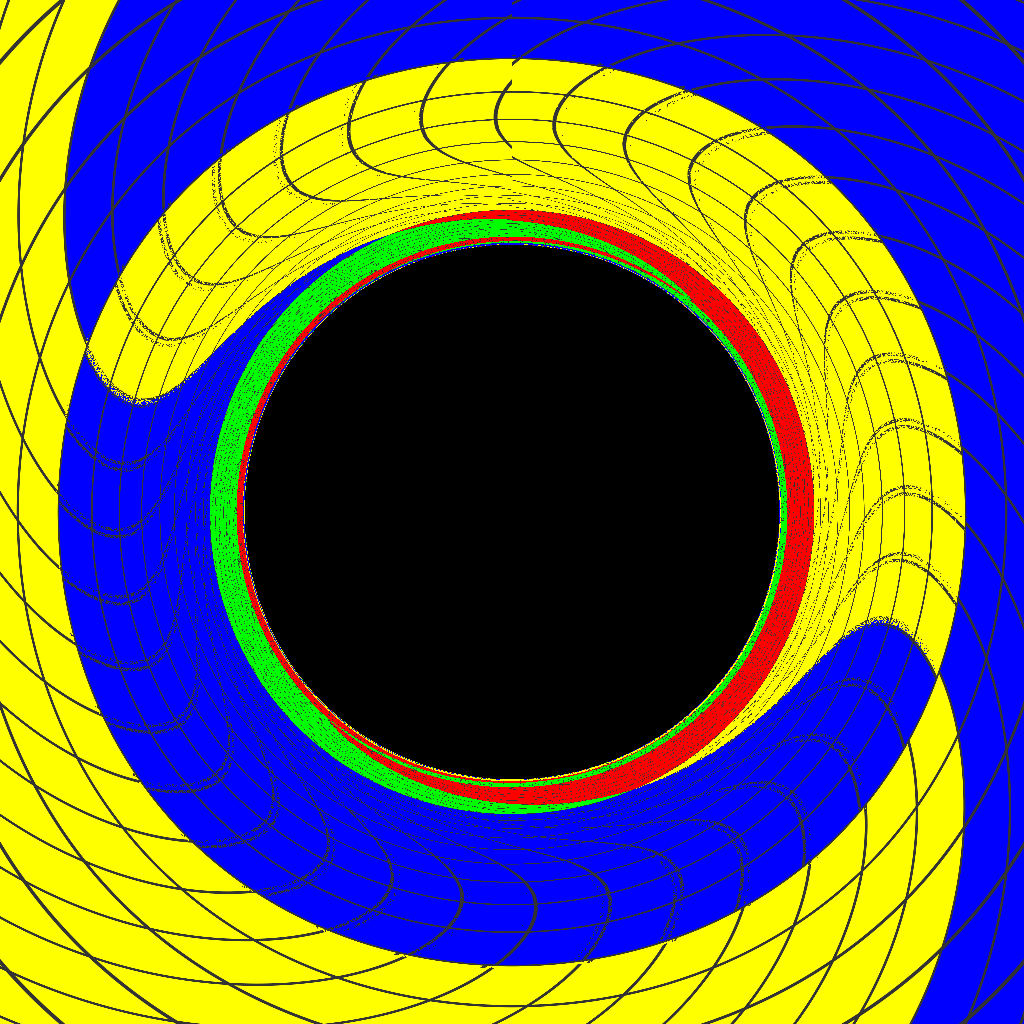}
        \caption{$\theta=10^{-8}$}
    \end{subfigure}
    \begin{subfigure}[t]{0.2\textwidth}
        \centering
        \includegraphics[width=\textwidth]{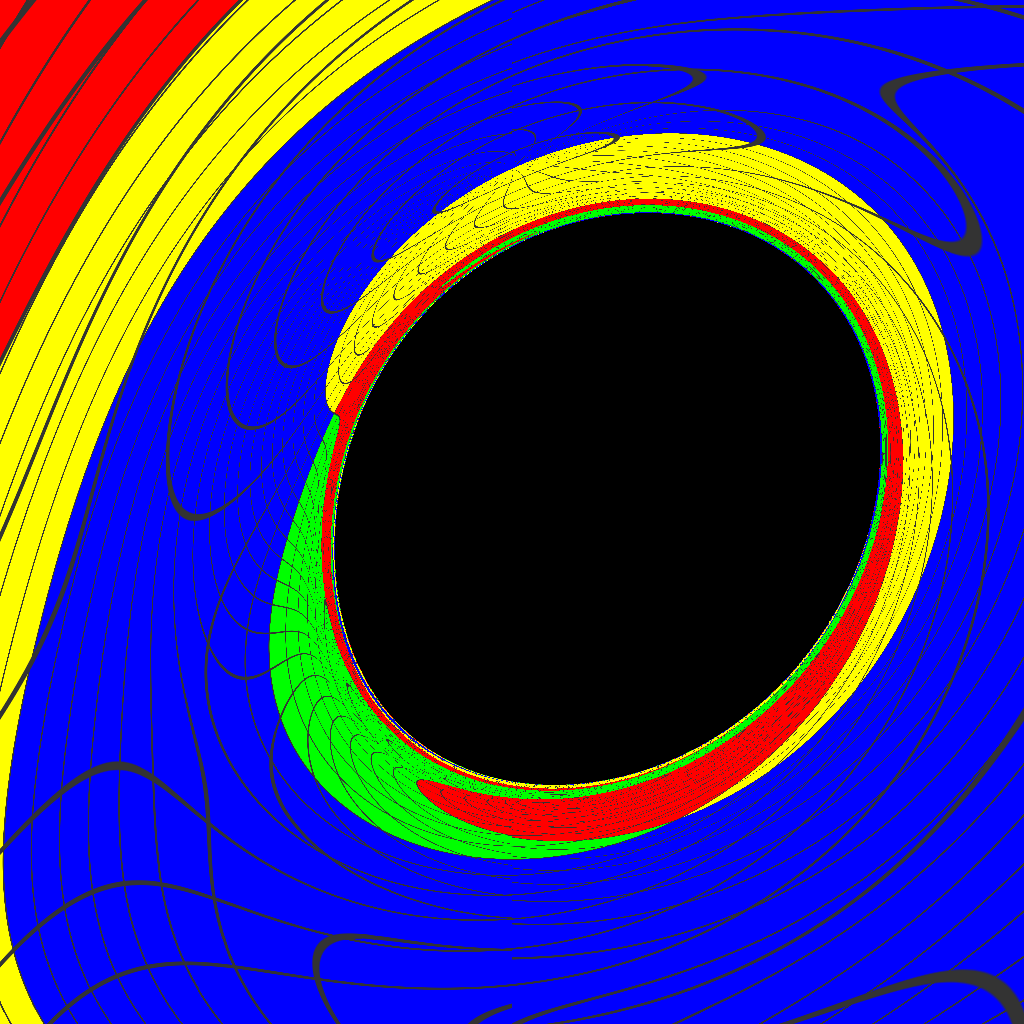}
        \caption{$\theta=\pi/4$}
    \end{subfigure} 
    \begin{subfigure}[t]{0.2\textwidth}
       \centering       \includegraphics[width=\textwidth]{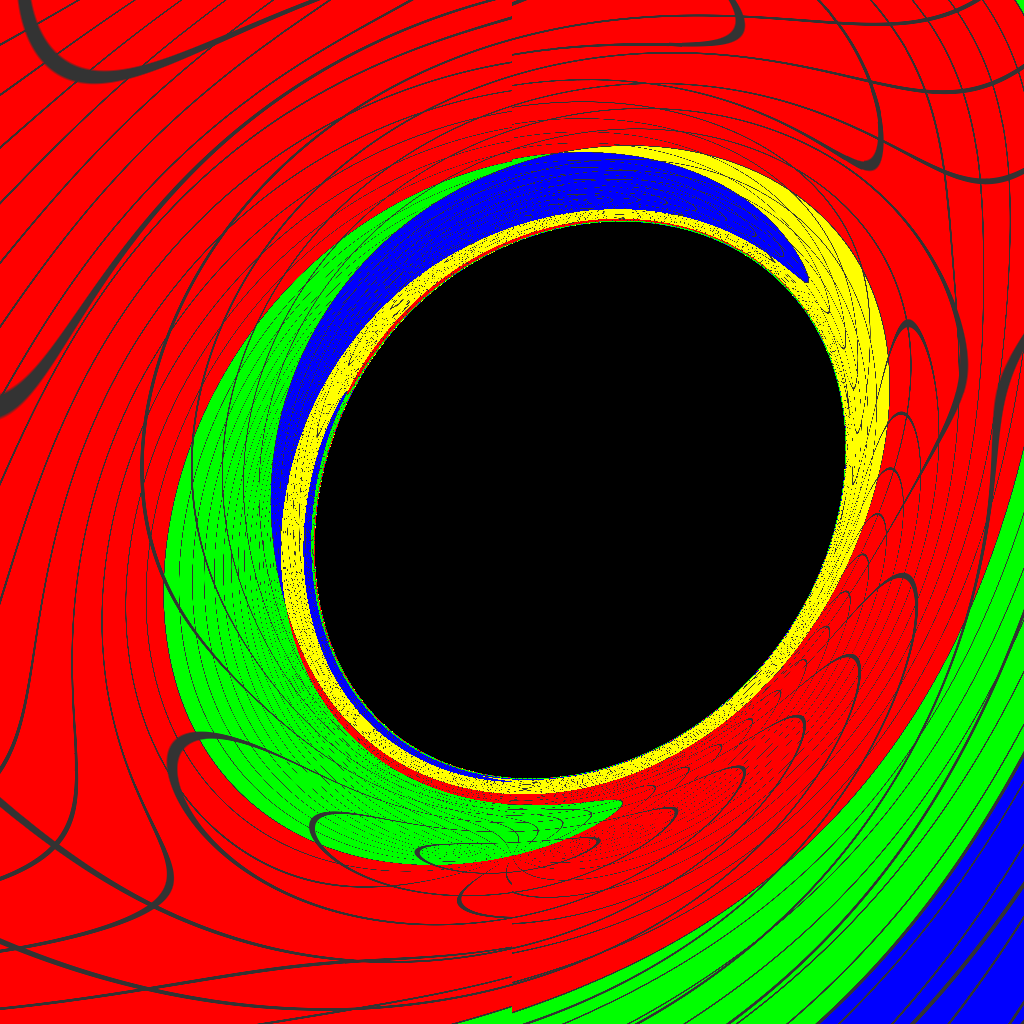}
      \caption{$\theta=3\pi/4$}     
    \end{subfigure}
    \begin{subfigure}[t]{0.2\textwidth}
       \centering
        \includegraphics[width=\textwidth]{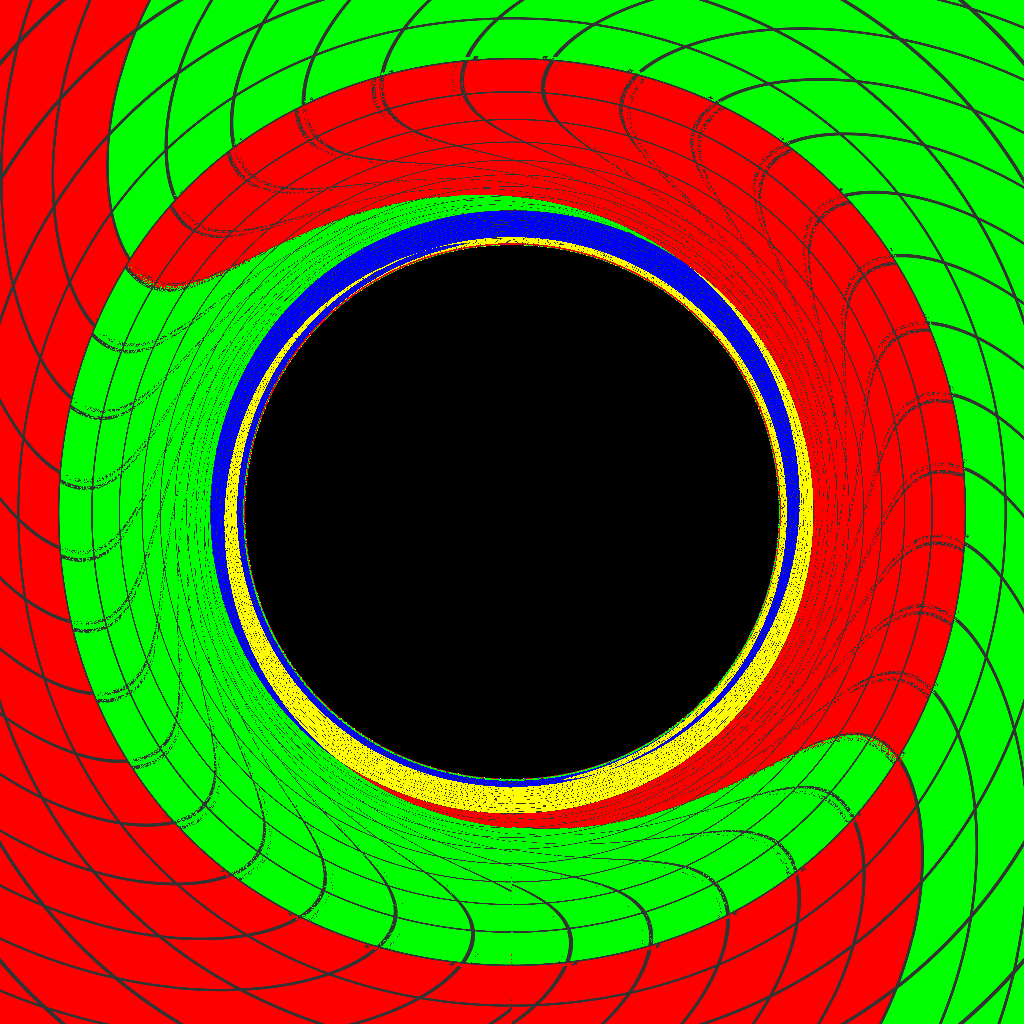}
        \caption{$\theta = \pi - 10^{-8}$}
    \end{subfigure}

    \caption{We show the lensing images for a Kerr swirling spacetime with $a=0.9M$, $j=0.0008$ and different observer's angles $\theta$. 
    The observer is set at $r_o = 15$. The lensing image for the observer at $\pi/2$ is given in Fig. \ref{Lensing_a_0.9_equatorial}
    }
    \label{fig:pyhole_a_09_j_0_008_off_equatorial}
\end{figure}

In the pure-Kerr case, since photons travelling in the direction of the frame dragging can orbit closer to the BH than photons travelling opposite to the frame dragging, this leads to a shadow that looks offset relative to the non-rotating case \cite{Bohn:2014xxa}. This effect is clearly observed for larger values of $a$. 
Once a non-zero value of the swirling parameter is considered, the shadow becomes \textit{twisted}, as already observed in the Schwarzschild-swirling case \cite{Moreira:2024sjq}. 
To the first set of our images, we place the observer on the equatorial plane ($\theta_o=\pi/2$) at $r_o = 15$, and fixed the BH rotational parameter while the swirling parameter varies in the range $\{ 0, 10^{-3} \}$, those are shown in Fig. \ref{Lensing_a_0.9_equatorial}. 
Within the considered range of the swirling parameter, we see that the BH shadow becomes more prolate and twisted as $j$ increases. 
We note, however, that the presence of both the angular momentum parameter $a$ and the swirling parameter $j$ breaks the odd $\mathbb{Z}_2$ symmetry present in the SBHSU case for observers in the equatorial plane.
For non-equatorial observers, the spacetime asymmetry is less attenuated in our images. The images for an off-equatorial observer are shown in Fig. \ref{fig:pyhole_a_09_j_0_008_off_equatorial} for $a=0.9M$, $j=0.0008$, $r_o=15$ and different values of $\theta_o$.

\begin{figure}[h!]
    \centering
 \begin{subfigure}[t]{0.2\textwidth}
        \centering
        \includegraphics[width=\textwidth]{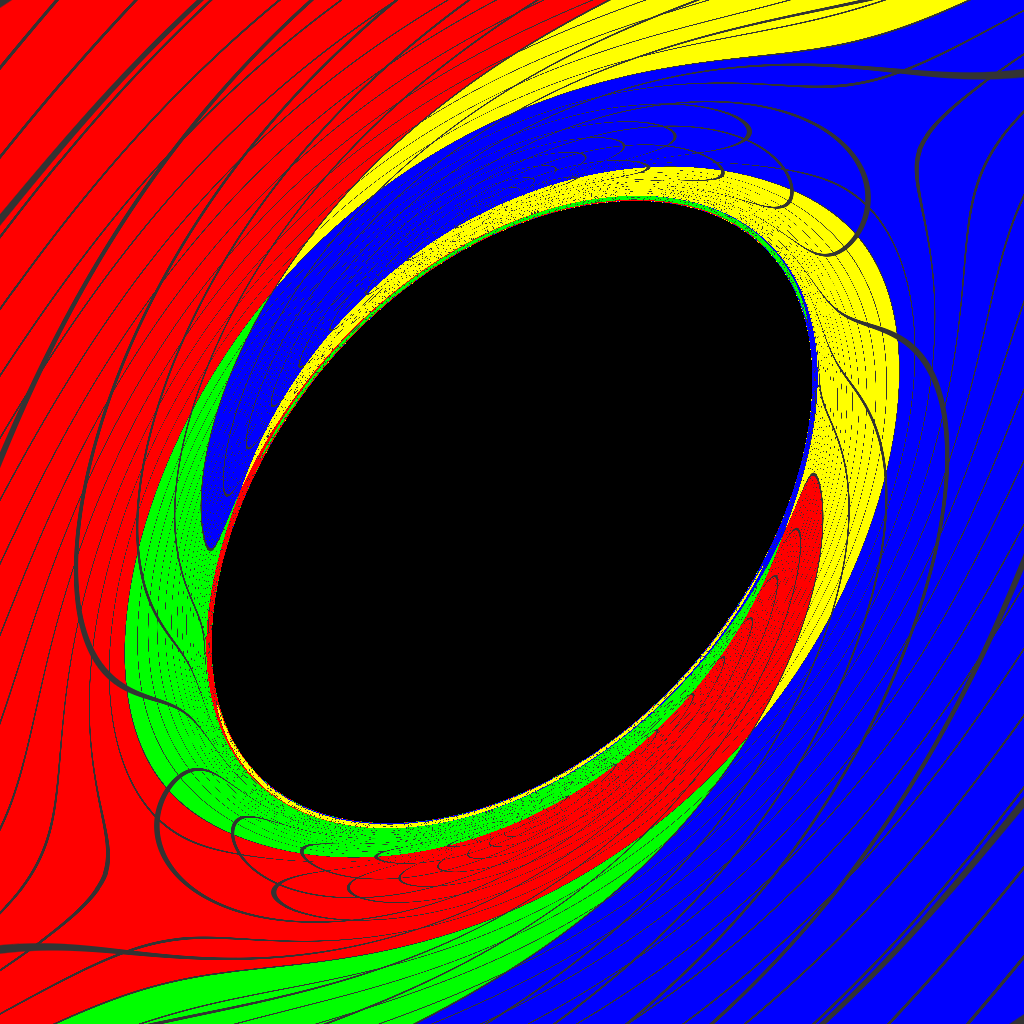}
        \caption{$ a = 0.0 $}
    \end{subfigure}
        \begin{subfigure}[t]{0.2\textwidth}
       \centering       \includegraphics[width=\textwidth]{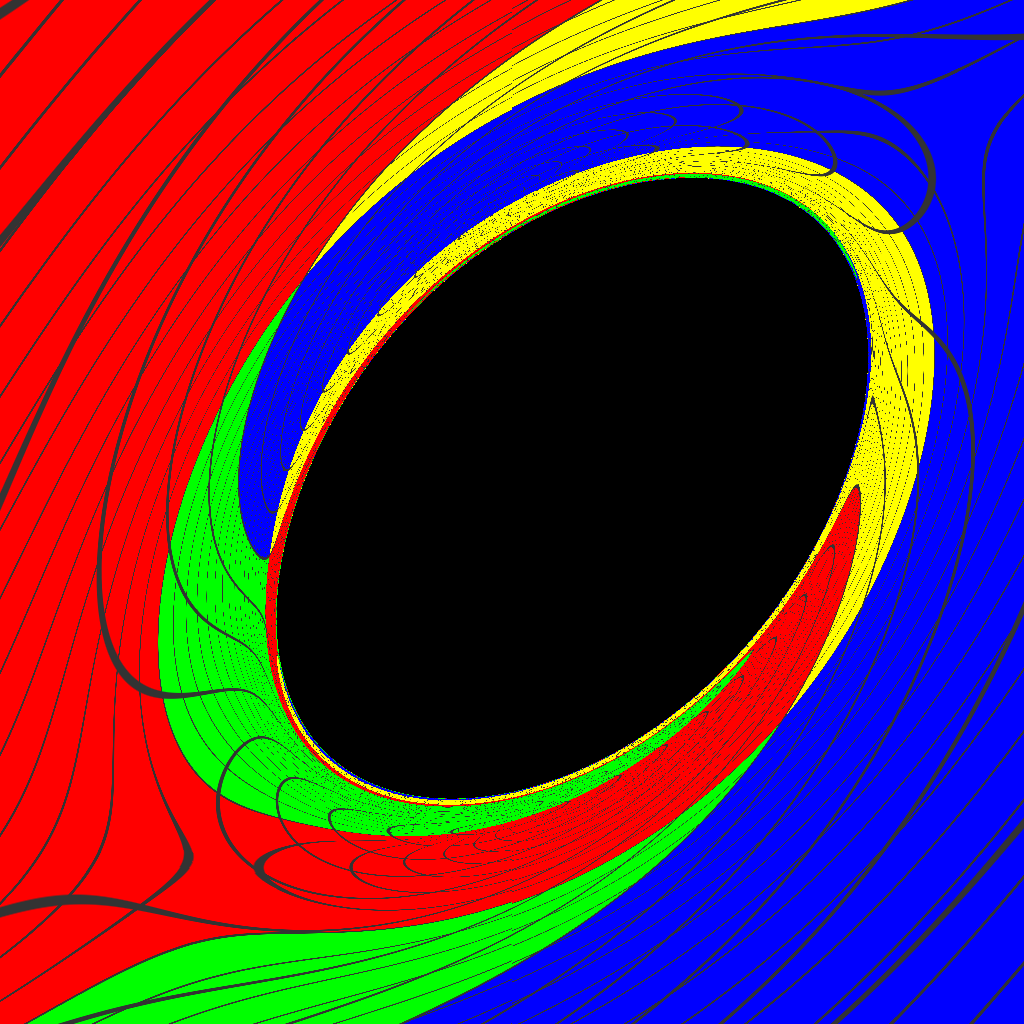}
    \caption{$a=0.5M$}     
    \end{subfigure}
    \begin{subfigure}[t]{0.2\textwidth}
       \centering
        \includegraphics[width=\textwidth]{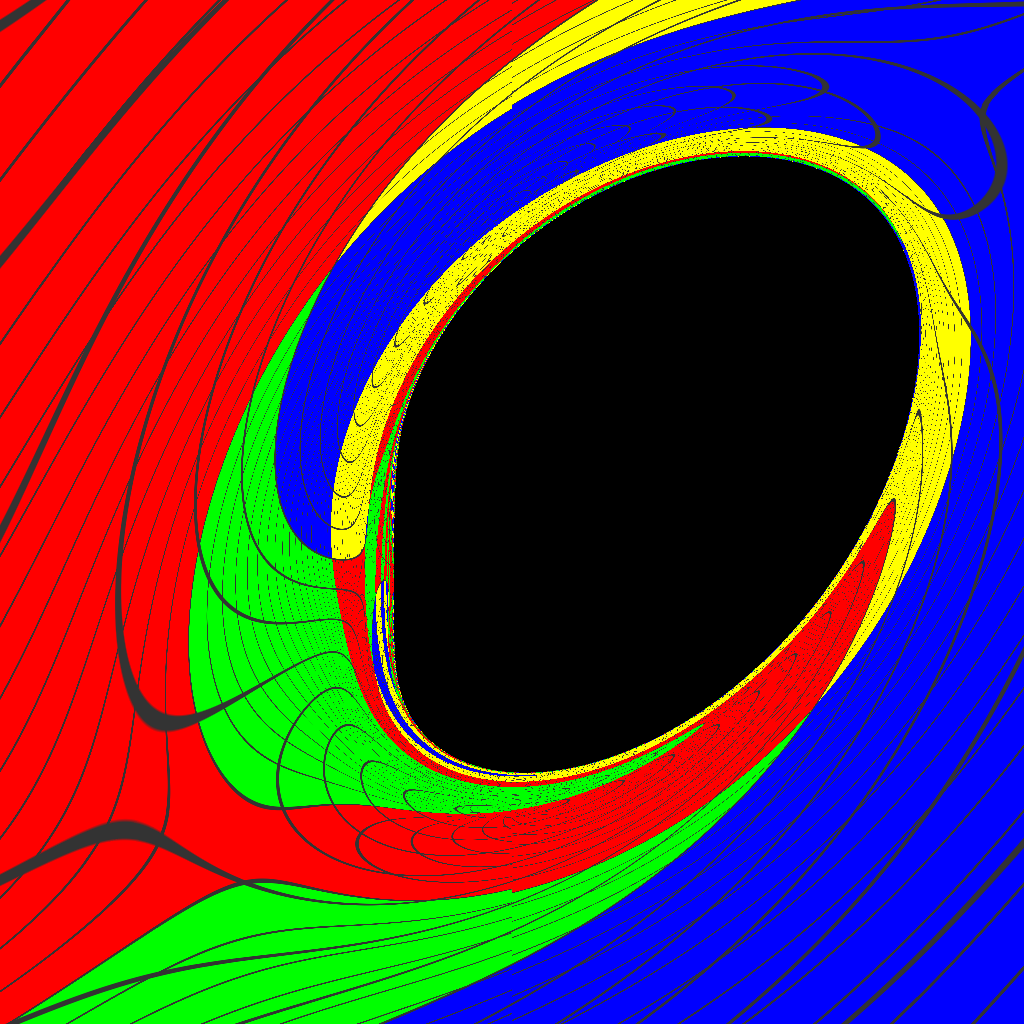}
        \caption{$a = 0.999M$}
        \label{offset_shadow_near_extreme}
    \end{subfigure}

    \caption{We show the lensing images for a Schwarzschild/Kerr swirling spacetime with a fixed value of $j=0.001$ and different values of the angular momentum parameter, $a$. 
    The observer is set on the equatorial plane at $r_o = 15$, $\theta_o=\pi/2$.
    }
    \label{fig:pyhole_comparison_a_fixed-j}
\end{figure}

Finally, we consider the effect of the Kerr rotational parameter $a$. To illustrate the effect, in Fig. \ref{fig:pyhole_comparison_a_fixed-j} we show the shadows for black holes with different rotational parameters $a$ while keeping the swirling parameter fixed at $j=0.001$. Again, we have chosen $r_o=15$ and $\theta_o=\pi/2$. As the value of $a$ increases, the shadow becomes more asymmetric, already seen in the pure Kerr case, stands still once the swirling background is considered; this effect is attenuated for slowly rotating BHs and thus is easily visualized for rapidly rotating compact objects. To demonstrate this, we show the lensing effect of a BH approaching extremality in Fig. \ref{offset_shadow_near_extreme}.   



\section{Conclusions}

The spacetime of a Kerr black hole in a swirling universe is a closed-form solution in General Relativity, obtained by well-known transformations from the Kerr black hole seed solution \cite{Astorino:2022aam}.
Unlike the Schwarzschild black holes in a swirling universe, however, the interaction between the black hole spin and the swirling parameter does not yield \sout{a} \bh{an} (odd) $\mathbb{Z}_2$ symmetry of the spacetime.
Thus the upper and lower hemispheres are not related by a simple symmetry transformation.

Here we have investigated some properties of this spacetime. 
First, we have recalled the conical singularities, which diverge at a critical value of the modulus of the parameter combination $ajM=0.25$, where both spin parameters enter. We have computed the Kretschmann scalar and demonstrated that this conical singularity is a physical singularity.
Next we have addressed the ergoregions.
These also exhibit a notable change in the vicinity of the critical value.
Whereas for small $ajM$, upper and lower hemispheres possess disconnected counter-rotating non-compact ergoregions in addition to the common ergosphere of the Kerr black hole, this changes at the critical value $ajM=0.25$.
For larger values of $ajM$, the three disconnected regions are merged to become one infinitely extended ergoregion with hence only one sense of direction of rotation.

We have extensively studied the light rings in the KBHSU spacetimes.
For the SBHSU spacetime the degenerate Schwarzschild LRs split up in a symmetrical way with increasing swirling parameter and move towards the horizon and the poles.
For the KBHSU, the symmetry is broken and the two LRs now possess different radii and no longer have angular variables $\theta$ and $\pi-\theta$.  Using a topological argument, we have demonstrated that two light rings should be present and that both are unstable. Furthermore, an interesting feature appears as the value of the swirling parameter is increased: there is a \textit{light point}. This is a special regime in which a light ring exists exactly as the two sections of ergoregions merge; this LR does not possess a rotational sense. To the author's knowledge, this is the first report of such a class of orbit.

Finally, we investigated the shadow of KBHSU spacetimes.
Since the solution is Petrov type I and hence the equations describing geodesic motion are not separable, we had to employ \sout{fully} numerical methods to obtain the shadows seen by observers outside the black hole.
The SBHSU shadows exhibit an odd-$\mathbb{Z}_2$ symmetry for observers in the equatorial plane.
The interaction between the spins of the KBHSU spacetime does not allow for such odd-$Z_2$ symmetric shadows.
All shadows are twisted, though, which would make the appearance of such a black hole most unusual.

While the spacetime may not appear very realistic given the well-known features of our universe, it may still have features that might be of relevance.
Here we would like to point out that the universe is threaded by huge structures, cosmic filaments, that rotate and thus might be modeled in part by the swirling universe solutions.
Moreover, Astorino et al.~\cite{Astorino:2022aam} considered the swirling universe solutions as possibly relevant in the description of the collision of rotating galaxies.
Finally, we would like to mention that it has recently been proposed that a rotating universe might solve the Hubble tension \cite{Szigeti:2025jxz}. 

The KBHSU is also interesting from another point of view~: it is an example of a spacetime in which test particle motion is chaotic \cite{Cao:2024pdb}. It would hence be an ideal system to quantify and qualify chaotic behaviour. This could lead to a better understanding of the underlying connection between chaotic test particle motion and the algebraic classification of solutions generated via suitable transformations in the Ernst formalism.\\
\\
\\
{\bf Acknowledgments}:  R.C. would like to thank CAPES for financial support under Grant No: 88887.371717/2019-00.  J.N. is supported by the FCT grant 10.54499/2021.06539.BD (https://doi.org/10.54499/2021.06539.BD).

\newpage

\end{document}